\PassOptionsToPackage{unicode}{hyperref}
\PassOptionsToPackage{hyphens}{url}
\PassOptionsToPackage{dvipsnames,svgnames,x11names}{xcolor}
\documentclass[
  12pt]{article}
\usepackage{url}
\usepackage{algorithm}
\usepackage{algpseudocode}
\usepackage{amsmath,amssymb}
\usepackage{comment}
\usepackage{bbm}
\usepackage{iftex}
\ifPDFTeX
  \usepackage[T1]{fontenc}
  \usepackage[utf8]{inputenc}
  \usepackage{textcomp} 
\else 
  \usepackage{unicode-math}
  \defaultfontfeatures{Scale=MatchLowercase}
  \defaultfontfeatures[\rmfamily]{Ligatures=TeX,Scale=1}
\fi
\usepackage{lmodern}
\ifPDFTeX\else  
\fi
\IfFileExists{upquote.sty}{\usepackage{upquote}}{}
\IfFileExists{microtype.sty}{
  \usepackage[]{microtype}
  \UseMicrotypeSet[protrusion]{basicmath} 
}{}
\makeatletter
\@ifundefined{KOMAClassName}{
  \IfFileExists{parskip.sty}{%
    \usepackage{parskip}
  }{
    \setlength{\parindent}{0pt}
    \setlength{\parskip}{6pt plus 2pt minus 1pt}}
}{
  \KOMAoptions{parskip=half}}
\makeatother
\usepackage{xcolor}
\makeatletter
\ifx\paragraph\undefined\else
  \let\oldparagraph\paragraph
  \renewcommand{\paragraph}{
    \@ifstar
      \xxxParagraphStar
      \xxxParagraphNoStar
  }
  \newcommand{\xxxParagraphStar}[1]{\oldparagraph*{#1}\mbox{}}
  \newcommand{\xxxParagraphNoStar}[1]{\oldparagraph{#1}\mbox{}}
\fi
\ifx\subparagraph\undefined\else
  \let\oldsubparagraph\subparagraph
  \renewcommand{\subparagraph}{
    \@ifstar
      \xxxSubParagraphStar
      \xxxSubParagraphNoStar
  }
  \newcommand{\xxxSubParagraphStar}[1]{\oldsubparagraph*{#1}\mbox{}}
  \newcommand{\xxxSubParagraphNoStar}[1]{\oldsubparagraph{#1}\mbox{}}
\fi
\makeatother

\usepackage{longtable,multirow,booktabs,array}
\usepackage{calc} 
\usepackage{etoolbox}
\makeatletter
\patchcmd\longtable{\par}{\if@noskipsec\mbox{}\fi\par}{}{}
\makeatother
\IfFileExists{footnotehyper.sty}{\usepackage{footnotehyper}}{\usepackage{footnote}}
\makesavenoteenv{longtable}
\usepackage{graphicx}
\makeatletter
\def\maxwidth{\ifdim\Gin@nat@width>\linewidth\linewidth\else\Gin@nat@width\fi}
\def\maxheight{\ifdim\Gin@nat@height>\textheight\textheight\else\Gin@nat@height\fi}
\makeatother
\setkeys{Gin}{width=\maxwidth,height=\maxheight,keepaspectratio}
\makeatletter
\def\fps@figure{htbp}
\makeatother

\makeatletter
\@ifpackageloaded{caption}{}{\usepackage{caption}}
\AtBeginDocument{%
\ifdefined\contentsname
  \renewcommand*\contentsname{Table of contents}
\else
  \newcommand\contentsname{Table of contents}
\fi
\ifdefined\listfigurename
  \renewcommand*\listfigurename{List of Figures}
\else
  \newcommand\listfigurename{List of Figures}
\fi
\ifdefined\listtablename
  \renewcommand*\listtablename{List of Tables}
\else
  \newcommand\listtablename{List of Tables}
\fi
\ifdefined\figurename
  \renewcommand*\figurename{Figure}
\else
  \newcommand\figurename{Figure}
\fi
\ifdefined\tablename
  \renewcommand*\tablename{Table}
\else
  \newcommand\tablename{Table}
\fi
}
\@ifpackageloaded{float}{}{\usepackage{float}}
\@ifpackageloaded{placeins}{}{\usepackage[section]{placeins}}
\floatstyle{ruled}
\@ifundefined{c@chapter}{\newfloat{codelisting}{h}{lop}}{\newfloat{codelisting}{h}{lop}[chapter]}
\floatname{codelisting}{Listing}

\makeatother
\makeatletter
\@ifpackageloaded{caption}{}{\usepackage{caption}}
\@ifpackageloaded{subcaption}{}{\usepackage{subcaption}}
\makeatother

\ifLuaTeX
  \usepackage{selnolig}  
\fi

\usepackage[]{natbib}
\bibpunct{(}{)}{;}{a}{}{,} 

\usepackage{bibunits}
\defaultbibliographystyle{accepo-nodoi}
\defaultbibliography{bibliography} 

\usepackage{bookmark}

\makeatletter
\renewcommand{\@extra@b@citeb}{\@bibunitname}
\renewcommand{\@extra@binfo}{\@bibunitname}
\makeatother

\IfFileExists{xurl.sty}{\usepackage{xurl}}{} 
\hypersetup{
  pdftitle={Two-Stage Randomization with a Learned Balancing Score: Efficiency and Robust Inference},
  pdfauthor={Siyu Zhang; Wei Ma; Jiahui Xin},
  pdfkeywords={Difference-in-means estimator; sequential allocation; learned balancing target; model misspecification; prior-data fitted networks.},
  colorlinks=true,
  linkcolor={blue},
  filecolor={Maroon},
  citecolor={Blue},
  urlcolor={Blue},
  pdfcreator={LaTeX via pandoc}}

\usepackage{etoc}

\usepackage{amsthm}

\theoremstyle{plain}
\newtheorem{theorem}{Theorem}[section]
\newtheorem{proposition}{Proposition}[section]
\newtheorem{corollary}{Corollary}[section]
\newtheorem{lemma}{Lemma}[section]

\theoremstyle{plain}
\newtheorem{assumption}{Assumption}

\theoremstyle{definition}

\theoremstyle{remark}
\newtheorem{remark}{Remark}[section]

\newcommand{\Var}{\mathrm{Var}}

\newcommand{\anon}{1}

\begin{document}
\etocdepthtag.toc{main}

\def\spacingset#1{\renewcommand{\baselinestretch}%
{#1}\small\normalsize} \spacingset{1}

\if1\anon
{
  \title{\bf Two-Stage Randomization with a Learned Balancing Score: Efficiency and Robust Inference}
  \author{
    Siyu Zhang$^1$,\quad Wei Ma$^1$,\quad Jiahui Xin$^2$\thanks{Corresponding author. Email: jxin@mdanderson.org
    }
    \\
    $^1$ Institute of Statistics and Big Data, Renmin University of China, China\\
    $^2$ Department of Biostatistics, The University of Texas\\MD Anderson Cancer Center, Texas, USA
  }
  \maketitle
} \fi

\if0\anon
{
  \bigskip
  \bigskip
  \bigskip
  \begin{center}
    {\LARGE\bf Two-Stage Randomization with a Learned Balancing Score: Efficiency and Robust Inference}
\end{center}
  \medskip
} \fi

\bigskip

\begin{abstract}
In clinical trials and other comparative studies, covariate balance is central to credible and efficient treatment effect estimation.
We consider the two-arm randomization and focus on a scalar balancing target, which is the \textbf{A}verage of the two \textbf{C}entered \textbf{C}onditional \textbf{E}xpectations of \textbf{P}otential \textbf{O}utcomes (ACCEPO).
When the imbalance of ACCEPO is reduced to $o_p(\sqrt{n})$, where $n$ is the sample size, the difference-in-means estimator achieves the optimal precision ``by design.''
However, ACCEPO is often unknown before the trial.
We therefore propose a two-stage design that learns it from the first stage data and uses the learned balancing score to guide sequential biased-coin allocation in the second stage.
Under $L_2$-consistent learning in the first stage, the second stage difference-in-means estimator achieves the oracle variance.
We further establish robust inference without requiring learning consistency, which supports our empirical use of tabular foundation models as learners.
Simulation studies and two clinical trial applications illustrate the performance of the proposed method.
\end{abstract}

\noindent%
{\it Keywords:} Difference-in-means estimator; sequential allocation; learned balancing target; model misspecification; prior-data fitted networks.
\vfill

\newpage
\spacingset{1.8} 

\begin{bibunit}

\section{Introduction}
\label{sec:introduction}
Covariate-adaptive randomization (CAR) procedures have been widely used in clinical trials to improve covariate balance by incorporating baseline covariate information into treatment assignment \citep{HuEtAl2014AdaptiveRandomization,RosenbergerLachin2016Randomization}.
The main objectives are to enhance treatment group comparability and improve treatment effect estimation precision.

For continuous covariates, balance is typically achieved by minimizing a prespecified imbalance measure \citep{NishiTakaichi2004,MaHu2013,JiangMaYin2018Kernel,LiZhouHu2019Continuous,QinEtAl2024ARM}.
Recently, \citet{MaJASA2024CAR} proposed a unified CAR framework that accommodates many existing CAR procedures and allows randomization to balance general functions of the covariates, represented by a prespecified feature map $\phi(X)$.
More importantly, they identify a scalar balancing target, which is the \textbf{A}verage of the two \textbf{C}entered \textbf{C}onditional \textbf{E}xpectations of \textbf{P}otential \textbf{O}utcomes (ACCEPO), denoted by $m(X)$.
When the design reduces the imbalance of $m(X)$ to $o_p(\sqrt{n})$, where $n$ is the sample size, the na\"{i}ve difference-in-means estimator achieves the optimal asymptotic variance, which takes the form of the semiparametric efficiency bound in \citet{Hahn1998}.
The same efficiency could be attained under alternative treatment assignment through \emph{ex post} nonparametric covariate adjustment \citep{rafi2023,TuMaLiu2024UnifiedAdjustment}.
In contrast, balancing ACCEPO achieves optimal precision ``by design,'' preserving the simplicity and transparency of the estimator.

However, $m(X)$ is unknown in practice, and balancing prespecified covariate features may be inadequate for controlling the imbalance of $m(X)$.
To address this, we propose a two-stage design that learns $m(X)$ from the first stage data and uses the learned balancing score to guide sequential biased-coin allocation \citep{efronForcingSequentialExperiment1971} in the second stage.

A further challenge is that $L_2$ consistency of the learned score may not be guaranteed for some learners, such as prior-data fitted networks (PFNs). PFNs were proposed by \citet{muller2022transformers} as a new machine learning paradigm. They amortize the computation of approximate Bayesian posterior predictive inference through pretraining, and make predictions on a new dataset through in-context learning, without updating parameters. Recent PFN models have achieved state-of-the-art performance on tabular prediction benchmarks \citep{HollmannEtAl2025TabPFN,ZhangEtAl2025TabPFN,qu2026tabiclv2}. However, no $L_2$ consistency guarantee has been established for PFNs, especially with fixed pretrained weights. 

In summary, it remains elusive whether valid inference can still be achieved when $m(X)$ is unknown and even consistent learning of $m(X)$ cannot be guaranteed. Within our two-stage framework, we provide an affirmative answer to this question by establishing valid inference both under learning consistency and without this requirement. 
Our contributions are summarized as follows:
\begin{enumerate}
\item[1)] Methodologically, we propose a two-stage design that learns $m(X)$ from first stage data and uses the learned balancing score to guide sequential biased-coin allocation in the second stage. Under $L_2$-consistent learning and suitable regularity conditions, the second stage difference-in-means estimator attains the oracle variance. By targeting a function directly related to estimation precision, our method achieves optimal precision ``by design.'' We further develop two extensions: pooled estimation and a multi-stage score updating design.
\item[2)] Theoretically, we conduct robust inference without requiring the learned balancing score to converge to $m(X)$ or any other deterministic function under the same regularity conditions. Using a martingale representation, we establish standardized asymptotic normality for the ATE estimator, with the asymptotic variance equal to the oracle variance plus four times the conditional mean squared difference between the $m(X)$ and a rescaled learned score, given the first stage data. We further provide a consistent closed-form variance estimator for this variance, addressing the variance estimation difficulty discussed in \citet[Remark~4.4]{MaJASA2024CAR}. 
Our conditional triangular-array approach establishes both the oracle efficiency and robust inference results without restricting the relative growth rates of the two stage sample sizes. Under $L_2$-consistent learning, using only the second stage data for ATE estimation incurs no first-order efficiency loss when the first stage sample proportion vanishes. The robust inference results hold even with a fixed first stage sample size.
\item[3)] Empirically, we implement the design using prior-data fitted networks (PFNs), with first stage data as context and pretrained weights held fixed. Through simulation studies and semi-synthetic analyses based on two clinical trials, we demonstrate that the PFN-based two-stage designs achieve precision gains over simple randomization and procedures that balance prespecified covariate moments, especially in nonlinear settings. Confidence intervals constructed using the variance estimator developed for robust inference achieve empirical coverage close to the nominal level, supporting the practical use of flexible learners without established consistency guarantees.
\end{enumerate} 

Three important studies are related to our work.
\citet{hahnAdaptiveExperimentalDesign2011} proposed a two-stage design that uses first stage data to estimate the optimal propensity score for treatment assignment in the second stage. 
Their framework relies on discretization of continuous covariates, whereas ours accommodates them directly.
\citet{tabordmeehanStratificationTreesAdaptive2023} uses the first stage data to learn the optimal stratification rule and treatment allocation proportions within strata, where the strata are represented as decision trees.
In contrast, our framework balances a learned score without requiring stratification and accommodates a broader class of learners.
\citet{Bai2022OptimalMatchedPairs} establishes the optimality of a matched-pair design that orders units by an oracle index function and pairs adjacent units. 
The oracle index function equals $2m(X)$ plus a constant, thus sorting units by this index function is equivalent to doing so using our $m(X)$.
However, matched-pair designs require covariate information for all units before randomization, which is not the case when units arrive sequentially and require immediate treatment assignment. In contrast, our proposed procedure assigns treatments adaptively, thus is more feasible in such settings.

The remainder of this paper is organized as follows. Section~\ref{sec:preliminaries} sets up the notation and gives a brief motivation discussion. Section~\ref{sec:baselineframework} describes the two-stage randomization framework. Section~\ref{sec:MainTheoreticalProperties} establishes theoretical results both with and without learning consistency. Section~\ref{sec:extensions} studies two extensions: pooled estimation and multi-stage score updating design. Sections~\ref{sec:simulations} and~\ref{sec:clinical trial example} report simulation studies and two clinical trial examples. Section~\ref{sec:Conclusion} summarizes the study and provides directions for future work. Implementation details, additional numerical results, and technical proofs are provided in the Supplementary Material.

\section{Preliminaries}
\label{sec:preliminaries}
We adopt the Neyman--Rubin potential outcomes framework to define the treatment effect. For each unit $i=1,\ldots,n$, let $Y_i(1)$ and $Y_i(0)$ denote the potential outcomes under treatment and control, $T_i\in\{0,1\}$ the treatment assignment, and $X_i\in\mathbb R^p$ the baseline covariates. The observed outcome is $Y_i=T_iY_i(1)+(1-T_i)Y_i(0)$, and the average treatment effect is $\tau=E\{Y_i(1)\}-E\{Y_i(0)\}$.

We study treatment effect estimation under a general potential outcomes model throughout this paper. For $a\in\{0,1\}$, denote the conditional expectations of potential outcomes (CEPO) by $\mu_a(X_i)=E\{Y_i(a)|X_i\}$, its centered version $m_a(X_i)=\mu_a(X_i)-E\{Y_i(a)\}$ and the residual $\varepsilon_{ia}=Y_i(a)-E\{Y_i(a)|X_i\}$. 
And we let
\begin{equation*}
\label{equa:define_mx}
    m(X_i) = \frac{m_1(X_i)+m_0(X_i)}{2}
\end{equation*}
denote the average of the two centered conditional expectations of potential outcomes (ACCEPO).

To estimate $\tau$, we consider the difference-in-means estimator
\begin{equation*}\label{eq:diff_in_means}
  \widehat{\tau} = \bar{Y}_1 - \bar{Y}_0=\frac{\sum_{i=1}^n T_i Y_i}{N_{1}}\;-\;\frac{\sum_{i=1}^n (1-T_i) Y_i}{N_{0}},
\end{equation*}
where $N_{1}=\sum_{i=1}^n T_i$, $N_{0}=\sum_{i=1}^n (1-T_i).$ We wish to provide a treatment assignment procedure with minimal asymptotic variance using the na\"{i}ve difference-in-means estimator without \emph{ex post} covariate adjustment at the analysis stage.

The sequential CAR framework of \citet{MaJASA2024CAR} balances prespecified covariate features $\phi(X_i)$, defined by a feature map $\phi(X_i):\mathbb R^p\to\mathbb R^q$ that maps $X_i$ into a $q$-dimensional feature space. 
Define the imbalance vector and the imbalance measure of $\phi(X)$ by 
$$\Delta_n(\phi)=\sum_{i=1}^n(2T_i-1)\phi(X_i),\qquad \mathrm{Imb}_n(\phi)=\|\Delta_n(\phi)\|_2^2,$$
where $\|\cdot\|_2$ is the Euclidean norm.

Their design assigns the first unit with equal probability to the treatment or the control, when unit $i\ge 2$ arrives,  
the design compares the two potential imbalance measures $$\mathrm{Imb}_i^{(1)}(\phi)=\|\Delta_{i-1}(\phi)+\phi(X_i)\|_2^2,\qquad \mathrm{Imb}_i^{(0)}(\phi)=\|\Delta_{i-1}(\phi)-\phi(X_i)\|_2^2,$$ corresponding to assigning that unit to treatment or control, respectively, 
and then assigns that unit to the arm that yields the smaller imbalance with probability $\rho\in(1/2,1]$ using Efron's biased-coin design \citep{efronForcingSequentialExperiment1971}.

The next proposition restates that, under their CAR framework, the only component of the difference-in-means estimator's asymptotic variance that can be reduced by choosing a suitable feature map $\phi(X)$ at the design stage arises from the imbalance of ACCEPO.
\begin{proposition}
\label{prop:MaJASA}
    For a fixed prespecified feature map $\phi(X)$, under the moment and irreducibility conditions of \citet{MaJASA2024CAR}, then $n^{-1/2}\Delta_n(\phi)=o_p(1)$. The difference-in-means estimator satisfies $\sqrt n(\widehat{\tau}-\tau)\xrightarrow{d}N(0,\sigma^2)$, where
    \begin{equation}
        \sigma^2= 2\Var(\varepsilon_{i1}) + 2\Var(\varepsilon_{i0}) + \Var\{m_1(X_i)-m_0(X_i)\} + 4\sigma_m^2.
        \label{eq:Ma_var_decomp}
    \end{equation}
    Here, $\sigma_m^2\ge0$ is defined by the limiting distribution of the normalized ACCEPO imbalance:
    \begin{equation*}
    \frac{1}{\sqrt{n}}\sum_{i=1}^n (2T_i-1)m(X_i)\;\xrightarrow{d}\; \mathcal{N}(0,\sigma_m^2),
    \label{eq:CLT_m}
    \end{equation*}
\end{proposition}

The first three terms in \eqref{eq:Ma_var_decomp} depend only on the outcome model and are invariant to the specific covariate adaptive randomization scheme or feature map $\phi(X)$, whereas $4\sigma_m^2$ measures the remaining contribution of ACCEPO imbalance, and hence the {\em only} part of the variance that can be improved by specifying the $\phi(X)$ at the design stage.

This motivates our design principle. If one could choose the feature map $\phi(X)$ to minimise the variability of $\sum_{i=1}^n(2T_i-1)m(X_i)$, in the favourable case where $m(X)\in\mathrm{span}\{\phi(X)\}$, then according to Proposition~\ref{prop:MaJASA}, $\sum_{i=1}^n (2T_i-1)m(X_i) = o_p(\sqrt{n})$ and $\sigma_m^2=0$, so \eqref{eq:Ma_var_decomp} reduces to the oracle variance
\begin{equation}
    V^\ast := 2\Var(\varepsilon_{i1}) + 2\Var(\varepsilon_{i0})
       + \Var\{m_1(X_i)-m_0(X_i)\},
    \label{eq:V_star}
\end{equation}
which is the minimal asymptotic variance attainable within this framework, and this variance takes the form of the semiparametric efficiency bound of \citet{Hahn1998} with equal treatment allocation. 

In practice, $m(X)$ is unknown and may contain nonlinearities and interactions that are difficult to capture with a prespecified feature map. 
We therefore propose a two-stage design, 
we use the first stage data to learn $m(X)$ and use the learned balancing score $\widehat{m}_{n_0}(X)$ to guide sequential biased-coin allocation in the second stage. 
Section~\ref{sec:baselineframework} defines the procedure. 

As discussed in \citet[Remark~4.4]{MaJASA2024CAR}, the construction of a sample analog variance estimator is not straightforward since $\sigma_m^2$ lacks a closed-form expression.
Within our two-stage framework, however, we establish consistent variance estimation both under learning consistency (Theorem~\ref{theo:plugin_consistency}) and without this requirement (Theorem~\ref{theo:general-score-inference}), with the latter analysis relying on a martingale representation.

\section{Two-stage Randomization Framework}
\label{sec:baselineframework}
Let $n_0$ and $n$ denote the sample sizes of the first stage and the second stage respectively, so that $N=n_0+n$ is the total sample size. We index the first stage units by $j=1,\ldots,n_0$ and second stage units by $i=1,\ldots,n$. 
Suppose that $(Y(1),Y(0),X)$ are i.i.d. draws from the same population across two stages; in particular, the second stage draws are independent of the first stage draws.

In the first stage, assign each unit independently to treatment with equal probability. After observing the first stage outcomes, we collect $\mathcal{D}_{n_0}=\{(Y_j,T_j,X_j)\}_{j=1}^{n_0}$. For $a\in\{0,1\}$, define the first stage arm size and outcome mean by
$$N_{n_0,a}=\sum_{j=1}^{n_0}\mathbf{1}\{T_j=a\},\qquad
\bar{Y}_a^{(1)}=\frac{1}{N_{n_0,a}}\sum_{j=1}^{n_0}\mathbf{1}\{T_j=a\}Y_j.$$

Then we use the first stage data $\mathcal{D}_{n_0}$ to estimate the conditional mean functions $\mu_a(x)=\mathbb{E}\{Y(a)\mid X=x\}$, with the resulting estimators denoted by $\widehat{\mu}_{n_0,a}(x)$.
We construct the learned balancing score by centering each fitted conditional mean function at its corresponding first stage arm mean $\bar{Y}_a^{(1)}$ and averaging the two centered functions:
$$\widehat{m}_{n_0,a}(x)=\widehat{\mu}_{n_0,a}(x)-\bar{Y}_a^{(1)}, 
\qquad 
\widehat{m}_{n_0}(x) =\frac{\widehat{m}_{n_0,1}(x)+\widehat{m}_{n_0,0}(x)}{2}.$$

In the second stage, we hold $\widehat{m}_{n_0}$ fixed and implement Efron’s biased-coin randomization based on the following imbalance measure:
\begin{equation*}
  \mathrm{Imb}_n(\widehat{m}_{n_0})=\|\Delta_n(\widehat{m}_{n_0})\|_2^2=\left\{\sum_{i=1}^n (2T_i-1)\,\widehat{m}_{n_0}(X_i)\right\}^2.
  \label{eq:Imb_mx}
\end{equation*}

The second stage sequential allocation proceeds as follows:
\begin{enumerate}
    \item[(a)] Randomly assign the first unit with equal probability to treatment or control, independently of its covariates.

    \item[(b)] Suppose that $(i-1)$ second stage units have been assigned (with $i>1$) and the $i$-th unit is to be allocated.  Compute the ``potential'' imbalance measures $\mathrm{Imb}_i^{(1)}(\widehat{m}_{n_0})$ and $\mathrm{Imb}_i^{(0)}(\widehat{m}_{n_0})$, corresponding to $T_i=1$ and $T_i=0$, respectively:
        $$\mathrm{Imb}_i^{(1)}(\widehat{m}_{n_0})=\left\|\Delta_{i-1}(\widehat{m}_{n_0})+\widehat{m}_{n_0}(X_i)\right\|_2^2, 
        \quad 
        \mathrm{Imb}_i^{(0)}(\widehat{m}_{n_0})=\left\|\Delta_{i-1}(\widehat{m}_{n_0})-\widehat{m}_{n_0}(X_i)\right\|_2^2.$$
        
    \item[(c)] Assign the $i$-th unit to the treatment with the probability
        \begin{equation*}
          \mathbb{P}(T_i = 1 \mid \mathcal{D}_{n_0},X_i,\ldots,X_1, T_{i-1},\ldots,T_1)=
            \begin{cases}
            \rho,   & \text{if } \mathrm{Imb}_i^{(1)}(\widehat{m}_{n_0})< \mathrm{Imb}_i^{(0)}(\widehat{m}_{n_0}),\\[0.35em]
            1-\rho, & \text{if } \mathrm{Imb}_i^{(1)}(\widehat{m}_{n_0})> \mathrm{Imb}_i^{(0)}(\widehat{m}_{n_0}),\\[0.35em]
            0.5,    & \text{if } \mathrm{Imb}_i^{(1)}(\widehat{m}_{n_0})= \mathrm{Imb}_i^{(0)}(\widehat{m}_{n_0}),
          \end{cases}
          \label{eq:EfronBCD}
        \end{equation*}
        where $0.5 < \rho < 1$ is a fixed biasing probability. 
        
   \item[(d)] Repeat the last two steps until all $n$ second stage units have been assigned.
\end{enumerate}

To estimate the average treatment effect $\tau$, we adopt the unpooled difference-in-means estimator based on the second stage data $\{(Y_i,T_i,X_i)\}_{i=1}^n$: 
\begin{equation}
\label{equ:tau_m}
    \widehat{\tau}(\widehat{m}_{n_0})= \bar{Y}_1 - \bar{Y}_0=\frac{\sum_{i=1}^n T_i Y_i}{N_{1}}\;-\;\frac{\sum_{i=1}^n (1-T_i) Y_i}{N_{0}},
\end{equation}
where $N_{1}=\sum_{i=1}^n T_i$, $N_{0}=\sum_{i=1}^n (1-T_i).$ 
The term ``unpooled'' means that the estimator uses only the second stage data; another ``pooled'' estimator that incorporates data from both stages is developed in Section~\ref{sec:pool-estimation}.

In our numerical studies, we use three pretrained foundation models \citep{HollmannEtAl2025TabPFN,balazadehCausalPFNAmortizedCausal2025,qu2026tabiclv2} to estimate $\mu_a(x)$, with the first stage data serving as context.
We denote the estimator as $\widehat{\mu}_{n_0,a}(x)=f_{\widehat{\theta},a}(x;\mathcal D_{n_0})$, where $\widehat{\theta}$ denotes the fixed pretrained parameters. 
A brief introduction to PFNs and details of our context construction are provided in Appendix~\ref{app:pfn-implementation}.

\begin{remark}
PROCOVA \citep{Schuler2022} uses historical control data to learn a prognostic score targeting $\mu_0(X)$, which is subsequently included as an additional covariate for analysis stage adjustment.
Our two-stage framework instead uses first stage data from both arms to learn ACCEPO and guide subsequent randomization.
The optimality result of \citet{Schuler2022} is established under a constant conditional treatment effect, whereas ours does not require this restriction.
Thus, our method incorporates treatment effect heterogeneity that linear adjustment for the control prognostic score and original covariates may not fully capture.
\end{remark}

\section{Theoretical Properties}\label{sec:MainTheoreticalProperties}
Section~\ref{sec:L2convergence results} establishes the balancing property of our proposed two-stage randomization procedure and the oracle efficiency of $\widehat{\tau}(\widehat{m}_{n_0})$ under $L_2$-consistent learning.
Section~\ref{sec:general-score-inference} extends the theoretical analysis without requiring learning consistency.

\subsection{Properties with Consistent Learning Guarantee}\label{sec:L2convergence results}
Let $Q$ denote the joint distribution of $(Y(1), Y(0), X)$. We assume that all units are i.i.d. draws from $Q$ and impose the following conditions on $Q$:
\begin{assumption}
\label{assum:Q-assumption}
$Q$ satisfies $E\{Y^2(a)\}<\infty$ and $E\!\left[\operatorname{var}\{Y(a)\mid X\}\right]>0$, for $a=0,1.$
\end{assumption}

We impose the following moment and $L_2$-consistency conditions on the arm-specific CEPO estimators $\widehat{\mu}_{n_0,a}(x)$, constructed from the first stage data $\mathcal{D}_{n_0}=\{(Y_j,T_j,X_j)\}_{j=1}^{n_0}$.
\begin{assumption}
    \label{assum:mu-moment}
    For $a\in\{0,1\}$, $\widehat{\mu}_{n_0,a}(x)$ is a jointly measurable function of $(\mathcal D_{n_0},x)$, and for some $\gamma>2$, as $n_0\to\infty$,
    $$E\!\left[|\widehat{\mu}_{n_0,a}(X)|^\gamma\mid\mathcal D_{n_0}\right]=O_p(1).$$
\end{assumption}

\begin{assumption}
    \label{assum:mu-l2-consistency}
    For $a\in\{0,1\}$, as $n_0\to\infty$,
    $$\left[E\!\left\{\bigl(\widehat{\mu}_{n_0,a}(X)-\mu_a(X)\bigr)^2\mid\mathcal D_{n_0}\right\}\right]^{1/2}=o_p(1).$$
\end{assumption}

The moment and consistency conditions on $\widehat{\mu}_{n_0,a}$ can be carried over to $\widehat{m}_{n_0}$ under the first stage simple randomization (Proposition~\ref{prop:hatm-moment} and Proposition~\ref{prop:hatm-l2-consistency} in the Supplementary Material). 
We further give a nonvanishing condition to prevent the learned balancing score from asymptotically collapsing toward zero.
\begin{assumption}
\label{assum:score-noncollapse}
The learned balancing score satisfies
$$\lim_{\kappa\downarrow0}\limsup_{n_0\to\infty}
P\!\left(E\{|\widehat{m}_{n_0}(X)|\mid\mathcal D_{n_0}\}\le\kappa\right)=0.$$
\end{assumption}
\begin{remark}
Assumption~\ref{assum:score-noncollapse} means that, for every
$\varepsilon>0$, there exists $\kappa_\varepsilon>0$ such that
\[
P(E\{|\widehat m_{n_0}(X)|\mid\mathcal D_{n_0}\}\le\kappa_\varepsilon)<\varepsilon
\]
for all sufficiently large $n_0$, i.e., $E\{|\widehat m_{n_0}(X)|\mid\mathcal D_{n_0}\}^{-1}=O_p(1)$.
For a fixed scalar score with finite first moment that is not almost surely zero, this condition holds automatically.
Unlike the spread-out condition
\citep[Assumption~4]{MaJASA2024CAR}, our condition does not require a convolution density lower bound, which is stronger.
A uniform spread-out condition among pilot data $D_{n_0}$ implies our nonvanishing condition here.
\end{remark}
We now state the first main result of the paper, the balance property of our two-stage design proposed in Section~\ref{sec:baselineframework}.
\begin{theorem}
\label{theorem:balancing property}
Suppose that Assumptions~\ref{assum:Q-assumption}, \ref{assum:mu-moment} and~\ref{assum:score-noncollapse} hold. Along every deterministic sequence $n_0,n\to\infty$:
\begin{enumerate}
\item[(i)] The imbalance of learned score satisfies
$$\Delta_n(\widehat{m}_{n_0})=o_p(\sqrt n).$$
\item[(ii)] If Assumption~\ref{assum:mu-l2-consistency} also holds, the imbalance of ACCEPO satisfies 
$$\Delta_n(m)=o_p(\sqrt n).$$
\end{enumerate}
\end{theorem}
\begin{remark}
For a fixed finite-dimensional feature map, \citet[Theorem~3.1]{MaJASA2024CAR} establish $o_p(\sqrt n)$ imbalance.
Theorem~\ref{theorem:balancing property}(i) extends this $\sqrt n$-scale balance conclusion to a score learned from pilot data, under the stated moment and nonvanishing conditions, without requiring learning consistency.
Part~(ii) further transfers this balance to the oracle target under conditional $L_2$ consistency.
\end{remark}

We next establish the asymptotic normality of the second stage difference-in-means estimator $\widehat{\tau}(\widehat{m}_{n_0})$. Part (ii) of Theorem~\ref{theorem:balancing property} shows that the imbalance of ACCEPO is asymptotically negligible on the $\sqrt{n}$ scale, allowing the estimator $\widehat{\tau}(\widehat{m}_{n_0})$ to attain the oracle variance introduced in Section~\ref{sec:preliminaries}. Theorem~\ref{theo:clt for tau_m} accounts for both the randomness of the learned score and the dependence induced by second stage sequential allocation.
\begin{theorem}
\label{theo:clt for tau_m}
Under Assumptions~\ref{assum:Q-assumption}--\ref{assum:score-noncollapse}, along every deterministic sequence $n_0,n\to\infty$,
$$\sqrt n\{\widehat{\tau}(\widehat{m}_{n_0})-\tau\}\xrightarrow{d}N(0,V^*),$$ where $V^*$ is defined as \eqref{eq:V_star}.
\end{theorem}

\begin{remark}
The proof of Theorem~\ref{theo:clt for tau_m} uses a conditional
Lindeberg--Feller central limit theorem for triangular arrays,
as in \citet{XinLiuMa2024ManyStrata}. Via the triangular array argument, the result does not require the specific growth rates of $n_0$ and $n$. It also accommodates sequences with $n_0\to\infty$ and $n_0=o(n)$ as $n\to\infty$. For example, writing $N=n_0+n$, where $n_0=\log n$, then we have 
$$\sqrt N\{\widehat{\tau}(\widehat{m}_{n_0})-\tau\}
=\sqrt{\frac Nn}\sqrt n\{\widehat{\tau}(\widehat{m}_{n_0})-\tau\}
\xrightarrow{d}N(0,V^*).$$
Thus, even when only second stage data are used in the difference-in-means estimator, it attains an asymptotic variance equal to the semiparametric efficiency bound of \citet{Hahn1998}.
\end{remark}

We now provide a consistent estimator for $V^*$. Recall that, for $a\in\{0,1\}$, $\mu_a(x)=E\{Y(a)\mid X=x\}$ denote the conditional expectations of potential outcomes, and $m_a(x)=\mu_a(x)-E\{Y(a)\}$ denote the centered conditional expectations of potential outcomes. 
Thus we have $m_1(x)-m_0(x)=\mu_1(x)-\mu_0(x)-\tau,$ and hence $\Var\{m_1(X)-m_0(X)\}=\Var\{\mu_1(X)-\mu_0(X)\}.$ 
For the second stage data $(Y_i,T_i,X_i)_{i=1}^n$, we consider constructing a plug-in variance estimator:  
$$\widehat{V}_n=2\widehat{\sigma}_{\varepsilon,1}^2+2\widehat{\sigma}_{\varepsilon,0}^2+\widehat{\mathrm{Var}}\{\widehat{m}_{n_0,1}(X)-\widehat{m}_{n_0,0}(X)\}, $$ where
\begin{align*}
    &\widehat{\sigma}_{\varepsilon,1}^{2}=\frac{1}{N_{1}}\sum_{i:T_{i}=1}\{Y_{i}-\widehat{\mu}_{n_0,1}(X_{i})\}^{2}, 
    \quad 
    \widehat{\sigma}_{\varepsilon,0}^{2}=\frac{1}{N_{0}}\sum_{i:T_{i}=0}\{Y_{i}-\widehat{\mu}_{n_0,0}(X_{i})\}^{2},\\
    &\widehat{\mathrm{Var}}\{\widehat{m}_{n_0,1}(X)-\widehat{m}_{n_0,0}(X)\}=\frac{1}{n}\sum_{i=1}^n\left([\widehat{\mu}_{n_0,1}(X_i)-\widehat{\mu}_{n_0,0}(X_i)]-\overline{\widehat{\tau}}_n\right)^2,\\
    &\overline{\widehat{\tau}}_n=\frac{1}{n}\sum_{i=1}^n\{\widehat{\mu}_{n_0,1}(X_i)-\widehat{\mu}_{n_0,0}(X_i)\}.
\end{align*}

The following result shows the consistency of the variance estimator.
\begin{theorem}
\label{theo:plugin_consistency}
Under Assumptions~\ref{assum:Q-assumption}--\ref{assum:score-noncollapse}, along every deterministic joint sequence $n_0,n\to\infty$, $$\widehat{V}_n \overset{p}{\longrightarrow} V^*.$$
Consequently,
\begin{equation*}
\frac{\sqrt n\{\widehat{\tau}(\widehat{m}_{n_0})-\tau\}}
{\sqrt{\widehat{V}_n}}
\xrightarrow{d}N(0,1).
\end{equation*}
For $\alpha\in(0,1)$, the confidence interval $$\widehat{\tau}(\widehat{m}_{n_0})
\ \pm\ z_{1-\alpha/2}\sqrt{\widehat{V}_n/n}$$ has asymptotic coverage $1-\alpha$, where $z_{1-\alpha/2}$ is the $1 - \alpha/2$ quantile of the standard normal random variable.
\end{theorem}

\subsection{Properties without Consistent Learning Guarantee}
\label{sec:general-score-inference}
The oracle balance and precision results in Section~\ref{sec:L2convergence results} are established under $L_2$-consistent learning.
For some advanced learners, such as PFNs, although their practical performances are very strong, such consistency results have not been derived or need further restricted assumptions. 
In this section, we study inference under the same two-stage design without requiring the learned balancing score $\widehat{m}_{n_0}$ to converge to $m$ or another deterministic target, even with a fixed pilot size (Remark~\ref{rmk:fixed-pilot}). 

Define the coefficient
\begin{equation}
\label{eq:general-score-coefficient}
d_{\widehat{m}_{n_0}}(m):=\frac{E[m(X)\operatorname{sgn}\{\widehat{m}_{n_0}(X)\}\mid\mathcal D_{n_0}]}{E[|\widehat{m}_{n_0}(X)|\mid\mathcal D_{n_0}]},
\end{equation}
with $d_{\widehat{m}_{n_0}}(m)=0$ when the denominator is zero and $\operatorname{sgn}(0)=0$. 
Specifically, $d_{\widehat{m}_{n_0}}(m)$ is chosen so that $E\!\left[\{m(X)-d_{\widehat{m}_{n_0}}(m)\widehat{m}_{n_0}(X)\}\operatorname{sgn}\{\widehat{m}_{n_0}(X)\}\mid\mathcal D_{n_0}\right]=0.$
Under the biased-coin rule, this property ensures that $E\!\left[(2T_i-1)\{m(X_i)-d_{\widehat{m}_{n_0}}(m)\widehat{m}_{n_0}(X_i)\}\mid\mathcal D_{n_0},(X_j,T_j)_{j<i}\right]=0,$ 
yielding the martingale representation used in the proof of Theorem~\ref{theo:general-score-ate}.

Theorem~\ref{theo:general-score-ate} characterizes how the residual $m-d_{\widehat{m}_{n_0}}(m)\widehat{m}_{n_0}$ affects the precision of the second stage difference-in-means estimator, and provides a basis for asymptotically valid inference without requiring learning consistency.
\begin{theorem}
\label{theo:general-score-ate}
Consider the two-stage design in Section~\ref{sec:baselineframework}. Suppose Assumptions~\ref{assum:Q-assumption}, \ref{assum:mu-moment} and~\ref{assum:score-noncollapse} hold. Along every deterministic joint sequence $n_0,n\to\infty$, the second stage difference-in-means estimator $\widehat{\tau}(\widehat{m}_{n_0})$ is consistent for $\tau$. Moreover,
\begin{equation*}
\label{eq:general-score-ate-clt}
\frac{\sqrt n\{\widehat{\tau}(\widehat{m}_{n_0})-\tau\}}{\sqrt{V_{n_0}}}\xrightarrow{d}N(0,1),
\end{equation*}
where
\begin{equation*}
\label{eq:general-score-variance}
V_{n_0}:=V^*+4E\!\left[\{m(X)-d_{\widehat{m}_{n_0}}(m)\widehat{m}_{n_0}(X)\}^2\mid\mathcal D_{n_0}\right],
\end{equation*}
and $V^*$ is defined in \eqref{eq:V_star}.
\end{theorem}

Thus without learning consistency, the additional variance beyond the oracle variance $V^*$ is four times the conditional mean square of the residual $m-d_{\widehat{m}_{n_0}}(m)\widehat{m}_{n_0}$ given the first stage data $\mathcal D_{n_0}$.

We now provide a consistent estimator for $V_{n_0}$. The following identities express the coefficient and variance into a form that can be consistently estimated from the observed data.
\begin{equation*}
\label{eq:general-score-variance-identity}
\begin{aligned}
&V_{n_0}=2\sum_{a=0}^{1}E\!\left[\{Y(a)-E\{Y(a)\}-d_{\widehat{m}_{n_0}}(m)\widehat{m}_{n_0}(X)\}^{2}\mid\mathcal D_{n_0}\right],\\
&E[m(X)\operatorname{sgn}\{\widehat{m}_{n_0}(X)\}\mid\mathcal D_{n_0}]=\frac12\sum_{a=0}^{1}E\!\left[\{Y(a)-E\{Y(a)\}\}\operatorname{sgn}\{\widehat{m}_{n_0}(X)\}\mid\mathcal D_{n_0}\right].
\end{aligned}
\end{equation*}
Let $\bar{Y}_a=N_a^{-1}\sum_{i:T_i=a}Y_i$. We define the coefficient and variance estimator by
\begin{equation}
\label{eq:general-score-estimator}
\begin{aligned}
\widehat{d}_{\widehat{m}_{n_0},n}(m):=&\frac{\displaystyle\frac12\sum_{a=0}^{1}\frac1{N_a}
\sum_{i:T_i=a}(Y_i-\bar{Y}_a)\operatorname{sgn}\{\widehat{m}_{n_0}(X_i)\}}
{\displaystyle\frac1n\sum_{i=1}^{n}|\widehat{m}_{n_0}(X_i)|},\\
\widehat{V}_{n_0,n}
:=&2\sum_{a=0}^{1}\frac1{N_a}\sum_{i:T_i=a}
\{Y_i-\bar{Y}_a-\widehat{d}_{\widehat{m}_{n_0},n}(m)\widehat{m}_{n_0}(X_i)\}^{2}.
\end{aligned}
\end{equation}
with $\widehat{d}_{\widehat{m}_{n_0},n}(m)=0$ when the denominator is zero.

\begin{theorem}
\label{theo:general-score-inference}
Suppose Assumptions~\ref{assum:Q-assumption}, \ref{assum:mu-moment} and~\ref{assum:score-noncollapse} hold. Along every deterministic joint sequence $n_0,n\to\infty$,
\begin{equation*}
\label{eq:general-score-estimator-consistency}
\widehat{d}_{\widehat{m}_{n_0},n}(m)-d_{\widehat{m}_{n_0}}(m)
\xrightarrow{p}0,
\qquad
\widehat{V}_{n_0,n}-V_{n_0}
\xrightarrow{p}0.
\end{equation*}
Consequently,
\begin{equation*}
\label{eq:general-score-feasible-clt}
\frac{\sqrt n\{\widehat{\tau}(\widehat{m}_{n_0})-\tau\}}
{\sqrt{\widehat{V}_{n_0,n}}}
\xrightarrow{d}N(0,1),
\end{equation*}
For $\alpha\in(0,1)$, the confidence interval
\begin{equation*}
\label{eq:general-score-wald-interval}
\widehat{\tau}(\widehat{m}_{n_0})
\ \pm\ z_{1-\alpha/2}\sqrt{\widehat{V}_{n_0,n}/n}
\end{equation*}
has asymptotic coverage $1-\alpha$.
\end{theorem}

\begin{remark}\label{rmk:fixed-pilot}
Theorems~\ref{theo:general-score-ate} and~\ref{theo:general-score-inference} also hold when $n_0=k$ is fixed. 
In this case, Assumptions \ref{assum:mu-moment} and~\ref{assum:score-noncollapse} reduce to that $\widehat m_k$ is well-defined and jointly measurable, with
$E\{|\widehat m_k(X)|^\gamma\mid\mathcal D_k\}<\infty$ for some $\gamma>2$ and $E\{|\widehat m_k(X)|\mid\mathcal D_k\}>0$ almost surely. 
To see this, repeat the same pilot and learned score in the triangular array argument for every pilot index. (Equivalently, you could only use the first $k$ pilot samples to learn the score.)
Thus, the same proofs apply, and the inference results remain valid even with a fixed pilot sample size. However, poorly learned balancing scores may hurt the efficiency.
\end{remark}

These results therefore support valid inference for some advanced learners which do not have consistency guarantees, such as PFNs. Simulation results in Table~\ref{tab:pfn_general_score_wald} provide empirical support.
We provide a detailed analysis of PFN learners in Appendix~\ref{app:pfn-pseudo-theory}, defining a pseudo-target by averaging predictions over first stage samples. Under a replacement stability condition, we establish concentration around this pseudo-target and characterize how its possible discrepancy from the oracle $m(X)$ leads to the corresponding efficiency loss of the second stage difference-in-means estimator.

\section{Extensions}
\label{sec:extensions}
Section~\ref{sec:pool-estimation} considers a ``pooled'' estimator which uses data from both the first and second stages.
Section~\ref{sec:dynamic-updating} considers a multi-stage version of design which allows the balancing score to be updated between stages.

\subsection{Pooled Estimation}
\label{sec:pool-estimation}
We now study a ``pooled'' estimator which uses data from both stages when estimating the ATE. Pooling may be particularly useful when the sample size of first stage is large relative to the total sample size. 

Let $\widehat{\tau}_{1}$ be an ATE estimator constructed from the first stage data, we impose the following high level assumption on $\widehat{\tau}_1$ and its variance estimator, which should be also constructed from the first stage data.
\begin{assumption}
    \label{assum:pilotate}
    $\widehat{\tau}_1$ is an asymptotically normal estimator for the ATE: $\sqrt{n_0}(\widehat{\tau}_1-\tau)\xrightarrow{d}N(0,V_1)$ as $n_0\to\infty$, for some $V_1\in(0,\infty)$. There exists a  variance estimator $\widehat{V}_1\in(0,\infty),$ such that $\widehat{V}_1\xrightarrow{p}V_1$ as $n_0\to\infty$.
\end{assumption}

A lot of covariate adjusted ATE estimators can satisfy Assumption~\ref{assum:pilotate} under simple randomization and suitable regularity conditions \citep{TsiatisEtAl2008CovariateAdjustment,MaTuLiu2022RegressionAnalysis,YeEtAl2023CovariateAdjustment,TuMaLiu2024UnifiedAdjustment}. We also impose a condition on the relative stage sizes.
\begin{assumption}
\label{assum:lambda}
The first stage proportion $\widehat{\lambda}:=n_0/(n_0+n)$ satisfies that $\widehat{\lambda}\to\lambda$ for some $\lambda\in[0,1]$.
\end{assumption}

Use the second stage difference-in-means estimator $\widehat{\tau}(\widehat{m}_{n_0})$ defined in \eqref{equ:tau_m} and the plug-in variance estimator $\widehat{V}_n>0$ proposed in Section~\ref{sec:L2convergence results}, we define the pooled estimator with inverse variance weights by
\begin{equation*}
\widehat{\tau}_{\mathrm{pool}}
:=\widehat{w}\widehat{\tau}_1+(1-\widehat{w})\widehat{\tau}(\widehat{m}_{n_0}),
\label{eq:pool-opt-estimator}
\end{equation*}
where
\begin{equation*}
\widehat{w}
:=\frac{n_0/\widehat{V}_1}{n_0/\widehat{V}_1+n/\widehat{V}_n}
=\frac{\widehat{\lambda}\widehat{V}_n}{\widehat{\lambda}\widehat{V}_n+(1-\widehat{\lambda})\widehat{V}_1}.
\label{eq:pool-opt-weight}
\end{equation*}
Define the estimated variance by
\begin{equation*}
\widehat{V}_{\mathrm{pool}}
:=\left\{\frac{\widehat{\lambda}}{\widehat{V}_1}+
\frac{1-\widehat{\lambda}}{\widehat{V}_n}\right\}^{-1}
=\frac{\widehat{V}_1\widehat{V}_n}{\widehat{\lambda}\widehat{V}_n+(1-\widehat{\lambda})\widehat{V}_1}.
\label{eq:pool-opt-variance-estimator}
\end{equation*}

Theorem~\ref{theo:pool} establishes the asymptotic properties of the pooled estimator.
\begin{theorem}
    \label{theo:pool}
    Under Assumptions~\ref{assum:Q-assumption}--\ref{assum:score-noncollapse} and~\ref{assum:pilotate}--\ref{assum:lambda}, along every deterministic sequence $n_0,n\to\infty$,
    $$\sqrt N(\widehat{\tau}_{\mathrm{pool}}-\tau)
    \xrightarrow{d}N(0,V_\lambda^*),\quad \widehat{V}_{\mathrm{pool}}\xrightarrow{p}V_\lambda^*,$$
    where
    \begin{equation*}
    V_\lambda^*
    :=\left\{\frac{\lambda}{V_1}+\frac{1-\lambda}{V^*}\right\}^{-1}
    =\frac{V_1V^*}{\lambda V^*+(1-\lambda)V_1}.
    \label{eq:pool-opt-limit-variance}
    \end{equation*}
    Among combinations $\widehat{\tau}(w):=w\widehat{\tau}_1+(1-w)\widehat{\tau}(\widehat{m}_{n_0})$ with fixed deterministic $w\in[0,1]$, 
    $V_\lambda^*$ is the minimum limiting variance of $\sqrt N\{\widehat{\tau}(w)-\tau\}$, 
    attained at $w_\lambda^*=\lambda V^*/\{\lambda V^*+(1-\lambda)V_1\}\in[0,1]$.
    Thus, for fixed $\alpha\in(0,1)$, the confidence interval $\widehat{\tau}_{\mathrm{pool}}\pm z_{1-\alpha/2}\sqrt{\widehat{V}_{\mathrm{pool}}/N}$ has asymptotic coverage $1-\alpha$.
\end{theorem}

Relative to the ``unpooled'' estimator $\widehat{\tau}(\widehat{m}_{n_0})$, the asymptotic variance ratio satisfies
\begin{equation*}
\frac{V_\lambda^*/N}{V^*/n}
=\frac nN\frac{V_1}{\lambda V^*+(1-\lambda)V_1}
\longrightarrow\frac{(1-\lambda)V_1}{\lambda V^*+(1-\lambda)V_1}\le1.
\label{eq:pool-opt-vs-unpooled}
\end{equation*}
When $0<\lambda<1$, the inequality is strict since $\lambda V^*>0$. Thus pooling improves first order asymptotic precision when the first stage sample proportion is nonvanishing, a more precise first stage estimator produces a larger variance reduction.

\subsection{Multi-stage Design}
\label{sec:dynamic-updating}
As second stage outcomes become available, they provide additional data for estimating $m(X)$. We therefore extend the two-stage design by dividing the second stage into a fixed number $K$ of stages and updating the balancing score $\widehat{m}_{n_0}$ between stages. 

To avoid confusion, in this subsection and the corresponding implementation and proofs in the appendix, we refer to the first and second stages of the original two-stage design as the pilot stage and the main stage, respectively.
As before, units in the pilot stage are assigned to treatment or control with equal probability.
We divide the main stage into $K$ stages, where $K$ is a fixed positive integer. 
When we call stage $k$, it refers to the $k$-th stage within the main stage, for $k=1,\ldots,K$.

For $k=1,\ldots,K$, denote the size of stage $k$ by $n_k$, with $\sum_{k=1}^K n_k=n$, and let $\mathcal I_k$ be the index set in stage $k$. Define $M_k=n_0+\sum_{\ell=1}^k n_\ell$ and $M_0=n_0$. Let $\mathcal D^{(k)}:=\mathcal D^{(0)}\cup \{(Y_i,T_i,X_i):i\in\mathcal I_1\cup\cdots\cup\mathcal I_k\}$ contain the pilot stage data and the observed data from the first $k$ completed stages within the main stage, with
$\mathcal D^{(0)}=\mathcal D_{n_0}$ and
$\mathcal H_k=\sigma(\mathcal D^{(k)})$.

At the beginning of stage $k$, use $\mathcal D^{(k-1)}$ to construct $\mathcal H_{k-1}$-measurable predictors $\widehat{\mu}_a^{[k-1]}$ for $\mu_a$
and arm mean estimators $\widehat{\nu}_a^{[k-1]}$ for $E\{Y(a)\}$, for $a\in\{0,1\}$.
Then the updated balancing score used in stage $k$ is $$\widehat{m}^{[k-1]}(x)=\frac12\left[\{\widehat{\mu}_1^{[k-1]}(x)-\widehat{\nu}_1^{[k-1]}\}+\{\widehat{\mu}_0^{[k-1]}(x)-\widehat{\nu}_0^{[k-1]}\}\right].$$
Hold these estimators fixed throughout stage $k$. For each arriving unit $j\in\mathcal I_k$, define the current within-stage imbalance by
$$\Delta_{k,j-1}(\widehat{m}^{[k-1]})=\sum_{\substack{i\in\mathcal I_k\\i<j}}(2T_i-1)\widehat{m}^{[k-1]}(X_i),$$
then apply the biased-coin rule to favor the assignment that yields the smaller potential within-stage imbalance. 
After all $n_k$ units in stage $k$ have been assigned and their outcomes have been observed, move on to the next stage, and each stage starts with zero imbalance.
The formal randomization procedure with total $K+1$ stages is given in Appendix~\ref{app:stage-local-dynamic-update}.

We use $\widehat{\tau}_{\mathrm{dyn}}$ defined in \eqref{eq:multi-stage-ate-estimator} to estimate ATE, which is the difference-in-means estimator based on the outcomes of the $n$ main stage units. 

Since later updates use data collected under adaptive treatment assignment, we impose learning conditions on each update. All predictors and centering estimators are real-valued and jointly measurable. The following conditions are understood along every deterministic joint sample size sequence with $n_0\to\infty$ and
$\min_{1\le k\le K}n_k\to\infty$.
\begin{assumption}
\label{assum:dynamic-learner-rate}
For some $\gamma>2$, for each $k=0,\ldots,K-1$ and $a\in\{0,1\}$,
\begin{equation*}
\begin{aligned}
&\left[E\!\left\{\bigl(\widehat{\mu}_a^{[k]}(X)-\mu_a(X)\bigr)^2\mid\mathcal H_k\right\}\right]^{1/2}=o_p(1),\\
&E\!\left[|\widehat{\mu}_a^{[k]}(X)|^\gamma\mid\mathcal H_k\right]=O_p(1),\\
&\widehat{\nu}_a^{[k]}-E\{Y(a)\}=o_p(1).
\end{aligned}
\end{equation*}
\end{assumption}

Under Assumption~\ref{assum:Q-assumption}, these conditions imply the corresponding moment bound and $L_2$ consistency of the
updated balancing score; see Proposition~\ref{prop:dynamic-stagewise-l2-rate}. Similar with Assumption~\ref{assum:score-noncollapse}, we require the updated balancing scores to remain nonvanishing.
\begin{assumption}
\label{assum:dynamic-feature-regularity}
For $k=0,\ldots,K-1$, assume $$\max_{0\le k<K}(E\{|\widehat{m}^{[k]}(X)|\mid\mathcal H_k\})^{-1}=O_p(1),$$ with \(1/0=+\infty\).
\end{assumption}

The following theorem combines the within-stage balance bounds to control the imbalance of $m(X)$ and establish the asymptotic distribution of $\widehat{\tau}_{\mathrm{dyn}}$, which attains the same oracle asymptotic variance $V^*$ as the two-stage design in Theorem~\ref{theo:clt for tau_m}.
\begin{theorem}
\label{theo:multi-stage-dynamic}
Suppose Assumptions~\ref{assum:Q-assumption}, \ref{assum:dynamic-learner-rate}, and \ref{assum:dynamic-feature-regularity} hold. Let $K$ be fixed. Along every deterministic sequence of sample sizes
$n_0,n_1,\ldots,n_K$ satisfying $n_0\to\infty$ and
$\min_{1\le k\le K}n_k\to\infty$, we have
\begin{equation}
\label{eq:dynamic-oracle-imbalance-small}
\Delta_n(m)=\sum_{k=1}^K\sum_{i\in\mathcal I_k}(2T_i-1)m(X_i)=o_p(\sqrt n), 
\quad 
\sqrt n(\widehat{\tau}_{\mathrm{dyn}}-\tau)\xrightarrow{d}N(0,V^*).
\end{equation}
\end{theorem}

Intuitively, updating the balancing score with accumulated observations may improve its approximation to $m(X)$ and thereby improve estimation precision in finite samples. The simulation results in Section~\ref{sec:multi-stage-simulation} suggest that updating can improve the balance of $m(X)$ and precision in finite samples, at the cost of greater computation from repeated score estimation.

\section{Simulation Studies}
\label{sec:simulations}
We consider five data generating models, all with an ATE of 2.0. 
Scenario~1 is a linear additive model with mixed covariates and constant treatment effect.
Scenario~2 is a univariate nonlinear model with treatment effect heterogeneity.
Scenario~3 considers a sparse nonlinear structure with interactions, nuisance covariates, and a constant treatment effect.
Scenario~4 incorporates nonlinear treatment effect heterogeneity through interactions and threshold effects, together with nuisance covariates.
Scenario~5 combines nonlinear treatment effect heterogeneity, heteroscedastic errors, and nuisance covariates.

We compare seven randomization procedures: simple randomization (SR), \texttt{COV(1,1,0)}, \texttt{COV(1,p,1)}, \texttt{ACCEPOCausalPFN}, \texttt{ACCEPOTabPFN}, \texttt{ACCEPOTabICL}, and \texttt{ACCEPOTrue}, where $p$ is the covariate dimension. 
The two COV procedures balance the constant, first order covariate moments, and second order covariate moments using weights $(1,1,0)$ and $(1,p,1)$, respectively.
The three PFN-based two-stage designs use their respective PFNs to estimate the CEPOs $\mu_a(X)$. \texttt{ACCEPOTrue} is an oracle benchmark that balances the true $m(X)$ implied by the data generating model. 

We perform the simulations with total sample sizes $N=n_0+n\in\{1000,2000,5000\}$, first stage proportions $\lambda=n_0/N\in\{0.10,0.15,0.20\}$, and 1,000 Monte Carlo replications per setting. Assignments in the first stage follow simple randomization with equal allocation, and all biased-coin procedures use $\rho=0.9$.
Detailed descriptions of the simulation scenarios and methods, along with additional simulation results, are provided in Appendix~\ref{app:additional-numerical}. 

\subsection{Convergence Rates of ACCEPO Imbalance}
\label{sec:balance-simulation}
We first evaluate the convergence rates of ACCEPO imbalance.  
Two levels of imbalance are investigated, 
including the estimated ACCEPO imbalance $\sum_{i=1}^{n}(2T_i-1)\widehat{m}_{n_0}(X_i)$, and true ACCEPO imbalance $\sum_{i=1}^{n}(2T_i-1)m(X_i)$. 
Table~\ref{tab:balance_property_lambda0.15} reports Monte Carlo mean of these two normalized absolute imbalances at $\lambda=0.15$.

For the imbalance of the estimated ACCEPO, all three PFN-based two-stage designs yield small imbalances that generally decrease toward zero as the sample size increases across all five scenarios. 
For the imbalance of the oracle ACCEPO, the same pattern occurs in the three PFN-based two-stage designs, although a gap from \texttt{ACCEPOTrue} remains, which can be attributed to learning error. In the nonlinear Scenarios~2–5 with larger sample size, the PFN-based two-stage designs reduce true ACCEPO imbalance relative to SR and COV designs. \texttt{COV(1,1,0)} retains an advantage in the linear setting Scenario~1, as expected.

\begin{table}[!htbp]
\centering
\caption{Monte Carlo means of absolute ACCEPO imbalances divided by $\sqrt{n}$ at $\lambda=0.15$. Dashes denote unavailable learned scores.}
\label{tab:balance_property_lambda0.15}
\resizebox{\textwidth}{!}{%
\begin{tabular}{@{}ll*{10}{c}@{}}
\toprule
& & \multicolumn{5}{c}{$n^{-1/2}\left|\sum_{i=1}^{n}(2T_i-1)\widehat{m}_{n_0}(X_i)\right|$} & \multicolumn{5}{c}{$n^{-1/2}\left|\sum_{i=1}^{n}(2T_i-1)m(X_i)\right|$} \\
\cmidrule(lr){3-7}\cmidrule(lr){8-12}
& & \multicolumn{5}{c}{Scenario} & \multicolumn{5}{c}{Scenario} \\
Method & $N$ & 1 & 2 & 3 & 4 & 5 & 1 & 2 & 3 & 4 & 5 \\
\midrule
SR & 1000 & -- & -- & -- & -- & -- & 1.048 & 0.526 & 3.042 & 0.585 & 0.823 \\
 & 2000 & -- & -- & -- & -- & -- & 1.038 & 0.539 & 2.997 & 0.578 & 0.878 \\
 & 5000 & -- & -- & -- & -- & -- & 1.056 & 0.543 & 3.019 & 0.573 & 0.836 \\
\addlinespace[2pt]
\texttt{COV(1,1,0)} & 1000 & -- & -- & -- & -- & -- & 0.153 & 0.293 & 2.496 & 0.540 & 0.687 \\
 & 2000 & -- & -- & -- & -- & -- & 0.110 & 0.299 & 2.615 & 0.497 & 0.732 \\
 & 5000 & -- & -- & -- & -- & -- & 0.070 & 0.302 & 2.619 & 0.548 & 0.665 \\
\addlinespace[2pt]
\texttt{COV(1,p,1)} & 1000 & -- & -- & -- & -- & -- & 0.236 & 0.295 & 1.239 & 0.395 & 0.642 \\
 & 2000 & -- & -- & -- & -- & -- & 0.184 & 0.291 & 1.057 & 0.351 & 0.640 \\
 & 5000 & -- & -- & -- & -- & -- & 0.143 & 0.285 & 0.898 & 0.306 & 0.651 \\
\addlinespace[2pt]
\texttt{ACCEPOCausalPFN} & 1000 & 0.034 & 0.017 & 0.083 & 0.015 & 0.019 & 0.268 & 0.213 & 1.565 & 0.435 & 0.705 \\
 & 2000 & 0.025 & 0.011 & 0.069 & 0.012 & 0.017 & 0.165 & 0.131 & 1.074 & 0.315 & 0.534 \\
 & 5000 & 0.015 & 0.007 & 0.046 & 0.008 & 0.012 & 0.103 & 0.076 & 0.724 & 0.226 & 0.408 \\
\addlinespace[2pt]
\texttt{ACCEPOTabPFN} & 1000 & 0.033 & 0.013 & 0.079 & 0.014 & 0.020 & 0.213 & 0.186 & 1.207 & 0.330 & 0.527 \\
 & 2000 & 0.024 & 0.010 & 0.070 & 0.012 & 0.017 & 0.158 & 0.127 & 0.787 & 0.235 & 0.449 \\
 & 5000 & 0.015 & 0.006 & 0.049 & 0.008 & 0.012 & 0.096 & 0.085 & 0.399 & 0.187 & 0.307 \\
\addlinespace[2pt]
\texttt{ACCEPOTabICL} & 1000 & 0.036 & 0.015 & 0.085 & 0.014 & 0.023 & 0.267 & 0.161 & 1.439 & 0.408 & 0.574 \\
 & 2000 & 0.024 & 0.010 & 0.064 & 0.011 & 0.017 & 0.181 & 0.111 & 0.995 & 0.286 & 0.458 \\
 & 5000 & 0.015 & 0.007 & 0.047 & 0.008 & 0.012 & 0.122 & 0.075 & 0.646 & 0.193 & 0.382 \\
\addlinespace[2pt]
\texttt{ACCEPOTrue} & 1000 & -- & -- & -- & -- & -- & 0.035 & 0.015 & 0.108 & 0.018 & 0.031 \\
 & 2000 & -- & -- & -- & -- & -- & 0.025 & 0.011 & 0.078 & 0.014 & 0.021 \\
 & 5000 & -- & -- & -- & -- & -- & 0.016 & 0.007 & 0.051 & 0.008 & 0.014 \\
\bottomrule
\end{tabular}
}
\end{table}

\subsection{Treatment Effect Estimation and Inference}
\label{sec:unpooled-simulation-results}
We now pay attention to the treatment effect estimation precision and validation of inference. 
Table~\ref{tab:ate_fixed_lambda_0.15} reports the Monte Carlo mean and standard deviation of the unpooled difference-in-means estimator at $\lambda=0.15$ across all scenarios.
All Monte Carlo means are close to the true ATE of 2.
The PFN-based two-stage designs consistently improve precision over SR and outperform both COV designs in the nonlinear Scenarios~2–5 at larger sample sizes, and the oracle \texttt{ACCEPOTrue} design achieves the lowest Monte Carlo standard deviations across all settings.

\begin{table}[!htbp]
\centering
\caption{Monte Carlo mean (standard deviation) of the unpooled difference-in-means estimator at $\lambda=0.15$.}
\label{tab:ate_fixed_lambda_0.15}
\resizebox{\textwidth}{!}{%
\begin{tabular}{@{}llccccc@{}}
\toprule
& & \multicolumn{5}{c}{Scenario} \\
\cmidrule(lr){3-7}
Method & $N$ & 1 & 2 & 3 & 4 & 5 \\
\midrule
SR & 1000 & 1.998 (0.117) & 1.998 (0.084) & 2.004 (0.289) & 2.001 (0.085) & 2.005 (0.110) \\
 & 2000 & 2.001 (0.077) & 2.002 (0.059) & 2.007 (0.208) & 1.999 (0.058) & 1.998 (0.081) \\
 & 5000 & 2.002 (0.049) & 2.001 (0.037) & 2.005 (0.135) & 2.000 (0.037) & 2.003 (0.049) \\
\addlinespace[2pt]
\texttt{COV(1,1,0)} & 1000 & 1.997 (0.070) & 1.999 (0.076) & 1.991 (0.263) & 2.002 (0.084) & 2.003 (0.106) \\
 & 2000 & 2.000 (0.048) & 2.002 (0.053) & 2.002 (0.187) & 2.002 (0.056) & 2.002 (0.074) \\
 & 5000 & 2.000 (0.031) & 2.000 (0.034) & 1.999 (0.114) & 2.002 (0.036) & 2.002 (0.046) \\
\addlinespace[2pt]
\texttt{COV(1,p,1)} & 1000 & 1.996 (0.073) & 2.001 (0.076) & 1.986 (0.174) & 2.003 (0.077) & 2.005 (0.100) \\
 & 2000 & 1.998 (0.051) & 2.003 (0.054) & 1.996 (0.115) & 2.001 (0.052) & 2.000 (0.071) \\
 & 5000 & 1.999 (0.031) & 2.001 (0.032) & 1.999 (0.071) & 2.002 (0.032) & 2.002 (0.047) \\
\addlinespace[2pt]
\texttt{ACCEPOCausalPFN} & 1000 & 2.002 (0.070) & 2.000 (0.075) & 2.002 (0.192) & 2.002 (0.078) & 2.003 (0.105) \\
 & 2000 & 2.004 (0.050) & 2.001 (0.049) & 2.004 (0.111) & 2.001 (0.051) & 2.000 (0.071) \\
 & 5000 & 2.000 (0.031) & 2.000 (0.031) & 2.002 (0.069) & 2.001 (0.031) & 2.001 (0.043) \\
\addlinespace[2pt]
\texttt{ACCEPOTabPFN} & 1000 & 1.999 (0.071) & 2.002 (0.072) & 1.997 (0.172) & 1.998 (0.072) & 2.002 (0.096) \\
 & 2000 & 2.000 (0.050) & 2.003 (0.051) & 2.001 (0.106) & 2.001 (0.051) & 1.999 (0.066) \\
 & 5000 & 1.999 (0.030) & 2.000 (0.031) & 2.001 (0.063) & 2.001 (0.030) & 2.002 (0.041) \\
\addlinespace[2pt]
\texttt{ACCEPOTabICL} & 1000 & 1.997 (0.071) & 2.002 (0.073) & 1.992 (0.189) & 2.000 (0.079) & 1.999 (0.099) \\
 & 2000 & 1.998 (0.049) & 2.003 (0.049) & 2.000 (0.119) & 2.002 (0.051) & 2.002 (0.067) \\
 & 5000 & 1.999 (0.031) & 1.999 (0.031) & 2.001 (0.069) & 2.000 (0.031) & 2.002 (0.041) \\
\addlinespace[2pt]
\texttt{ACCEPOTrue} & 1000 & 1.997 (0.067) & 2.002 (0.072) & 1.999 (0.137) & 2.000 (0.068) & 2.004 (0.087) \\
 & 2000 & 1.999 (0.048) & 2.001 (0.049) & 2.001 (0.094) & 2.001 (0.048) & 1.998 (0.060) \\
 & 5000 & 1.999 (0.030) & 2.001 (0.031) & 2.001 (0.061) & 2.001 (0.030) & 2.003 (0.038) \\
\bottomrule
\end{tabular}
}
\end{table}

For the three PFN-based two-stage designs, Table~\ref{tab:pfn_general_score_wald} reports the empirical confidence interval coverage probability proposed in Theorem~\ref{theo:general-score-inference}, based on variance estimator $\widehat{V}_{n_0,n}$, with $\alpha=0.05$. 
Coverage is generally close to the nominal level of $0.95$ across the settings considered. 
These results provide empirical support for valid inference for our two-stage design even when we use the learners without consistent learning guarantee.

\begin{table}[!htbp]
\centering
\renewcommand{\baselinestretch}{1}\small
\setlength{\tabcolsep}{3pt}
\renewcommand{\arraystretch}{1.02}
\caption{Empirical coverage (\%) of confidence intervals based on the variance estimator from Section~\ref{sec:general-score-inference}, with $\alpha=0.05$.}
\label{tab:pfn_general_score_wald}
\resizebox{\textwidth}{!}{%
\begin{tabular}{@{}ll*{15}{c}@{}}
\toprule
& & \multicolumn{5}{c}{$\lambda=0.10$} & \multicolumn{5}{c}{$\lambda=0.15$} & \multicolumn{5}{c}{$\lambda=0.20$} \\
\cmidrule(lr){3-7}\cmidrule(lr){8-12}\cmidrule(lr){13-17}
& & \multicolumn{5}{c}{Scenario} & \multicolumn{5}{c}{Scenario} & \multicolumn{5}{c}{Scenario} \\
Method & $N$ & 1 & 2 & 3 & 4 & 5 & 1 & 2 & 3 & 4 & 5 & 1 & 2 & 3 & 4 & 5 \\
\midrule
\texttt{ACCEPOCausalPFN} & 1000 & 93.9 & 93.3 & 94.4 & 94.8 & 94.4 & 95.3 & 94.1 & 94.8 & 95.0 & 94.7 & 94.6 & 94.3 & 95.2 & 94.6 & 94.8 \\
 & 2000 & 95.4 & 93.8 & 94.4 & 94.3 & 95.4 & 95.1 & 95.7 & 96.3 & 96.1 & 93.1 & 95.7 & 95.2 & 94.0 & 94.0 & 94.6 \\
 & 5000 & 94.4 & 95.0 & 94.7 & 95.4 & 94.8 & 95.1 & 95.3 & 93.7 & 96.4 & 93.9 & 94.2 & 93.3 & 95.8 & 95.9 & 94.0 \\
\addlinespace[2pt]
\texttt{ACCEPOTabPFN} & 1000 & 96.1 & 94.1 & 94.3 & 93.5 & 95.0 & 94.9 & 94.7 & 94.7 & 95.3 & 95.0 & 95.0 & 94.7 & 94.9 & 95.4 & 94.5 \\
 & 2000 & 95.1 & 94.4 & 95.9 & 94.5 & 94.2 & 94.3 & 94.7 & 95.1 & 94.1 & 93.6 & 95.5 & 94.2 & 94.6 & 94.7 & 94.4 \\
 & 5000 & 94.6 & 94.2 & 94.9 & 95.4 & 93.6 & 96.1 & 95.4 & 95.1 & 95.6 & 94.1 & 94.4 & 95.0 & 95.7 & 95.8 & 95.5 \\
\addlinespace[2pt]
\texttt{ACCEPOTabICL} & 1000 & 95.1 & 94.7 & 94.8 & 94.8 & 94.4 & 95.1 & 93.3 & 93.6 & 93.5 & 94.5 & 96.3 & 94.2 & 94.8 & 93.9 & 95.4 \\
 & 2000 & 96.3 & 95.0 & 94.6 & 94.8 & 94.8 & 95.3 & 95.1 & 94.0 & 94.8 & 94.7 & 95.5 & 94.6 & 93.5 & 95.6 & 94.6 \\
 & 5000 & 95.5 & 95.2 & 93.8 & 95.3 & 94.7 & 95.0 & 95.7 & 93.8 & 95.3 & 94.1 & 94.6 & 94.1 & 94.6 & 96.5 & 94.5 \\
\bottomrule
\end{tabular}
}
\end{table}

\section{Clinical Trial Examples}\label{sec:clinical trial example}
We use two clinical trial examples, ACTG~175 \citep{Hammer1996} and AWARE \citep{Metsch2013AWARE}, to illustrate the estimation precision gains of our proposed two-stage design. 

ACTG~175 is a double-blind trial comparing zidovudine or didanosine monotherapy with combination antiretroviral therapy. 
We restrict attention to two arms: monotherapy with zidovudine as the control arm and combination therapy with zidovudine and didanosine as the treatment arm. 
This yields $N=1054$ units, with 532 in the control arm and 522 in the treatment arm. 
We use week-20 CD4 T-cell count as the outcome and baseline CD4 count, age, Karnofsky score, and past-treatment stratification as covariates.

For AWARE, a randomized trial evaluating rapid HIV testing with risk reduction counseling,
we use the 6-month risky sexual behavior count as the outcome. 
After excluding missing follow-up outcomes and extreme values above 360, the analysis includes $N=3414$ units, with 1717 in the control arm and 1697 in the treatment arm. 
We use 16 baseline covariates covering demographic, behavioral, clinical, and social characteristics.

We perform our analysis using synthetic data generated based on the trial. For each dataset, we analyzed the data using separate Bayesian additive regression tree (BART) models for the two arms \citep{Chipman2010BART}. In each replication, we generate potential outcomes by adding independent Gaussian errors to the BART fitted means and reassign patients according to each randomization procedure. The error standard deviations are set to 85.60 for ACTG~175 and 15.09 for AWARE, with corresponding average treatment effects of 72.17 and -3.31, respectively.

For each first stage proportion $\lambda=n_0/N\in\{0.15,0.20,0.30\}$, we compare seven randomization procedures as in Section~\ref{sec:simulations} over 1,000 replications. 
Table~\ref{tab:real-data-combined} reports the Monte Carlo mean and standard deviation of the unpooled difference-in-means estimator.
In both applications, the PFN-based two-stage designs have lower Monte Carlo standard deviations than SR and generally improve on designs that balance covariate moments.
\begin{table}[!htbp]
\centering
\caption{Monte Carlo mean and standard deviation (SD) of the unpooled ATE estimator over 1,000 replications based on the two clinical trial examples.}
\label{tab:real-data-combined}
\resizebox{\textwidth}{!}{%
\begin{tabular}{@{}l*{12}{r}@{}}
\toprule
\multirow{3}{*}{Method}
& \multicolumn{6}{c}{ACTG~175}
& \multicolumn{6}{c}{AWARE} \\
\cmidrule(lr){2-7}
\cmidrule(lr){8-13}
& \multicolumn{2}{c}{$\lambda=0.15$}
& \multicolumn{2}{c}{$\lambda=0.20$}
& \multicolumn{2}{c}{$\lambda=0.30$}
& \multicolumn{2}{c}{$\lambda=0.15$}
& \multicolumn{2}{c}{$\lambda=0.20$}
& \multicolumn{2}{c}{$\lambda=0.30$} \\
\cmidrule(lr){2-3}
\cmidrule(lr){4-5}
\cmidrule(lr){6-7}
\cmidrule(lr){8-9}
\cmidrule(lr){10-11}
\cmidrule(lr){12-13}
& Mean & SD & Mean & SD & Mean & SD
& Mean & SD & Mean & SD & Mean & SD \\
\midrule
SR
& 72.21 & 8.04 & 72.28 & 8.53 & 72.31 & 8.67
& -3.30 & 0.80 & -3.31 & 0.81 & -3.27 & 0.87 \\
\texttt{COV(1,1,0)}
& 72.27 & 6.31 & 72.43 & 6.52 & 72.34 & 7.32
& -3.28 & 0.60 & -3.30 & 0.65 & -3.28 & 0.67 \\
\texttt{COV(1,p,1)}
& 72.29 & 7.93 & 72.35 & 8.06 & 72.12 & 8.93
& -3.30 & 0.60 & -3.33 & 0.62 & -3.28 & 0.69 \\
\texttt{ACCEPOCausalPFN}
& 72.10 & 5.80 & 72.36 & 6.22 & 72.14 & 6.54
& -3.28 & 0.58 & -3.31 & 0.58 & -3.30 & 0.65 \\
\texttt{ACCEPOTabPFN}
& 71.88 & 5.99 & 72.47 & 6.12 & 72.26 & 6.51
& -3.30 & 0.56 & -3.29 & 0.62 & -3.27 & 0.65 \\
\texttt{ACCEPOTabICL}
& 71.81 & 5.94 & 72.44 & 6.24 & 72.50 & 6.62
& -3.29 & 0.58 & -3.32 & 0.62 & -3.31 & 0.63 \\
\texttt{ACCEPOTrue}
& 72.15 & 5.71 & 71.99 & 5.97 & 72.08 & 6.44
& -3.28 & 0.56 & -3.29 & 0.57 & -3.28 & 0.62 \\
\bottomrule
\end{tabular}}
\end{table}

\FloatBarrier

\section{Conclusion}\label{sec:Conclusion}
This paper develops a two-stage randomization that uses the first stage data to estimate ACCEPO then uses the learned balancing score to guide sequential biased-coin randomization in the second stage. 
We establish the asymptotic theory both with $L_2$ consistency guarantee of the learners and without this requirement.
These robust inference results open the door to using advanced learners such as PFNs within our randomization framework, without requiring consistent learning guarantee.
Simulations and clinical trial examples support our main theoretical properties of this paper.

The framework could be extended in several directions.
First, the randomization framework could be generalized to multiple treatments and more general allocation functions \citep{Wei1978Adaptive,Bugni2019}. 
Second, transportability across stages deserves further study when the population, treatment environment, or outcome mechanism changes between stages. 
Finally, broadening the scope of our framework to multiple endpoints and treatment effect estimands beyond the ATE is a natural direction for future research.

\putbib
\end{bibunit}

\clearpage
\thispagestyle{empty}
\vspace*{\fill}
\begin{center}
\begin{minipage}{0.9\textwidth}
\centering
{\LARGE\bfseries Supplementary Material for ``Two-Stage Randomization with a Learned Balancing Score: Efficiency and Robust Inference''}
\vspace{2em}
\end{minipage}
\end{center}
\vspace*{\fill}

\clearpage
\begin{bibunit}

\appendix
\etocdepthtag.toc{appendices}

\begingroup
\etocsettagdepth{main}{none}
\etocsettagdepth{appendices}{subsection}
\etocsettocstyle{\section*{Contents}}{}
\tableofcontents
\endgroup

\clearpage

\numberwithin{equation}{section}
\numberwithin{table}{section}
\numberwithin{figure}{section}
\numberwithin{assumption}{section}

\section{Additional Results and Implementation Details}
\label{app:additional-results-implementation-details}
\subsection{Asymptotic Properties of Pseudo-target Learning}
\label{app:pfn-pseudo-theory}
In this section, we consider the regime in which the CEPO estimator $\widehat{\mu}_{n_0,a}(x)$ converges to a pseudo-target $\mu_{n_0,a}^\dagger(x)$ that need not coincide with the CEPO $\mu_a(x)$, for $a\in\{0,1\}$. We retain the two-stage design in Section~\ref{sec:baselineframework}, using the pilot score $\widehat{m}_{n_0}$ as the balancing target throughout the main stage. Define the pseudo-target of $\widehat{m}_{n_0}$ and its approximation residual by
\begin{equation*}
\label{eq:pseudo-target-definition}
m_{n_0}^\dagger(x):=\frac12\sum_{a=0}^1\left[\mu_{n_0,a}^\dagger(x)-E\{Y(a)\}\right],
\qquad
b_{n_0}^\dagger(x):=m(x)-m_{n_0}^\dagger(x).
\end{equation*}

Thus $m=m_{n_0}^\dagger+b_{n_0}^\dagger$. Assumption~\ref{assum:pseudo-target-learning} replaces consistency for the true CEPOs with concentration around the pseudo-targets. 
\begin{assumption}
\label{assum:pseudo-target-learning}
The functions $\mu_{n_0,a}^\dagger$ are square-integrable, as $n_0\to\infty$
\begin{equation*}
\label{eq:pseudo-target-rate}
\left[E\!\left\{\bigl(\widehat{\mu}_{n_0,a}(X)-\mu_{n_0,a}^\dagger(X)\bigr)^2\mid\mathcal D_{n_0}\right\}\right]^{1/2}
=o_p(1),
\qquad a\in\{0,1\}.
\end{equation*}
\end{assumption}

Similar with the definition in~\eqref{eq:general-score-coefficient}, define
\begin{equation*}
\label{eq:pseudo-row-variance}
d_{\widehat{m}_{n_0}}(b_{n_0}^\dagger):=\frac{E[b_{n_0}^\dagger(X)\operatorname{sgn}\{\widehat{m}_{n_0}(X)\}\mid\mathcal D_{n_0}]}{E[|\widehat{m}_{n_0}(X)|\mid\mathcal D_{n_0}]},
\end{equation*}
with $d_{\widehat{m}_{n_0}}(b_{n_0}^\dagger)=0$ when the denominator is zero. 
Theorem~\ref{theo:pseudo-target-ate} shows that balancing the learned score $\widehat{m}_{n_0}$ also reduces the imbalance of its pseudo-target $m^\dagger_{n_0}$ to $o_p(\sqrt n)$, while the adjusted residual of the approximation error $b_{n_0}^\dagger-d_{\widehat{m}_{n_0}}(b_{n_0}^\dagger)\widehat{m}_{n_0}$ affects the precision of the main stage difference-in-means estimator. When $b_{n_0}^\dagger\equiv0$, the oracle precision of Theorem~\ref{theo:clt for tau_m} is recovered.

\begin{theorem}
\label{theo:pseudo-target-ate}
Consider the two-stage design in Section~\ref{sec:baselineframework}, and suppose Assumptions~\ref{assum:Q-assumption}, \ref{assum:mu-moment}, \ref{assum:score-noncollapse}, and~\ref{assum:pseudo-target-learning} hold.
Along every deterministic joint sequence $n_0,n\to\infty$, the learned score and the pseudo-target satisfy
\begin{equation}
\label{eq:pseudo-score-balance}
\Delta_n(\widehat{m}_{n_0})=o_p(\sqrt n),
\qquad
\Delta_n(m_{n_0}^\dagger)=o_p(\sqrt n),
\end{equation}
Moreover,
\begin{equation}
\label{eq:pseudo-ate-clt}
\frac{\sqrt n\{\widehat{\tau}(\widehat{m}_{n_0})-\tau\}}
{\left\{V^*+4E\!\left[\{b_{n_0}^\dagger(X)-d_{\widehat{m}_{n_0}}(b_{n_0}^\dagger)\widehat{m}_{n_0}(X)\}^2\mid\mathcal D_{n_0}\right]\right\}^{1/2}}
\xrightarrow{d}N(0,1).
\end{equation}
\end{theorem}

Inference can be done directly from the results in Section~\ref{sec:general-score-inference}.
\begin{corollary}
\label{cor:pseudo-target-variance}
Suppose Assumptions~\ref{assum:Q-assumption}, \ref{assum:mu-moment}, \ref{assum:score-noncollapse} and~\ref{assum:pseudo-target-learning} hold. Then along every deterministic joint sequence $n_0,n\to\infty$,
$$V_{n_0}-\left\{V^*+4E\!\left[\{b_{n_0}^\dagger(X)-d_{\widehat{m}_{n_0}}(b_{n_0}^\dagger)\widehat{m}_{n_0}(X)\}^2\mid\mathcal D_{n_0}\right]\right\}\xrightarrow{p}0.$$
Consequently, the variance estimator in \eqref{eq:general-score-estimator} satisfies
$$\widehat{V}_{n_0,n}-\left\{V^*+4E\!\left[\{b_{n_0}^\dagger(X)-d_{\widehat{m}_{n_0}}(b_{n_0}^\dagger)\widehat{m}_{n_0}(X)\}^2\mid\mathcal D_{n_0}\right]\right\}\xrightarrow{p}0.$$
\end{corollary}

For the PFN predictors $\widehat{\mu}_{n_0,a}(x)
=f_{\widehat{\theta},a}(x;\mathcal D_{n_0})$, with pretrained parameters held fixed, repeated context samples (here is the first stage data) induce variability in the predictions. From a frequentist perspective, we define the pseudo-target by taking expectation
over first stage data, and use replacement stability to control concentration around it. Thus the  pseudo-target for PFN predictor is defined as:
\begin{equation}
\label{eq:pfn-pilot-mean-target}
\mu_{n_0,a}^\dagger(x):=E_{\mathcal D_{n_0}}\!\left\{f_{\widehat{\theta},a}(x;\mathcal D_{n_0})\right\},\qquad a\in\{0,1\}.
\end{equation}

Write $\mathcal D_{n_0}=(Z_1,\ldots,Z_{n_0})$, where $Z_j=(Y_j,T_j,X_j)$. For any $j\in\{1,\ldots,n_0\}$, let
$\mathcal D_{n_0}^{\prime(j)}$ be obtained by replacing $Z_j$
with an independent copy $Z'_j$, leaving all other observations unchanged. 
Let $X\sim P_X$ be a fresh covariate vector independent of $\mathcal D_{n_0}$ and all replacement observations. Expectations below are taken over the first stage data, their replacements, and this independent query covariate, as applicable. 
We wish the influence of individual elements should diminish as the first stage sample size grows. 
\begin{assumption}
\label{assum:pfn-replacement-stability}
The PFN predictors are jointly measurable and satisfy $E\{f_{\widehat{\theta},a}^2(X;\mathcal D_{n_0})\}<\infty$ for each $n_0$ and $a\in\{0,1\}$. There exist constants $C<\infty$ and $\alpha>1/2$ such that,
for all sufficiently large $n_0$,
\begin{equation}
\label{eq:pfn-replacement-stability}
\max_{\substack{a\in\{0,1\}\\1\le j\le n_0}}E\!\left[\left\{f_{\widehat{\theta},a}(X;\mathcal D_{n_0})-f_{\widehat{\theta},a}(X;\mathcal D_{n_0}^{\prime(j)})\right\}^{2}\right]\le C^2 n_0^{-2\alpha}.
\end{equation}
\end{assumption}

The Efron--Stein inequality \citep{EfronStein1981} gives the following learning guarantee.
\begin{proposition}
\label{prop:pfn-pseudo-target}
Under Assumption~\ref{assum:pfn-replacement-stability}, for all sufficiently large $n_0$,
\begin{equation}
\label{eq:pfn-efron-stein-bound}
E\!\left[\bigl\{f_{\widehat{\theta},a}(X;\mathcal D_{n_0})-\mu_{n_0,a}^\dagger(X)\bigr\}^2\right]
\le\frac{C^2}{2}n_0^{1-2\alpha},\qquad a\in\{0,1\}.
\end{equation}
Consequently, Assumption~\ref{assum:pseudo-target-learning} holds since $n_0\to\infty$ and $1-2\alpha<0$. 
\end{proposition}

As discussed in \citet{Nagler2023PFN}, the quality of a deployed PFN depends on the task prior, the distribution of context sizes used in pretraining, the expressiveness of the architecture, and the Monte Carlo and optimization errors involved in pretraining. For a one-layer classification transformer, \citet[Theorems~6.2--6.3]{Nagler2023PFN} establishes vanishing sampling fluctuations and a population limit under specified input and parameter conditions. The replacement-stability condition in Assumption~\ref{assum:pfn-replacement-stability} has not been verified here for the pretrained PFNs used in our simulations. Proofs of the results in this subsection are provided in Appendix~\ref{app:pseudo-target-proofs}.

\subsection{Implementation of Multi-stage Design}
\label{app:stage-local-dynamic-update}
This subsection gives the formal randomization procedure for the multi-stage design in Section~\ref{sec:dynamic-updating}. 

The total $K+1$ stages randomization procedure is performed as follows:
\begin{enumerate}
    \item[(1)] Assign each unit in the pilot stage independently to treatment or control with equal probability. After the pilot stage, we use pilot stage data $\mathcal{D}^{(0)}=\{(Y_j,T_j,X_j)\}_{j=1}^{n_0}$ to get estimator $\widehat{m}^{[0]}(\cdot)$.
    \item[(2)] For $k=1,\ldots,K$, repeat steps (3)-(4) given below until all $K$ stages of units are assigned.
    \item[(3)] During stage $k$, for each arriving unit $j\in\mathcal I_k$, let
    $$\Delta_{k,j-1}(\widehat{m}^{[k-1]})=\sum_{\substack{i\in\mathcal I_k\\ i<j}}(2T_i-1)\widehat{m}^{[k-1]}(X_i)$$ be the current within-stage imbalance before assigning unit $j$. Then the two potential imbalance measures are
    $$
    \mathrm{Imb}_{k,j}^{(1)}
    =\left\|\Delta_{k,j-1}(\widehat{m}^{[k-1]})+\widehat{m}^{[k-1]}(X_j)\right\|^2, 
    \mathrm{Imb}_{k,j}^{(0)}
    =\left\|\Delta_{k,j-1}(\widehat{m}^{[k-1]})-\widehat{m}^{[k-1]}(X_j)\right\|^2.$$
    Assign the $j$-th unit to treatment with probability
    $$
    \mathbb{P}(T_j=1\mid  \mathcal{H}_{k-1},\{X_i,T_i:i\in \mathcal{I}_k,i<j\},X_j)
    =
    \begin{cases}
    \rho, & \text{if } \mathrm{Imb}_{k,j}^{(1)}<\mathrm{Imb}_{k,j}^{(0)},\\
    1-\rho, & \text{if } \mathrm{Imb}_{k,j}^{(1)}>\mathrm{Imb}_{k,j}^{(0)},\\
    1/2, & \text{if } \mathrm{Imb}_{k,j}^{(1)}=\mathrm{Imb}_{k,j}^{(0)},
    \end{cases}
    $$ where $0.5<\rho<1$ is the biasing probability.
    \item[(4)] After all $n_k$ units in stage $k$ have been assigned and their outcomes are observed, update the available data to $\mathcal{D}^{(k)}$ and $\mathcal H_k$. If $k<K$, update the predictors and centering estimators to obtain $\widehat{m}^{[k]}$. Proceed to the next stage until all $n$ main stage units have been assigned.
\end{enumerate}
Index the $n$ main stage units chronologically by $i=1,\ldots,n$, so that $\mathcal I_1,\ldots,\mathcal I_K$ partition $\{1,\ldots,n\}$. We only use the $n$ main stage outcomes to estimate ATE using difference-in-means estimator:
\begin{equation}
\label{eq:multi-stage-ate-estimator}
\widehat{\tau}_{\mathrm{dyn}}
:=\frac{\sum_{i=1}^nT_iY_i}{N_1}
-\frac{\sum_{i=1}^n(1-T_i)Y_i}{N_0},
\quad
N_a:=\sum_{i=1}^n\mathbf1\{T_i=a\},
\quad a=0,1.
\end{equation}

\section{Additional Numerical Results and Details}
\label{app:additional-numerical}

\subsection{Data generating Models}
\label{app:simulation-models}
We provide the five data generating models used in Section~\ref{sec:simulations}. We write $X^{(r)}$ for the $r$-th covariate coordinate, $\mu_0$ and $\mu_1$ denote scalar intercepts in this subsection.

\textbf{Scenario 1: }
\begin{align*}
Y(1) &= \mu_1 + \alpha^\top X + \varepsilon_1, \\
Y(0) &= \mu_0 + \alpha^\top X + \varepsilon_0,
\end{align*}
where \(X=(X^{(1)},\ldots,X^{(5)})^\top\), \(X^{(1)} \sim \mathrm{Unif}[-2,2]\), \(X^{(2)} \sim N(2,2)\), \(X^{(3)} \sim t_5\), \(X^{(4)} \sim \mathrm{Beta}(2,2)\), \(X^{(5)} \in \{1,2,3\}\) with $P(X^{(5)}=1)=0.3, P(X^{(5)}=2)=0.4, P(X^{(5)}=3)=0.3,$ \(\mu_0 \sim \mathrm{Unif}[-10,10]\), \(\mu_1 = \mu_0 + 2\), \(\alpha=(\alpha_1,\ldots,\alpha_5)^\top\) with \(\alpha_j \stackrel{\mathrm{i.i.d.}}{\sim} \mathrm{Unif}[-1,1]\) for \(j=1,\ldots,5\), and \(\varepsilon_0,\varepsilon_1 \stackrel{\mathrm{i.i.d.}}{\sim} N(0,1)\).

\textbf{Scenario 2: }
\begin{align*}
Y(1) &= \mu_1 + 0.70\sin(\pi X^{(1)}) - 0.55\left((X^{(1)})^2-\frac{1}{3}\right) + 0.35X^{(1)} + \varepsilon_1, \\
Y(0) &= \mu_0 + 0.95\sin(\pi X^{(1)}) + 0.55\left((X^{(1)})^2-\frac{1}{3}\right) + \varepsilon_0,
\end{align*}
where \(X^{(1)} \sim \mathrm{Unif}[-1,1]\), \(\mu_0 \sim \mathrm{Unif}[0,10]\), \(\mu_1 = \mu_0 + 2\), and \(\varepsilon_0,\varepsilon_1 \stackrel{\mathrm{i.i.d.}}{\sim} N(0,1)\).

\textbf{Scenario 3: }
\begin{align*}
Y(1) &= \mu_1 + 2.5\sin\!\left(\frac{\pi X^{(1)}}{2}\right) + 2(X^{(2)})^2 + 1.5\max\{X^{(1)}+0.5X^{(2)}-1,0\} + 2X^{(3)}X^{(4)} + 2 \varepsilon_1, \\
Y(0) &= \mu_0 + 2.5\sin\!\left(\frac{\pi X^{(1)}}{2}\right) + 2(X^{(2)})^2 + 1.5\max\{X^{(1)}+0.5X^{(2)}-1,0\} + 2X^{(3)}X^{(4)} + 2 \varepsilon_0,
\end{align*}
where \(X=(X^{(1)},\ldots,X^{(10)})^\top\), \(X^{(1)} \sim \mathrm{Unif}[-2,2]\), \(X^{(2)} \sim N(0,1)\), \(X^{(3)} \sim N(0,1)\), \(X^{(4)} \sim \mathrm{Bernoulli}(0.5)\), \(X^{(5)} \sim \mathrm{Unif}[0,1]\), \(X^{(6)},\ldots,X^{(10)} \stackrel{\mathrm{i.i.d.}}{\sim} N(0,1)\), \(\mu_0 \sim \mathrm{Unif}[-5,5]\), \(\mu_1 = \mu_0 + 2\),  and \(\varepsilon_0,\varepsilon_1 \stackrel{\mathrm{i.i.d.}}{\sim} N(0,1)\).

\textbf{Scenario 4:}
\begin{align*}
Y(1) &= \mu_1 + \delta(X) + h(X) + \varepsilon_1,\\
Y(0) &= \mu_0 + h(X) + \varepsilon_0,
\end{align*}
where
\begin{align*}
h(X)&=0.25X^{(1)}+0.20\left((X^{(2)})^2-\frac{1}{3}\right),\\
\delta(X)&=2\sin(\pi X^{(1)}X^{(2)})+1.4\left[\mathbf{1}\{X^{(3)}>0.75\}-\bar{\Phi}(0.75)\right]+(2X^{(5)}-1)\left((X^{(4)})^2-\frac{1}{3}\right),
\end{align*}
Where \(X=(X^{(1)},\ldots,X^{(10)})^\top\), \(X^{(1)},X^{(2)},X^{(4)}\stackrel{\mathrm{i.i.d.}}{\sim}\mathrm{Unif}[-1,1]\), \(X^{(3)}\sim N(0,1)\), \(X^{(5)}\sim\mathrm{Bernoulli}(0.5)\), \(X^{(6)},\ldots,X^{(10)}\stackrel{\mathrm{i.i.d.}}{\sim}N(0,1)\), \(\mu_0 \sim \mathrm{Unif}[-5,5]\), \(\mu_1 = \mu_0 + 2\), and \(\varepsilon_0,\varepsilon_1\stackrel{\mathrm{i.i.d.}}{\sim}N(0,0.7^2)\). Let \(\bar{\Phi}(t)=1-\Phi(t)\) denote the upper tail probability of a standard normal distribution.

\textbf{Scenario 5:}
\begin{align*}
Y(1) &= \mu_0 + 2 + \delta(X) + h(X) + \sigma(X)\varepsilon_1,\\
Y(0) &= \mu_0 + h(X) + \sigma(X)\varepsilon_0,
\end{align*}
where
\begin{align*}
h(X)&=0.20X^{(1)}+0.15((X^{(2)})^2-1),\\
\delta(X)&=2\left[\mathbf{1}\{X^{(3)}+X^{(4)}>1\}-\bar{\Phi}\left(\frac{1}{\sqrt{2}}\right)\right]
+1.5\sin(X^{(5)}X^{(6)})+1.2\,\operatorname{sgn}(X^{(7)})((X^{(7)})^2-1),\\
\sigma(X)&=0.6+0.2\,\mathbf{1}\{|X^{(8)}|>1\},
\end{align*}
Where \(X=(X^{(1)},\ldots,X^{(10)})^\top\), \(X^{(1)},\ldots,X^{(10)}\stackrel{\mathrm{i.i.d.}}{\sim}N(0,1)\), \(\mu_0\sim\mathrm{Unif}[-5,5]\), and \(\varepsilon_0,\varepsilon_1\stackrel{\mathrm{i.i.d.}}{\sim}N(0,1)\).

\subsection{Competing Randomization Procedures}
\label{app:simulation-methods}
The seven randomization procedures used in Section~\ref{sec:simulations} and Section~\ref{sec:clinical trial example} are specified below.
\begin{itemize}
    \item \textbf{Simple Randomization (SR):} Here each unit in the second stage is independently assigned to treatment group with probability \(1/2\), without using any covariate information. 
    \item \textbf{Moment-based CAR:} Here treatment assignments are generated by a covariate adaptive biased-coin rule that balances constant and the first two moments covariate features, where the feature map is
    $\phi\left(X_{i}\right)=\left(\sqrt{w_{0}},\sqrt{w_{1}}X_{i}^{\mathrm{T}},\sqrt{w_{2}}\mathrm{vec}(X_{i}X_{i}^{\mathrm{T}})^{\mathrm{T}}\right)^{\mathrm{T}},$ and $w_{0},w_{1},w_{2}\geq0$ are the corresponding weights.
    We test the performance of  \texttt{COV}\((1,1,0)\) and \texttt{COV}\((1,p,1)\). 
    In the former method, we balance the constant to control the overall difference $\sum_{i=1}^n(2T_i-1)=N_1-N_0$, and the first moments of covariate, with weights $(w_{0},w_{1},w_{2})=(1,1,0)$. 
    In the latter method, we simultaneously consider balancing the constant, the first moments and second moments of covariate, with weights $(w_{0},w_{1},w_{2})=(1,p,1)$, where $p$ is the covariate dimension, and  $w_{1}/w_{2}=p$ was set to make the magnitudes of the first moment and second moment imbalances more comparable.
    \item \texttt{ACCEPOCausalPFN:} Here CausalPFN is used to estimate $\mu_a(X)$ and then treatment assignments are generated by balancing this estimated ACCEPO score $\widehat{m}_{n_0}(X)$.
    \item \texttt{ACCEPOTabPFN}: Here TabPFN3.5 is used to estimate $\mu_a(X)$ and then treatment assignments are generated by balancing this estimated ACCEPO score $\widehat{m}_{n_0}(X)$.
    \item \texttt{ACCEPOTabICL}: Here TabICLv2 is used to estimate $\mu_a(X)$ and then treatment assignments are generated by balancing this estimated ACCEPO score $\widehat{m}_{n_0}(X)$.
    \item \texttt{ACCEPOTrue:} Here treatment assignments are generated by balancing the true  $m(X)$ implied by the data generating model. 
\end{itemize}

Implementation details for the PFN-based methods are provided in
Appendix~\ref{app:pfn-implementation}.

\subsection{PFN Implementation Details}
\label{app:pfn-implementation}

\paragraph*{PFN-based prediction.}
Prior-data fitted networks (PFNs) use transformers to approximate the posterior predictive distributions (PPDs) induced by a prior over data generating mechanisms \citep{muller2022transformers,Nagler2023PFN,stith2026causalfoundations}. A PFN is first pretrained offline on many synthetic datasets sampled from a prior over tasks. Let $\theta$ denote the network parameters and $\widehat{\theta}$ their values obtained after pretraining. At deployment, the pretrained network maps a context dataset and a query input to a predictive distribution in a single forward pass, without refitting $\widehat{\theta}$ to the new dataset. We use TabPFN3.5, TabICLv2, and CausalPFN to estimate the CEPOs for each arm, using the first stage data as the context \citep{HollmannEtAl2025TabPFN,balazadehCausalPFNAmortizedCausal2025,qu2026tabiclv2}. 
These predictions are then combined to construct the ACCEPO balancing score in Section~\ref{sec:baselineframework}.

\paragraph*{Context encoding.}
For a tabular predictor, treatment identity can be represented
through different context constructions, such as:
\begin{itemize}
    \item \emph{Disjoint encoding:} use a separate predictor instance for each arm. Each instance receives only the first stage data from its own arm.

    \item \emph{Joint encoding:} use one predictor with first stage data from both arms. Observations from both arms form a shared context, with treatment identity included as an additional input through one-hot encoding.
\end{itemize}

We use disjoint encoding for TabPFN and TabICLv2. Specifically, the context for arm $a$ is $\mathcal D_{n_0,a}=\{(X_j,Y_j):1\leq j\leq n_0,\ T_j=a\}, a\in\{0,1\}.$ Each arm has its own instance of the same pretrained regressor. The treatment indicator is used to partition the first stage samples and not included as an input feature.

CausalPFN directly incorporates treatment indicators and uses factual observations as context. During its  pretraining phase, supervision is provided by CEPO values obtained from simulated data generating mechanisms, enabling the model to approximate posterior distributions over the CEPOs \citep{balazadehCausalPFNAmortizedCausal2025}. In our implementation, the full first stage sample $\mathcal D_{n_0}=\{(Y_j,T_j,X_j):1\leq j\leq n_0\}$ is supplied as the shared context for a pretrained CausalPFN model. For the same query covariate vector $x$, we query the model under treatment levels $a=0$ and $a=1$ to obtain the CEPO predictions for the two arms.

\paragraph*{Point predictions and score construction.}
We use the mean point predictions returned by the model interfaces and denote them by $f_{\widehat{\theta},a}(x;\mathcal D_{n_0})$, we let $\widehat{\mu}_{n_0,a}(x)=f_{\widehat{\theta},a}(x;\mathcal D_{n_0}), a\in\{0,1\},$ and construct $\widehat{m}_{n_0}(x)=\frac12\sum_{a=0}^1\left\{\widehat{\mu}_{n_0,a}(x)-\bar{Y}_a^{(1)}\right\},$ where $\bar{Y}_a^{(1)}$ is the mean of the observed first stage outcomes in arm $a$. 

\subsection{Additional Results on ACCEPO Balance}
\label{app:additional-balance}
Tables~\ref{tab:balance_property_lambda0.1} and~\ref{tab:balance_property_lambda0.2} supplement Table~\ref{tab:balance_property_lambda0.15} at first stage proportions $\lambda=0.10$ and $0.20$.
All entries are Monte Carlo means of the absolute imbalance sums divided by $\sqrt n$  based on 1,000 replications per setting.

\begin{table}[!htbp]
\centering
\caption{Monte Carlo means of absolute ACCEPO imbalances divided by $\sqrt{n}$ at $\lambda=0.1$. Dashes denote unavailable learned scores.}
\label{tab:balance_property_lambda0.1}
\resizebox{\textwidth}{!}{%
\begin{tabular}{@{}ll*{10}{c}@{}}
\toprule
& & \multicolumn{5}{c}{$n^{-1/2}\left|\sum_{i=1}^{n}(2T_i-1)\widehat{m}_{n_0}(X_i)\right|$} & \multicolumn{5}{c}{$n^{-1/2}\left|\sum_{i=1}^{n}(2T_i-1)m(X_i)\right|$} \\
\cmidrule(lr){3-7}\cmidrule(lr){8-12}
& & \multicolumn{5}{c}{Scenario} & \multicolumn{5}{c}{Scenario} \\
Method & $N$ & 1 & 2 & 3 & 4 & 5 & 1 & 2 & 3 & 4 & 5 \\
\midrule
SR & 1000 & -- & -- & -- & -- & -- & 1.036 & 0.503 & 2.934 & 0.567 & 0.862 \\
 & 2000 & -- & -- & -- & -- & -- & 1.050 & 0.510 & 2.984 & 0.576 & 0.889 \\
 & 5000 & -- & -- & -- & -- & -- & 1.034 & 0.534 & 3.136 & 0.591 & 0.859 \\
\addlinespace[2pt]
\texttt{COV(1,1,0)} & 1000 & -- & -- & -- & -- & -- & 0.153 & 0.297 & 2.507 & 0.545 & 0.716 \\
 & 2000 & -- & -- & -- & -- & -- & 0.102 & 0.303 & 2.515 & 0.544 & 0.735 \\
 & 5000 & -- & -- & -- & -- & -- & 0.067 & 0.295 & 2.631 & 0.531 & 0.715 \\
\addlinespace[2pt]
\texttt{COV(1,p,1)} & 1000 & -- & -- & -- & -- & -- & 0.242 & 0.309 & 1.222 & 0.390 & 0.640 \\
 & 2000 & -- & -- & -- & -- & -- & 0.178 & 0.295 & 1.036 & 0.333 & 0.639 \\
 & 5000 & -- & -- & -- & -- & -- & 0.134 & 0.295 & 0.910 & 0.300 & 0.674 \\
\addlinespace[2pt]
\texttt{ACCEPOCausalPFN} & 1000 & 0.035 & 0.018 & 0.073 & 0.014 & 0.018 & 0.370 & 0.264 & 1.951 & 0.496 & 0.744 \\
 & 2000 & 0.024 & 0.012 & 0.063 & 0.011 & 0.015 & 0.225 & 0.168 & 1.306 & 0.374 & 0.627 \\
 & 5000 & 0.015 & 0.007 & 0.044 & 0.008 & 0.011 & 0.123 & 0.095 & 0.842 & 0.255 & 0.444 \\
\addlinespace[2pt]
\texttt{ACCEPOTabPFN} & 1000 & 0.030 & 0.012 & 0.059 & 0.008 & 0.013 & 0.262 & 0.213 & 1.556 & 0.436 & 0.658 \\
 & 2000 & 0.022 & 0.009 & 0.063 & 0.010 & 0.015 & 0.191 & 0.146 & 0.969 & 0.276 & 0.487 \\
 & 5000 & 0.015 & 0.006 & 0.046 & 0.007 & 0.011 & 0.120 & 0.100 & 0.523 & 0.211 & 0.356 \\
\addlinespace[2pt]
\texttt{ACCEPOTabICL} & 1000 & 0.035 & 0.015 & 0.078 & 0.013 & 0.021 & 0.317 & 0.208 & 1.759 & 0.500 & 0.646 \\
 & 2000 & 0.023 & 0.010 & 0.061 & 0.011 & 0.016 & 0.232 & 0.140 & 1.213 & 0.359 & 0.514 \\
 & 5000 & 0.015 & 0.006 & 0.043 & 0.007 & 0.011 & 0.149 & 0.089 & 0.826 & 0.228 & 0.415 \\
\addlinespace[2pt]
\texttt{ACCEPOTrue} & 1000 & -- & -- & -- & -- & -- & 0.034 & 0.015 & 0.104 & 0.018 & 0.030 \\
 & 2000 & -- & -- & -- & -- & -- & 0.025 & 0.011 & 0.077 & 0.013 & 0.021 \\
 & 5000 & -- & -- & -- & -- & -- & 0.016 & 0.007 & 0.050 & 0.008 & 0.013 \\
\bottomrule
\end{tabular}
}
\end{table}

\begin{table}[!htbp]
\centering
\caption{Monte Carlo means of absolute ACCEPO imbalances divided by $\sqrt{n}$ at $\lambda=0.2$. Dashes denote unavailable learned scores.}
\label{tab:balance_property_lambda0.2}
\resizebox{\textwidth}{!}{%
\begin{tabular}{@{}ll*{10}{c}@{}}
\toprule
& & \multicolumn{5}{c}{$n^{-1/2}\left|\sum_{i=1}^{n}(2T_i-1)\widehat{m}_{n_0}(X_i)\right|$} & \multicolumn{5}{c}{$n^{-1/2}\left|\sum_{i=1}^{n}(2T_i-1)m(X_i)\right|$} \\
\cmidrule(lr){3-7}\cmidrule(lr){8-12}
& & \multicolumn{5}{c}{Scenario} & \multicolumn{5}{c}{Scenario} \\
Method & $N$ & 1 & 2 & 3 & 4 & 5 & 1 & 2 & 3 & 4 & 5 \\
\midrule
SR & 1000 & -- & -- & -- & -- & -- & 1.024 & 0.538 & 3.033 & 0.599 & 0.826 \\
 & 2000 & -- & -- & -- & -- & -- & 1.055 & 0.510 & 3.063 & 0.578 & 0.841 \\
 & 5000 & -- & -- & -- & -- & -- & 1.062 & 0.539 & 3.102 & 0.586 & 0.869 \\
\addlinespace[2pt]
\texttt{COV(1,1,0)} & 1000 & -- & -- & -- & -- & -- & 0.155 & 0.302 & 2.522 & 0.525 & 0.686 \\
 & 2000 & -- & -- & -- & -- & -- & 0.106 & 0.294 & 2.613 & 0.535 & 0.680 \\
 & 5000 & -- & -- & -- & -- & -- & 0.068 & 0.296 & 2.530 & 0.520 & 0.703 \\
\addlinespace[2pt]
\texttt{COV(1,p,1)} & 1000 & -- & -- & -- & -- & -- & 0.243 & 0.304 & 1.311 & 0.409 & 0.646 \\
 & 2000 & -- & -- & -- & -- & -- & 0.186 & 0.300 & 1.037 & 0.349 & 0.657 \\
 & 5000 & -- & -- & -- & -- & -- & 0.147 & 0.297 & 0.926 & 0.308 & 0.681 \\
\addlinespace[2pt]
\texttt{ACCEPOCausalPFN} & 1000 & 0.036 & 0.017 & 0.090 & 0.017 & 0.021 & 0.219 & 0.167 & 1.359 & 0.364 & 0.613 \\
 & 2000 & 0.025 & 0.012 & 0.073 & 0.013 & 0.018 & 0.137 & 0.108 & 0.917 & 0.273 & 0.505 \\
 & 5000 & 0.016 & 0.007 & 0.049 & 0.008 & 0.012 & 0.089 & 0.064 & 0.653 & 0.215 & 0.392 \\
\addlinespace[2pt]
\texttt{ACCEPOTabPFN} & 1000 & 0.035 & 0.014 & 0.093 & 0.016 & 0.024 & 0.184 & 0.160 & 1.031 & 0.269 & 0.511 \\
 & 2000 & 0.024 & 0.010 & 0.076 & 0.012 & 0.019 & 0.136 & 0.111 & 0.621 & 0.224 & 0.372 \\
 & 5000 & 0.016 & 0.007 & 0.050 & 0.008 & 0.012 & 0.093 & 0.076 & 0.359 & 0.165 & 0.273 \\
\addlinespace[2pt]
\texttt{ACCEPOTabICL} & 1000 & 0.036 & 0.016 & 0.092 & 0.015 & 0.024 & 0.231 & 0.142 & 1.259 & 0.363 & 0.504 \\
 & 2000 & 0.024 & 0.010 & 0.067 & 0.012 & 0.018 & 0.156 & 0.102 & 0.951 & 0.252 & 0.418 \\
 & 5000 & 0.016 & 0.007 & 0.049 & 0.008 & 0.012 & 0.107 & 0.064 & 0.565 & 0.186 & 0.353 \\
\addlinespace[2pt]
\texttt{ACCEPOTrue} & 1000 & -- & -- & -- & -- & -- & 0.036 & 0.016 & 0.113 & 0.019 & 0.032 \\
 & 2000 & -- & -- & -- & -- & -- & 0.026 & 0.011 & 0.079 & 0.014 & 0.022 \\
 & 5000 & -- & -- & -- & -- & -- & 0.017 & 0.007 & 0.051 & 0.008 & 0.014 \\
\bottomrule
\end{tabular}
}
\end{table}

\subsection{Additional Results on Treatment Effect Estimation}
\label{app:additional-estimation}
Tables~\ref{tab:estimation_fixed_lambda_10} and~\ref{tab:estimation_fixed_lambda_20} report the Monte Carlo mean and standard deviation of the unpooled difference-in-means estimator at $\lambda=0.10,0.20$, respectively. 

\begin{table}[!htbp]
\centering
\caption{Monte Carlo mean (standard deviation) of the unpooled difference-in-means estimator at $\lambda=0.10$.}
\label{tab:estimation_fixed_lambda_10}
\resizebox{\textwidth}{!}{%
\begin{tabular}{@{}llccccc@{}}
\toprule
& & \multicolumn{5}{c}{Scenario} \\
\cmidrule(lr){3-7}
Method & $N$ & 1 & 2 & 3 & 4 & 5 \\
\midrule
SR & 1000 & 1.998 (0.111) & 2.001 (0.081) & 2.007 (0.278) & 2.002 (0.080) & 2.004 (0.113) \\
 & 2000 & 1.999 (0.080) & 2.002 (0.055) & 2.004 (0.201) & 2.003 (0.057) & 2.000 (0.077) \\
 & 5000 & 1.999 (0.049) & 1.999 (0.036) & 2.000 (0.131) & 2.000 (0.036) & 2.002 (0.050) \\
\addlinespace[2pt]
\texttt{COV(1,1,0)} & 1000 & 1.999 (0.069) & 1.999 (0.073) & 1.983 (0.250) & 2.000 (0.082) & 2.003 (0.097) \\
 & 2000 & 1.999 (0.047) & 2.003 (0.052) & 2.013 (0.177) & 1.999 (0.059) & 1.999 (0.073) \\
 & 5000 & 1.999 (0.030) & 2.000 (0.032) & 1.999 (0.117) & 2.002 (0.036) & 2.001 (0.046) \\
\addlinespace[2pt]
\texttt{COV(1,p,1)} & 1000 & 1.999 (0.069) & 2.002 (0.075) & 1.983 (0.168) & 2.002 (0.074) & 2.005 (0.098) \\
 & 2000 & 2.001 (0.046) & 2.003 (0.051) & 2.004 (0.113) & 2.000 (0.052) & 1.998 (0.069) \\
 & 5000 & 2.000 (0.030) & 2.001 (0.033) & 2.001 (0.069) & 2.001 (0.031) & 2.000 (0.046) \\
\addlinespace[2pt]
\texttt{ACCEPOCausalPFN} & 1000 & 2.000 (0.073) & 1.997 (0.077) & 1.998 (0.210) & 2.000 (0.078) & 2.004 (0.103) \\
 & 2000 & 2.002 (0.048) & 2.003 (0.051) & 2.000 (0.125) & 1.998 (0.053) & 2.002 (0.069) \\
 & 5000 & 2.000 (0.031) & 1.999 (0.031) & 1.999 (0.069) & 2.001 (0.030) & 2.001 (0.041) \\
\addlinespace[2pt]
\texttt{ACCEPOTabPFN} & 1000 & 1.999 (0.067) & 2.000 (0.073) & 1.997 (0.195) & 1.999 (0.077) & 2.003 (0.102) \\
 & 2000 & 2.000 (0.047) & 2.002 (0.049) & 2.000 (0.108) & 2.000 (0.051) & 2.002 (0.067) \\
 & 5000 & 2.000 (0.030) & 1.999 (0.031) & 2.002 (0.063) & 2.000 (0.030) & 2.002 (0.040) \\
\addlinespace[2pt]
\texttt{ACCEPOTabICL} & 1000 & 2.000 (0.069) & 2.001 (0.071) & 1.997 (0.198) & 2.002 (0.079) & 2.003 (0.100) \\
 & 2000 & 2.001 (0.048) & 2.001 (0.048) & 2.005 (0.122) & 2.000 (0.052) & 2.001 (0.066) \\
 & 5000 & 2.000 (0.030) & 1.999 (0.031) & 1.998 (0.069) & 2.001 (0.031) & 2.002 (0.040) \\
\addlinespace[2pt]
\texttt{ACCEPOTrue} & 1000 & 1.999 (0.064) & 2.002 (0.067) & 1.994 (0.132) & 2.000 (0.068) & 2.003 (0.084) \\
 & 2000 & 2.000 (0.048) & 2.002 (0.049) & 1.998 (0.096) & 2.001 (0.047) & 2.001 (0.058) \\
 & 5000 & 2.000 (0.029) & 2.000 (0.030) & 2.002 (0.057) & 2.001 (0.029) & 2.002 (0.037) \\
\bottomrule
\end{tabular}
}
\end{table}

\begin{table}[!htbp]
\centering
\caption{Monte Carlo mean (standard deviation) of the unpooled difference-in-means estimator at $\lambda=0.20$.}
\label{tab:estimation_fixed_lambda_20}
\resizebox{\textwidth}{!}{%
\begin{tabular}{@{}llccccc@{}}
\toprule
& & \multicolumn{5}{c}{Scenario} \\
\cmidrule(lr){3-7}
Method & $N$ & 1 & 2 & 3 & 4 & 5 \\
\midrule
SR & 1000 & 1.995 (0.120) & 1.999 (0.089) & 1.992 (0.304) & 2.003 (0.090) & 2.007 (0.112) \\
 & 2000 & 2.002 (0.085) & 2.002 (0.058) & 2.001 (0.212) & 1.999 (0.061) & 2.003 (0.080) \\
 & 5000 & 1.999 (0.054) & 1.999 (0.039) & 2.002 (0.140) & 2.000 (0.038) & 1.999 (0.054) \\
\addlinespace[2pt]
\texttt{COV(1,1,0)} & 1000 & 1.999 (0.071) & 1.998 (0.079) & 1.997 (0.262) & 2.004 (0.083) & 2.008 (0.108) \\
 & 2000 & 1.999 (0.049) & 2.003 (0.053) & 2.004 (0.196) & 2.000 (0.060) & 2.001 (0.075) \\
 & 5000 & 1.999 (0.032) & 2.000 (0.034) & 2.003 (0.118) & 2.001 (0.037) & 2.003 (0.048) \\
\addlinespace[2pt]
\texttt{COV(1,p,1)} & 1000 & 1.998 (0.073) & 2.002 (0.078) & 1.985 (0.179) & 2.000 (0.077) & 2.004 (0.105) \\
 & 2000 & 1.999 (0.053) & 1.999 (0.054) & 1.997 (0.116) & 2.002 (0.054) & 2.001 (0.075) \\
 & 5000 & 2.000 (0.034) & 2.000 (0.035) & 2.001 (0.074) & 2.001 (0.033) & 2.000 (0.047) \\
\addlinespace[2pt]
\texttt{ACCEPOCausalPFN} & 1000 & 2.000 (0.072) & 1.999 (0.076) & 1.989 (0.182) & 1.999 (0.078) & 2.004 (0.106) \\
 & 2000 & 2.000 (0.051) & 2.001 (0.052) & 2.002 (0.118) & 2.001 (0.053) & 2.002 (0.069) \\
 & 5000 & 1.999 (0.033) & 2.000 (0.034) & 2.000 (0.066) & 2.001 (0.032) & 2.001 (0.043) \\
\addlinespace[2pt]
\texttt{ACCEPOTabPFN} & 1000 & 1.998 (0.072) & 2.000 (0.074) & 1.998 (0.166) & 2.003 (0.072) & 2.006 (0.100) \\
 & 2000 & 2.003 (0.049) & 2.002 (0.053) & 2.007 (0.110) & 2.000 (0.051) & 2.001 (0.065) \\
 & 5000 & 2.000 (0.033) & 2.000 (0.033) & 1.999 (0.065) & 2.000 (0.032) & 2.002 (0.040) \\
\addlinespace[2pt]
\texttt{ACCEPOTabICL} & 1000 & 1.995 (0.072) & 2.001 (0.073) & 1.994 (0.177) & 2.001 (0.079) & 2.005 (0.099) \\
 & 2000 & 2.001 (0.050) & 2.003 (0.052) & 2.005 (0.121) & 2.002 (0.052) & 2.001 (0.068) \\
 & 5000 & 2.000 (0.032) & 2.000 (0.034) & 2.002 (0.067) & 2.000 (0.031) & 2.001 (0.043) \\
\addlinespace[2pt]
\texttt{ACCEPOTrue} & 1000 & 2.000 (0.068) & 1.999 (0.073) & 1.987 (0.139) & 2.001 (0.068) & 2.004 (0.088) \\
 & 2000 & 2.002 (0.048) & 2.002 (0.052) & 2.004 (0.103) & 2.001 (0.050) & 2.000 (0.062) \\
 & 5000 & 2.000 (0.032) & 1.999 (0.032) & 2.001 (0.064) & 2.001 (0.030) & 2.001 (0.039) \\
\bottomrule
\end{tabular}
}
\end{table}

\FloatBarrier

\subsection{Additional Inference Results Using the Section~\ref{sec:L2convergence results} Variance Estimator}
\label{app:working-inference}
Table~\ref{tab:pfn_plugin_wald_working} reports coverage of nominal 95\% unpooled confidence intervals using the Section~\ref{sec:L2convergence results} variance estimator. Coverage remains generally close to the nominal level, but deviations tend to be slightly larger than those obtained with the variance estimator from Section~\ref{sec:general-score-inference}.

\begin{table}[!htbp]
\centering
\caption{Empirical coverage (\%) of confidence intervals based on the variance estimator from Section~\ref{sec:L2convergence results}.}
\label{tab:pfn_plugin_wald_working}
\resizebox{\textwidth}{!}{%
\begin{tabular}{@{}ll*{15}{c}@{}}
\toprule
& & \multicolumn{5}{c}{$\lambda=0.10$} & \multicolumn{5}{c}{$\lambda=0.15$} & \multicolumn{5}{c}{$\lambda=0.20$} \\
\cmidrule(lr){3-7}\cmidrule(lr){8-12}\cmidrule(lr){13-17}
& & \multicolumn{5}{c}{Scenario} & \multicolumn{5}{c}{Scenario} & \multicolumn{5}{c}{Scenario} \\
Method & $N$ & 1 & 2 & 3 & 4 & 5 & 1 & 2 & 3 & 4 & 5 & 1 & 2 & 3 & 4 & 5 \\
\midrule
\texttt{ACCEPOCausalPFN} & 1000 & 96.3 & 96.6 & 95.9 & 96.0 & 95.2 & 95.9 & 95.4 & 95.5 & 95.7 & 95.1 & 95.4 & 95.4 & 95.9 & 95.3 & 94.7 \\
 & 2000 & 96.1 & 95.1 & 95.3 & 94.4 & 95.4 & 95.1 & 95.9 & 96.8 & 96.0 & 93.2 & 96.0 & 95.3 & 94.3 & 93.7 & 93.9 \\
 & 5000 & 94.6 & 95.2 & 95.8 & 95.2 & 94.6 & 95.4 & 95.4 & 94.4 & 96.5 & 93.5 & 94.2 & 93.4 & 96.3 & 96.1 & 93.4 \\
\addlinespace[2pt]
\texttt{ACCEPOTabPFN} & 1000 & 97.3 & 95.3 & 97.6 & 93.7 & 94.7 & 95.9 & 96.2 & 97.3 & 95.1 & 94.4 & 96.2 & 95.2 & 96.7 & 95.4 & 93.8 \\
 & 2000 & 95.9 & 95.0 & 97.4 & 94.3 & 93.2 & 95.1 & 94.8 & 96.5 & 94.2 & 92.8 & 96.0 & 94.4 & 96.8 & 94.3 & 93.9 \\
 & 5000 & 95.7 & 94.7 & 96.1 & 95.8 & 93.2 & 96.5 & 95.5 & 95.7 & 95.6 & 94.0 & 94.7 & 95.1 & 96.3 & 95.9 & 94.9 \\
\addlinespace[2pt]
\texttt{ACCEPOTabICL} & 1000 & 97.5 & 96.6 & 98.3 & 96.2 & 96.0 & 96.7 & 94.8 & 96.1 & 94.6 & 95.0 & 97.4 & 95.2 & 97.4 & 94.4 & 95.2 \\
 & 2000 & 97.0 & 96.3 & 97.2 & 95.5 & 94.9 & 96.0 & 95.4 & 95.9 & 94.6 & 94.5 & 95.7 & 95.3 & 95.1 & 95.6 & 94.4 \\
 & 5000 & 96.2 & 95.2 & 95.4 & 95.2 & 94.3 & 95.5 & 96.4 & 95.6 & 95.3 & 93.6 & 94.8 & 94.2 & 96.2 & 96.5 & 94.5 \\
\bottomrule
\end{tabular}
}
\end{table}

\FloatBarrier

\subsection{Simulation Results for Pooling}
\label{sec:pooled-unpooled-simulation}
We compare the unpooled estimator $\widehat{\tau}(\widehat{m}_{n_0})$ with the pooled estimator $\widehat{\tau}_{\mathrm{pool}}$ proposed in Section~\ref{sec:pool-estimation}. In this simulation, the first stage ATE estimator $\widehat{\tau}_1$ is obtained by fully interacted OLS \citep{MaTuLiu2022RegressionAnalysis}.
For the first stage data, we use the working model
\[
Y_j=\alpha+\tau T_j+\beta^\top (X_j-\bar{X}_{\mathrm{first}})+\gamma^\top\{T_j(X_j-\bar{X}_{\mathrm{first}})\}+\varepsilon_j,
\]
where $X_j$ contains all baseline covariates used in the simulation and $\bar{X}_{\mathrm{first}}$ is their first stage sample mean. The estimator $\widehat{\tau}_1$ is the OLS coefficient on $T_j$.

Let $\widehat{b}_{\mathrm{adj}}
=\widehat{\beta}+\widehat{\gamma}/2$, for $a=0,1$, let $n_0^{a}=\sum_{j=1}^{n_0}\mathbf1\{T_j=a\}$, and let $\bar{Y}_a$ and $\bar{X}_a$ be the first stage sample means in arm $a$.
We estimate the asymptotic variance of $\widehat{\tau}_1$ by $$\widehat{V}_1=2\sum_{a=0}^1\frac{1}{n_0^a}\sum_{j:T_j=a}\{Y_j-\bar{Y}_a-\widehat{b}_{\mathrm{adj}}^\top(X_j-\bar{X}_a)\}^2.$$

Under simple randomization with equal allocation and the regularity conditions of \citet[Theorem~3]{MaTuLiu2022RegressionAnalysis},
$\widehat{\tau}_1$ is asymptotically normal and $\widehat{V}_1$ consistently estimates its asymptotic variance, as required by Assumption~\ref{assum:pilotate}.
We combine $\widehat{V}_1$ with the second stage variance estimator $\widehat{V}_{n}$ from Section~\ref{sec:L2convergence results} to obtain the pooled estimator weighted by inverse variances and the corresponding variance estimator of the pooled estimator.

Table~\ref{tab:pooling_plugin_N2000} reports the Monte Carlo mean of the pooled estimator, the percentage reduction in Monte Carlo variance relative to the unpooled estimator, and confidence interval coverage probability at $N=2000$. The variance reduction is defined by 
$\left\{1-\frac{\widehat{\operatorname{Var}}_{\mathrm{MC}}(\widehat{\tau}_{\mathrm{pool}})}{\widehat{\operatorname{Var}}_{\mathrm{MC}}(\widehat{\tau})}\right\}\times100\%$.
All variance reductions are positive, confirming the precision gains from pooling.

\begin{table}[!htbp]
\centering
\renewcommand{\baselinestretch}{1}\small
\setlength{\tabcolsep}{3pt}
\renewcommand{\arraystretch}{1.02}
\caption{Monte Carlo mean of the pooled estimator, percentage variance reduction, and pooled confidence interval coverage probability with $\alpha=0.05$. }
\label{tab:pooling_plugin_N2000}
\resizebox{\textwidth}{!}{%
\begin{tabular}{@{}ll*{10}{c}@{}}
\toprule
& & \multicolumn{5}{c}{Pooled ATE mean (Variance Reduction\%)} & \multicolumn{5}{c}{Coverage (\%)} \\
\cmidrule(lr){3-7}\cmidrule(lr){8-12}
& & \multicolumn{5}{c}{Scenario} & \multicolumn{5}{c}{Scenario} \\
Method & $\lambda$ & 1 & 2 & 3 & 4 & 5 & 1 & 2 & 3 & 4 & 5 \\
\midrule
\texttt{ACCEPOCausalPFN} & 0.10 & 2.001 (10.7) & 2.003 (10.4) & 2.000 (6.9) & 1.998 (10.7) & 2.001 (6.3) & 96.7 & 94.8 & 95.4 & 94.1 & 94.5 \\
 & 0.15 & 2.003 (15.1) & 2.000 (11.1) & 2.004 (7.4) & 2.001 (11.6) & 2.000 (11.0) & 95.0 & 95.8 & 96.6 & 95.7 & 93.7 \\
 & 0.20 & 1.999 (18.6) & 2.001 (18.5) & 2.002 (10.5) & 2.000 (15.9) & 2.001 (15.8) & 95.0 & 95.3 & 94.6 & 94.3 & 94.5 \\
\addlinespace[2pt]
\texttt{ACCEPOTabPFN} & 0.10 & 1.999 (10.2) & 2.002 (6.4) & 2.000 (2.7) & 1.999 (8.8) & 2.002 (8.4) & 95.6 & 94.9 & 97.1 & 93.6 & 93.6 \\
 & 0.15 & 1.999 (15.1) & 2.002 (14.9) & 2.001 (9.7) & 2.000 (13.5) & 1.999 (12.2) & 94.9 & 95.1 & 97.0 & 94.1 & 93.7 \\
 & 0.20 & 2.001 (16.7) & 2.001 (17.2) & 2.006 (9.8) & 1.999 (17.2) & 2.001 (14.7) & 95.0 & 93.6 & 95.9 & 95.0 & 94.8 \\
\addlinespace[2pt]
\texttt{ACCEPOTabICL} & 0.10 & 2.000 (9.5) & 2.001 (8.3) & 2.007 (5.7) & 2.000 (10.7) & 2.001 (9.0) & 96.8 & 95.9 & 97.1 & 95.3 & 95.1 \\
 & 0.15 & 1.997 (13.2) & 2.002 (13.0) & 2.002 (9.2) & 2.001 (12.8) & 2.002 (14.6) & 95.8 & 95.4 & 96.0 & 95.8 & 94.1 \\
 & 0.20 & 1.999 (17.4) & 2.002 (19.1) & 2.005 (8.2) & 2.001 (17.3) & 2.000 (17.8) & 95.7 & 95.3 & 95.2 & 95.0 & 95.1 \\
\addlinespace[2pt]
\texttt{ACCEPOTrue} & 0.10 & 2.000 (9.8) & 2.001 (9.1) & 1.999 (4.4) & 2.001 (8.2) & 2.001 (8.3) & 94.3 & 94.5 & 94.5 & 95.9 & 94.7 \\
 & 0.15 & 1.999 (13.4) & 2.001 (14.5) & 2.002 (2.3) & 2.000 (11.2) & 1.999 (11.2) & 94.8 & 94.9 & 95.4 & 95.0 & 94.7 \\
 & 0.20 & 2.001 (17.3) & 2.001 (19.8) & 2.004 (8.3) & 2.000 (14.0) & 1.999 (16.1) & 95.3 & 95.0 & 93.9 & 94.9 & 94.4 \\
\bottomrule
\end{tabular}
}
\end{table}

\FloatBarrier

\subsection{Simulation Results for Multi-stage Design}
\label{sec:multi-stage-simulation}
We now compare the two-stage design procedures with multi-stage design procedures proposed in Section~\ref{sec:dynamic-updating}. We compare seven methods: the two-stage and multi-stage versions of \texttt{ACCEPOCausalPFN}, \texttt{ACCEPOTabPFN}, and \texttt{ACCEPOTabICL}, together with the oracle benchmark \texttt{ACCEPOTrue}. We set $N=5000$, $\lambda=0.10$, $K=9$. 
Results are based on 1,000 Monte Carlo replications per scenario. 
Table~\ref{tab:dynamic_update} reports the Monte Carlo mean and standard deviation of the treatment effect estimator defined in Section~\ref{sec:dynamic-updating}, the Monte Carlo mean of the normalized absolute ACCEPO imbalance and the runtime.

\begin{table}[!htbp]
\centering
\renewcommand{\baselinestretch}{1}\small
\caption{Monte Carlo mean and standard deviation of the ATE estimator, Imbalance is the Monte Carlo mean of the normalized absolute ACCEPO imbalance $n^{-1/2}|\sum_{i=1}^{n}(2T_i-1)m(X_i)|$, and the mean wall clock time per replication.}
\label{tab:dynamic_update}
\resizebox{\textwidth}{!}{%
\begin{tabular}{@{}clcccc@{}}
\toprule
Scenario & Method & ATE mean & ATE SD & Imbalance & Runtime (s) \\
\midrule
\multirow{7}{*}{1}& \texttt{ACCEPOCausalPFN} & 2.000 & 0.031 & 0.123 & 12.64 \\
&\texttt{ACCEPOCausalPFN(multi)} & 1.999 & 0.029 & 0.092 & 37.22 \\
&\texttt{ACCEPOTabPFN} & 2.000 & 0.030 & 0.120 & 4.42 \\
&\texttt{ACCEPOTabPFN(multi)}  & 1.999 & 0.030 & 0.087 & 27.76 \\
&\texttt{ACCEPOTabICL} & 2.000 & 0.030 & 0.149 & 2.87 \\
&\texttt{ACCEPOTabICL(multi)} & 2.000 & 0.030 & 0.106 & 17.85 \\
&\texttt{ACCEPOTrue} & 2.000 & 0.029 & 0.016 & 0.10 \\
\midrule
\multirow{7}{*}{2}& \texttt{ACCEPOCausalPFN} & 1.999 & 0.031 & 0.095 & 12.37 \\
&\texttt{ACCEPOCausalPFN(multi)}  & 1.999 & 0.031 & 0.057 & 33.18 \\
&\texttt{ACCEPOTabPFN} & 1.999 & 0.031 & 0.100 & 4.26 \\
&\texttt{ACCEPOTabPFN(multi)}  & 2.001 & 0.032 & 0.065 & 26.86 \\
&\texttt{ACCEPOTabICL} & 1.999 & 0.031 & 0.089 & 1.97 \\
&\texttt{ACCEPOTabICL(multi)}  & 2.000 & 0.030 & 0.059 & 15.93 \\
&\texttt{ACCEPOTrue} & 2.000 & 0.030 & 0.007 & 0.10 \\
\midrule
\multirow{7}{*}{3}& \texttt{ACCEPOCausalPFN} & 1.999 & 0.069 & 0.842 & 12.76 \\
&\texttt{ACCEPOCausalPFN(multi)}  & 2.000 & 0.063 & 0.565 & 43.48 \\
&\texttt{ACCEPOTabPFN} & 2.002 & 0.063 & 0.523 & 4.51 \\
&\texttt{ACCEPOTabPFN(multi)}  & 2.000 & 0.061 & 0.334 & 28.19 \\
&\texttt{ACCEPOTabICL} & 1.998 & 0.069 & 0.826 & 3.06 \\
&\texttt{ACCEPOTabICL(multi)}  & 2.002 & 0.062 & 0.450 & 18.33 \\
&\texttt{ACCEPOTrue} & 2.002 & 0.057 & 0.050 & 0.10 \\
\midrule
\multirow{7}{*}{4}& \texttt{ACCEPOCausalPFN} & 2.001 & 0.030 & 0.255 & 12.56 \\
&\texttt{ACCEPOCausalPFN(multi)}  & 2.001 & 0.030 & 0.195 & 43.43 \\
&\texttt{ACCEPOTabPFN} & 2.000 & 0.030 & 0.211 & 4.51 \\
&\texttt{ACCEPOTabPFN(multi)}  & 2.000 & 0.030 & 0.136 & 28.19 \\
&\texttt{ACCEPOTabICL} & 2.001 & 0.031 & 0.228 & 3.06 \\
&\texttt{ACCEPOTabICL(multi)}  & 2.001 & 0.030 & 0.170 & 18.34 \\
&\texttt{ACCEPOTrue} & 2.001 & 0.029 & 0.008 & 0.10 \\
\midrule
\multirow{7}{*}{5}& \texttt{ACCEPOCausalPFN} & 2.001 & 0.041 & 0.444 & 12.85 \\
&\texttt{ACCEPOCausalPFN(multi)}  & 2.002 & 0.040 & 0.373 & 44.69 \\
&\texttt{ACCEPOTabPFN} & 2.002 & 0.040 & 0.356 & 4.43 \\
&\texttt{ACCEPOTabPFN(multi)}  & 2.001 & 0.039 & 0.219 & 27.80 \\
&\texttt{ACCEPOTabICL} & 2.002 & 0.040 & 0.415 & 3.07 \\
&\texttt{ACCEPOTabICL(multi)}  & 2.002 & 0.039 & 0.317 & 18.23 \\
&\texttt{ACCEPOTrue} & 2.002 & 0.037 & 0.013 & 0.10 \\
\bottomrule
\end{tabular}
}
\end{table}

The results suggest that updating the balancing score through multi-stage design can improve finite sample precision and achieve a smaller ACCEPO imbalance relative to the two-stage design. These gains are consistent with improved balancing score estimation from accumulating data. The multi-stage procedure also incurs higher computational because the balancing score is repeatedly updated using the accumulated data. Its practical value therefore depends on whether the finite sample gain in precision justifies the additional computational and operational cost.

\FloatBarrier

\subsection{Comparison with PROCOVA}
\label{app:procova-implementation}
In this section, we compare our method with PROCOVA. PROCOVA learns the control prognostic score $\mu_0(X)$ from historical control data, and uses it as an additional covariate for interacted regression adjustment under second stage simple randomization \citep{Schuler2022}.
Let $\mathcal I_1$ and $\mathcal I_2$ denote the first stage and second stage samples, with sample sizes $n_0$ and $n$. In our simulation, the method using PROCOVA is implemented as follows \citep{Schuler2022}.
\begin{enumerate}
    \item[(1)]  Units in $\mathcal I_1$ are assigned by simple randomization, then after observing their outcomes, we collect the first stage data $\mathcal D_{n_0}=\{(Y_j,T_j,X_j):j\in\mathcal I_1\}.$
    \item[(2)]  Estimate the control prognostic score $\mu_0(X)$ from the control arm data in the first stage $\mathcal D_{n_0,0}=\{(Y_j,X_j):j\in\mathcal I_1,\ T_j=0\}$
    and the estimated score is denoted by $\widehat{\mu}_{n_0,0}(x).$
    \item[(3)]  Units in $\mathcal I_2$ are assigned by simple randomization. After observing their outcomes, use the learned prognostic scores $\widehat{\mu}_{n_0,0}(X_i), i\in\mathcal I_2$ in the final regression adjustment. Let$$
    \widetilde{T}_i=T_i-\bar{T},\qquad
    \widetilde{X}_i=X_i-\bar{X},\qquad
    \widetilde{\mu}_{n_0,0,i}=\widehat{\mu}_{n_0,0}(X_i)-\bar{\mu}_{n_0,0},\qquad i\in\mathcal I_2,$$ denote the empirically centered versions, where $\bar{T}=n^{-1}\sum_{i\in\mathcal I_2}T_i$, $\bar{X}=n^{-1}\sum_{i\in\mathcal I_2}X_i$, and $\bar{\mu}_{n_0,0}=n^{-1}\sum_{i\in\mathcal I_2}\widehat{\mu}_{n_0,0}(X_i)$. Then we fit the interacted regression
    \begin{equation}
    \label{eq:procova-regression}
    Y_i=\alpha+\tau \widetilde{T}_i+\beta^\top \widetilde{X}_i+\eta \widetilde{\mu}_{n_0,0,i}
    +\gamma^\top(\widetilde{T}_i\widetilde{X}_i)+\xi \widetilde{T}_i\widetilde{\mu}_{n_0,0,i}+e_i,\qquad i\in\mathcal I_2.
    \end{equation}
    Then we denote the ATE estimator using PROCOVA by $\widehat{\tau}_{\mathrm{PROCOVA}}$, which is the ordinary least squares coefficient on $\widetilde{T}_i$ in \eqref{eq:procova-regression}. 
\end{enumerate}

We consider comparing eight methods:
\begin{itemize}
    \item \texttt{ACCEPOCausalPFN}, \texttt{ACCEPOTabPFN}, \texttt{ACCEPOTabICL}, \texttt{ACCEPOTrue}: The same as in Section~\ref{sec:simulations}.
    \item \texttt{PROCOVACausalPFN}: Use CausalPFN to estimate the control prognostic score $\widehat{\mu}_{n_0,0}(x)$.
    \item \texttt{PROCOVATabPFN}: Use TabPFN~3.5 to estimate the control prognostic score $\widehat{\mu}_{n_0,0}(x)$.
    \item \texttt{PROCOVATabICL}: Use TabICLv2 to estimate the control prognostic score $\widehat{\mu}_{n_0,0}(x)$.
    \item \texttt{PROCOVATrue}: Use the true control prognostic score $\mu_{0}(x)$ implied by the data generating model.
\end{itemize}

We use total sample sizes $N=1000$, first stage proportions $\lambda=0.2$, and 1000 Monte Carlo replications. Results are reported in Table~\ref{tab:compareProcova}. 

\begin{table}[!htbp]
\centering
\small
\caption{Monte Carlo comparison between PROCOVA and our two-stage randomization procedure.  Each scenario reports the Monte Carlo mean and empirical variance of the final ATE estimator, with variances multiplied by $10^3$.}
\label{tab:compareProcova}
\resizebox{\textwidth}{!}{
\begin{tabular}{l*{5}{cc}}
\toprule
& \multicolumn{2}{c}{Scenario 1} & \multicolumn{2}{c}{Scenario 2} & \multicolumn{2}{c}{Scenario 3} & \multicolumn{2}{c}{Scenario 4} & \multicolumn{2}{c}{Scenario 5} \\
\cmidrule(lr){2-3}\cmidrule(lr){4-5}\cmidrule(lr){6-7}\cmidrule(lr){8-9}\cmidrule(lr){10-11}
Method & Mean & $10^3\times\mathrm{Var}$ & Mean & $10^3\times\mathrm{Var}$ & Mean & $10^3\times\mathrm{Var}$ & Mean & $10^3\times\mathrm{Var}$ & Mean & $10^3\times\mathrm{Var}$ \\
\midrule
\texttt{PROCOVACausalPFN} & 2.002 & 4.809 & 2.000 & 5.681 & 1.992 & 33.572 & 2.000 & 6.660 & 2.002 & 12.271 \\
\texttt{PROCOVATabPFN} & 1.997 & 5.090 & 1.999 & 5.922 & 1.996 & 30.881 & 1.997 & 7.338 & 2.003 & 12.189 \\
\texttt{PROCOVATabICL} & 2.000 & 4.748 & 2.002 & 5.133 & 1.994 & 31.690 & 2.002 & 6.944 & 2.003 & 10.742 \\
\texttt{PROCOVATrue} & 1.998 & 4.848 & 2.000 & 5.219 & 1.994 & 20.656 & 2.002 & 7.483 & 2.004 & 12.423 \\
\midrule
\texttt{ACCEPOCausalPFN} & 2.000 & 5.219 & 1.999 & 5.716 & 1.989 & 33.149 & 1.999 & 6.104 & 2.004 & 11.144 \\
\texttt{ACCEPOTabPFN} & 1.998 & 5.231 & 2.000 & 5.476 & 1.998 & 27.519 & 2.003 & 5.252 & 2.006 & 10.034 \\
\texttt{ACCEPOTabICL} & 1.995 & 5.171 & 2.001 & 5.305 & 1.994 & 31.205 & 2.001 & 6.292 & 2.005 & 9.809 \\
\texttt{ACCEPOTrue} & 2.000 & 4.600 & 1.999 & 5.367 & 1.987 & 19.371 & 2.001 & 4.622 & 2.004 & 7.758 \\
\bottomrule
\end{tabular}
}
\end{table}

Table~\ref{tab:compareProcova} shows that the Monte Carlo means are all close to the true ATE of 2.0. PROCOVA and our method have comparable variances in Scenarios~1--2, whereas our method yields lower empirical variances than PROCOVA in Scenarios~3--5. Overall, both approaches effectively exploit first stage information, while our proposed design provides more efficiency gain beyond the regime with a constant treatment effect.

\section{Auxiliary Lemmas With Proofs}
\subsection{Lemma~\ref{lem:pilot_arm_means}}
\label{app:conditional-conventions}
The following lemma controls the centering error from using first stage arm means in the learned balancing score, as used in the proof of Proposition~\ref{prop:hatm-l2-consistency}.

\begin{lemma}
\label{lem:pilot_arm_means}
Suppose $T_j\overset{\mathrm{i.i.d.}}{\sim}\operatorname{Bernoulli}(1/2)$, $j=1,\ldots,n_0$. For $a=0,1$, if $E|Y(a)|<\infty$, then $\bar{Y}_a^{(1)}\xrightarrow{p}E\{Y(a)\}$ and hence $\bar{Y}_a^{(1)}=O_p(1)$. If, in addition, $E\{Y^2(a)\}<\infty$, then
$$\bar{Y}_a^{(1)}-E\{Y(a)\}=O_p(n_0^{-1/2}),\qquad a=0,1.$$
\end{lemma}

\begin{proof}
Fix $a\in\{0,1\}$, define $\bar{Y}_a^{(1)}$ arbitrarily when $N_{n_0,a}=0$, an event of probability $2^{-n_0}$. By independence and the weak law of large numbers, $N_{n_0,a}/n_0\xrightarrow{p}1/2$, 
and on $\{N_{n_0,a}>0\}$,
$$\bar{Y}_a^{(1)}-E\{Y(a)\}=\frac{n_0^{-1}\sum_{j=1}^{n_0}\mathbf1\{T_j=a\}\{Y_j(a)-E\{Y(a)\}\}}{N_{n_0,a}/n_0}.$$
The numerator converges in probability to zero by the weak law of large numbers. If $E\{Y^2(a)\}<\infty$, it is $O_p(n_0^{-1/2})$ by Chebyshev's inequality. 
Both conclusions follow from Slutsky's theorem.
\end{proof}

\subsection{Lemmas~\ref{lem:conditional-small-to-unconditional} and~\ref{lem:conditional-big-to-unconditional}}
The following lemmas transfer conditional convergence and stochastic bounds to unconditional conclusions under joint asymptotics for the first and second stages. For the convergence result, see also \citet[Lemma~B.8]{XinLiuMa2024ManyStrata}.

\begin{lemma}
\label{lem:conditional-small-to-unconditional}
For $n_0,n\ge1$, let $U_{n_0,n}$ be real-valued random variables and let $\mathcal F_{n_0}$ be a sigma field for each $n_0$. Suppose
$$P\left(|U_{n_0,n}-c|\ge\varepsilon\mid\mathcal F_{n_0}\right)\xrightarrow{p}0$$
for any $\varepsilon>0$ and some constant $c$. Then $P(|U_{n_0,n}-c|\ge\varepsilon)\to0$ for any $\varepsilon>0$ and the same constant $c$.
\end{lemma}

\begin{proof}
For any $\varepsilon>0$, the conditional probability is bounded by one and converges to zero in probability, hence also in $L_1$.
The tower property therefore gives
$$P(|U_{n_0,n}-c|\ge\varepsilon)
=E\left[P\left(|U_{n_0,n}-c|\ge\varepsilon\mid\mathcal F_{n_0}\right)\right]\to E(0)=0.$$
\end{proof}

\begin{lemma}
\label{lem:conditional-big-to-unconditional}
Let $b_{n_0,n}>0$ be deterministic and let $U_{n_0,n}$ be real-valued random variables. Suppose $U_{n_0,n}=O_{p(\cdot\mid\mathcal F_{n_0,n})}(b_{n_0,n}),$
that is, for every $\eta>0$,
$$\lim_{C\to\infty}\limsup_{n_0,n\to\infty}P\!\left[P\{|U_{n_0,n}|>C b_{n_0,n}\mid\mathcal F_{n_0,n}\}>\eta\right]=0.$$
Then $U_{n_0,n}=O_p(b_{n_0,n})$.
\end{lemma}

\begin{proof}
For any $C,\eta>0$, the tower property and boundedness of conditional probabilities give
\begin{align*}
P(|U_{n_0,n}|>C b_{n_0,n})
&=E\!\left[P\{|U_{n_0,n}|>C b_{n_0,n}\mid\mathcal F_{n_0,n}\}\right]\\
&\le \eta+
P\!\left[P\{|U_{n_0,n}|>C b_{n_0,n}\mid\mathcal F_{n_0,n}\}>\eta\right].
\end{align*}
Taking $\limsup_{n_0,n\to\infty}$ and then letting $C\to\infty$
and $\eta\downarrow0$ gives
$U_{n_0,n}=O_p(b_{n_0,n})$.
\end{proof}

\subsection{Lemma~\ref{lem:conditional-lindeberg-feller}}
The following lemma gives the conditional normal approximation used in the proof of Theorem~\ref{theo:clt for tau_m}.
A related argument for conditional triangular array appears in \citet[Lemmas~B.4--B.5]{XinLiuMa2024ManyStrata}.

\begin{lemma}
\label{lem:conditional-lindeberg-feller}
Let $\mathcal F_r\subseteq\mathcal H_r$ be sigma fields on a common probability space. For each $r$, let $X_{r,1},\ldots,X_{r,k_r}$ be square-integrable real random variables that are conditionally independent given $\mathcal H_r$, with $E(X_{r,i}\mid\mathcal H_r)=0$. Suppose that, for a deterministic $\sigma^2\in(0,\infty)$,
$$V_r:=\sum_{i=1}^{k_r}E(X_{r,i}^2\mid\mathcal H_r)\xrightarrow{p}\sigma^2,$$
and that, for every $\eta>0$,
$$L_r(\eta):=\sum_{i=1}^{k_r}E\!\left[X_{r,i}^2\mathbf1\{|X_{r,i}|>\eta\}\mid\mathcal H_r\right]\xrightarrow{p}0.$$
Write $S_r:=\sum_{i=1}^{k_r}X_{r,i}$ and $\psi_r(u):=E(e^{iuS_r}\mid\mathcal H_r)$. Then, for every fixed $u\in\mathbb R$,
$$\psi_r(u)\xrightarrow{p}\exp(-u^2\sigma^2/2), \quad E\!\left[\left|\psi_r(u)-\exp(-u^2\sigma^2/2)\right|\middle|\mathcal F_r\right]\xrightarrow{p}0.$$
\end{lemma}

\begin{proof}
For each $r$, let $Q_r(\omega,\cdot)$ be a regular conditional distribution of $(X_{r,1},\ldots,X_{r,k_r})$ given $\mathcal H_r$, and use it to define the conditional moments and characteristic functions. By conditional independence, centering, and square integrability, each $Q_r$ is almost surely a product distribution with centered coordinates and finite second moments. Taking the intersection over the countably many rows gives an event $\Omega_*$ of probability one on which these properties hold for every $r$. On this event, $L_r(\eta)(\omega)$ is nonincreasing in $\eta$.

Start from an arbitrary deterministic subsequence $(r_m)$. For each fixed $j$, the assumed convergence in probability and a union bound over finitely many terms allow us to choose a deterministic $m_j>m_{j-1}$, with $m_0=0$, such that, writing $s_j=r_{m_j}$,
$$A_j:=\{|V_{s_j}-\sigma^2|>j^{-1}\}
\cup\left\{\max_{1\le\ell\le j}L_{s_j}(1/\ell)>j^{-1}\right\},
\quad P(A_j)\le2^{-j}.$$
Since $\sum_jP(A_j)<\infty$, the Borel--Cantelli lemma \citep[Theorem~2.3.1]{Durrett2019PTE} gives $P(\limsup_{j\to\infty}A_j)=0$. Define
$\Omega_0:=\Omega_*\setminus\limsup_{j\to\infty}A_j,$ then $P(\Omega_0)=1.$
For every $\omega\in\Omega_0$, $\omega\notin A_j$ for all sufficiently large $j$. Thus, for each fixed integer $\ell\ge1$, also $j\ge\ell$,
$$|V_{s_j}(\omega)-\sigma^2|\le j^{-1},
\qquad 0\le L_{s_j}(1/\ell)(\omega)\le j^{-1}.$$
Thus $V_{s_j}(\omega)\to\sigma^2$ and $L_{s_j}(1/\ell)(\omega)\to0$ for every $\ell$ on the same event $\Omega_0$. For any $\eta>0$, choose $\ell$ such that $1/\ell<\eta$. Monotonicity then gives $0\le L_{s_j}(\eta)(\omega)\le L_{s_j}(1/\ell)(\omega)\longrightarrow0.$
This verifies every Lindeberg threshold on $\Omega_0$ along the same further subsequence.

Fix $\omega\in\Omega_0$. Under $Q_{s_j}(\omega,\cdot)$, the coordinates in each row are independent and centered, with total variance $V_{s_j}(\omega)$ and Lindeberg quantity $L_{s_j}(\eta)(\omega)$. 
The limits above verify the conditions of the Lindeberg--Feller theorem \citep[Theorem~3.4.10]{Durrett2019PTE}; no independence between rows is required. 
Thus the distributions of the row sums converge weakly to $N(0,\sigma^2)$, 
and $$\psi_{s_j}(u)(\omega)\longrightarrow\exp(-u^2\sigma^2/2) \quad\text{for every }u\in\mathbb R.$$
For each fixed $u$, since the subsequence $(r_m)$ was arbitrary, every subsequence of $\psi_r(u)$ has a further subsequence converging almost surely to $\exp(-u^2\sigma^2/2)$. Applying the subsequence principle \citep[Theorem~2.3.2]{Durrett2019PTE} to the absolute difference yields $\psi_r(u)\xrightarrow{p}\exp(-u^2\sigma^2/2)$ for the full sequence.

Finally, for fixed $u$, $E|\psi_r(u)-e^{-u^2\sigma^2/2}|\to0$ since the absolute error is bounded by $2$. For every $a>0$, Markov's inequality and the tower property now give
\begin{equation*}
P\!\left\{E\!\left[|\psi_r(u)-e^{-u^2\sigma^2/2}|\mid\mathcal F_r\right]>a\right\}
\le a^{-1}E|\psi_r(u)-e^{-u^2\sigma^2/2}|\longrightarrow0.
\end{equation*}
\end{proof}

\subsection{Lemmas~\ref{lem:uniform-learned-feature-imbalance} and~\ref{lem:uniform-target-imbalance}}
The next two lemmas control the imbalance of a learned scalar score and of other targets under the same biased-coin rule, the latter through an exact martingale decomposition. 
Related results for prespecified targets appear in \citet[Theorems~3.1 and~3.5]{MaJASA2024CAR}.
Here, the score may depend on the training data, and the conditional bounds hold under joint growth of the training and subsequent sample sizes, without restrictions on their relative rates.

For integers $t,n\ge1$, let $t$ index the training history, $n$ the subsequent sample size, and $j$ arrivals within a row. 
Write $\mathcal F_t$ for the training history sigma field, and let the learned balancing target $\phi_t(x)$ be jointly measurable in the training history and $x$. 
For the two-stage design in the main text, $t=n_0$ is the first stage sample size, $n$ is the second stage sample size, and $\phi_{n_0}(x)=\widehat{m}_{n_0}(x)$.
Conditional on $\mathcal F_t$, the future covariates $X_{t,j}$ are i.i.d. with the fixed law $P_X$. Generate $T_{t,j}$ by the scalar biased-coin rule in Section~\ref{sec:baselineframework}, with fixed $1/2<\rho<1$. For a measurable target function $h$, define its imbalance by
$$\Delta_{t,n}(h)=\sum_{j=1}^n(2T_{t,j}-1)h(X_{t,j}).$$
Although the learned target and transition laws may vary with $t$, we derive conditional bounds within each row and localize them using constants independent of the row. Lemma~\ref{lem:uniform-learned-feature-imbalance} first establishes the balancing property of the learned balancing target $\phi_t$ used for treatment assignment.

\begin{lemma}
\label{lem:uniform-learned-feature-imbalance}
For $r>1$, write $\beta_{r,t}:=E\{|\phi_t(X)|^r\mid\mathcal F_t\}$. Suppose that, for some $\gamma>2$, $\beta_{\gamma,t}$ is finite almost surely for each $t$ and $\beta_{\gamma,t}=O_p(1)$.
Let$$\delta_\gamma:=\left(\frac{\gamma-2}{\gamma-1}\right)^2.$$
For every $0<B<\infty$, there exists $K_B<\infty$, depending only on $B$, $\gamma$, and $\rho$, such that, on $\{\beta_{\gamma,t}\le B\}$,$$E\left[\left\{\frac{\Delta_{t,n}(\phi_t)}{\sqrt n}\right\}^2\middle|\mathcal F_t\right] \le K_Bn^{-\delta_\gamma}$$
for every $n\ge1$. Consequently, along every sequence $t,n\to\infty$,$$\frac{\Delta_{t,n}(\phi_t)}{\sqrt n}=o_{p(\cdot\mid\mathcal F_t)}(1).$$
\end{lemma}

\begin{proof}
Fix $t$ and condition on $\mathcal F_t$. Write
$$Z_j:=\phi_t(X_{t,j}),\qquad \Lambda_j:=\Delta_{t,j}(\phi_t),\qquad \Lambda_0=0.$$
Then $\Lambda_j=\Lambda_{j-1}+(2T_{t,j}-1)Z_j$. 
Set $c_t:=E(|Z_j|\mid\mathcal F_t)$ and $\kappa_\rho:=2\rho-1$, with $\operatorname{sgn}(0)=0$. 
Conditional on $\mathcal F_t$, the $Z_j$ are i.i.d., and $Z_j$ is independent of the previous covariates and assignments. 
Since $|\Lambda_j|\le\sum_{\ell=1}^j|Z_\ell|$, all conditional moments below are finite up to order $\gamma$.
If $c_t=0$, then $Z_j=0$ conditionally almost surely and the result is immediate. Hence suppose $c_t>0$.

For every $p\ge2$, a Taylor expansion gives
$$|x+y|^p-|x|^p\le pxy|x|^{p-2}+C_p\{|y|^p+|y|^2|x|^{p-2}\},\qquad x,y\in\mathbb R,$$
where $C_p>0$ is a finite constant depending only on $p$. The biased-coin rule gives
$$E\{(2T_{t,j}-1)Z_j\mid\mathcal F_t,\Lambda_{j-1},Z_j\} =-\kappa_\rho\,\operatorname{sgn}(\Lambda_{j-1}Z_j)Z_j.$$
Apply the Taylor inequality with $x=\Lambda_{j-1}$ and $y=(2T_{t,j}-1)Z_j$. Using this conditional mean and then averaging over $Z_j$ gives, for $2\le p\le\gamma$,
\begin{equation}
\label{eq:uniform-learned-drift} E\{|\Lambda_j|^p-|\Lambda_{j-1}|^p\mid\mathcal F_t,\Lambda_{j-1}\} \le-p\kappa_\rho c_t|\Lambda_{j-1}|^{p-1} +C_p\beta_{p,t}+C_p\beta_{2,t}|\Lambda_{j-1}|^{p-2}.
\end{equation}

Let $\alpha_\gamma:=(\gamma-2)/(\gamma-1)$. Conditional H\"older's inequality yields
\begin{equation}
\label{eq:uniform-learned-interpolation} \beta_{2,t} =E\left\{|Z_j|^{\alpha_\gamma}|Z_j|^{\gamma/(\gamma-1)} \middle|\mathcal F_t\right\} \le c_t^{\alpha_\gamma}\beta_{\gamma,t}^{1/(\gamma-1)},\qquad \beta_{2,t}^{\gamma-1}\le c_t^{\gamma-2}\beta_{\gamma,t}.
\end{equation}
Taking $p=\gamma$ in \eqref{eq:uniform-learned-drift}, maximizing the two terms involving $|\Lambda_{j-1}|$, and using \eqref{eq:uniform-learned-interpolation}, we obtain
$$\sup_{x\ge0}\{-\gamma\kappa_\rho c_tx^{\gamma-1} +C_\gamma\beta_{2,t}x^{\gamma-2}\} \le C_{\gamma,\rho}^{\mathrm{drift}} \frac{\beta_{2,t}^{\gamma-1}}{c_t^{\gamma-2}} \le C_{\gamma,\rho}^{\mathrm{drift}}\beta_{\gamma,t},$$
where
$$C_{\gamma,\rho}^{\mathrm{drift}}:= \frac{(\gamma-2)^{\gamma-2}}{(\gamma-1)^{\gamma-1}} \frac{C_\gamma^{\gamma-1}}{(\gamma\kappa_\rho)^{\gamma-2}}$$
is a finite positive constant depending only on $\gamma,\rho$.
Fix $B>0$ and set $K_B^{(\gamma)}:=(C_\gamma+C_{\gamma,\rho}^{\mathrm{drift}})B$.
On $\{\beta_{\gamma,t}\le B\}$, \eqref{eq:uniform-learned-drift} and the preceding bound give
$$E(|\Lambda_j|^\gamma\mid\mathcal F_t) \le E(|\Lambda_{j-1}|^\gamma\mid\mathcal F_t)+K_B^{(\gamma)}.$$
Iterating this inequality from $\Lambda_0=0$ yields
\begin{equation}
\label{eq:uniform-learned-gamma-moment} E(|\Lambda_j|^\gamma\mid\mathcal F_t)\le K_B^{(\gamma)}j,\qquad j\ge0.
\end{equation}
In particular, $K_B^{(\gamma)}$ is independent of $c_t$ and the training row.

For $p=2$, direct expansion gives the exact identity
\begin{equation}
\label{eq:uniform-learned-second-drift} E(\Lambda_j^2-\Lambda_{j-1}^2\mid\mathcal F_t) =\beta_{2,t}-2\kappa_\rho c_tE(|\Lambda_{j-1}|\mid\mathcal F_t).
\end{equation}
For fixed $n$, define
$$j_*:=\max\left\{0\le j\le n-1: 2\kappa_\rho c_tE(|\Lambda_j|\mid\mathcal F_t)\le\beta_{2,t}\right\}.$$
The set is nonempty because $\Lambda_0=0$. The index $j_*$ is $\mathcal F_t$-measurable and is fixed once the training information is given.
For $j_*<j\le n-1$, its definition and \eqref{eq:uniform-learned-second-drift} give
$E(\Lambda_{j+1}^2\mid\mathcal F_t)\le E(\Lambda_j^2\mid\mathcal F_t)$.
Applying \eqref{eq:uniform-learned-second-drift} once more at $j_*+1$ therefore gives
\begin{equation}
\label{eq:uniform-learned-last-index} E(\Lambda_n^2\mid\mathcal F_t) \le E(\Lambda_{j_*+1}^2\mid\mathcal F_t) \le E(\Lambda_{j_*}^2\mid\mathcal F_t)+\beta_{2,t}.
\end{equation}

The same conditional H\"older inequality used in \eqref{eq:uniform-learned-interpolation}, now applied to $|\Lambda_{j_*}|$, gives
\begin{equation}
\label{eq:uniform-learned-second-interpolation} E(\Lambda_{j_*}^2\mid\mathcal F_t) \le \{E(|\Lambda_{j_*}|\mid\mathcal F_t)\}^{\alpha_\gamma} \{E(|\Lambda_{j_*}|^\gamma\mid\mathcal F_t)\}^{1/(\gamma-1)} \le \left(\frac{\beta_{2,t}}{2\kappa_\rho c_t}\right)^{\alpha_\gamma} (K_B^{(\gamma)}n)^{1/(\gamma-1)}.
\end{equation}
The second inequality uses the defining bound for $j_*$ and
\eqref{eq:uniform-learned-gamma-moment}, with $j_*\le n-1$.
On $\{\beta_{\gamma,t}\le B\}$, \eqref{eq:uniform-learned-interpolation} also gives
$$\frac{\beta_{2,t}}{c_t} \le c_t^{\alpha_\gamma-1}B^{1/(\gamma-1)} =c_t^{-1/(\gamma-1)}B^{1/(\gamma-1)}.$$
Define the intermediate constant
$$L_B:=\max\left\{ B^{2/\gamma},\ \frac{(K_B^{(\gamma)})^{1/(\gamma-1)} B^{(\gamma-2)/(\gamma-1)^2}}{(2\kappa_\rho)^{\alpha_\gamma}} \right\}.$$
It depends only on $B,\gamma,\rho$. Substituting the preceding bound for $\beta_{2,t}/c_t$
into \eqref{eq:uniform-learned-second-interpolation}, and using
$\alpha_\gamma/(\gamma-1)=(\gamma-2)/(\gamma-1)^2$, yields
$$E(\Lambda_{j_*}^2\mid\mathcal F_t) \le L_Bn^{1/(\gamma-1)}c_t^{-(\gamma-2)/(\gamma-1)^2}.$$
Consequently, \eqref{eq:uniform-learned-last-index} implies
\begin{equation}
\label{eq:uniform-learned-second-prelocal} E(\Lambda_n^2\mid\mathcal F_t) \le L_Bn^{1/(\gamma-1)}c_t^{-(\gamma-2)/(\gamma-1)^2}+\beta_{2,t}.
\end{equation}

To remove the dependence on $c_t$, consider two cases.
If $c_t\ge n^{-1}$, conditional Jensen's inequality gives
$\beta_{2,t}\le\beta_{\gamma,t}^{2/\gamma}\le B^{2/\gamma}\le L_B$.
Dividing \eqref{eq:uniform-learned-second-prelocal} by $n$ and using
$c_t^{-(\gamma-2)/(\gamma-1)^2}\le n^{(\gamma-2)/(\gamma-1)^2}$, we obtain
$$\frac1nE(\Lambda_n^2\mid\mathcal F_t) \le L_Bn^{1/(\gamma-1)-1+(\gamma-2)/(\gamma-1)^2}+L_Bn^{-1} =L_Bn^{-\delta_\gamma}+L_Bn^{-1}.$$
If $c_t<n^{-1}$, use $|\Lambda_n|\le\sum_{j=1}^n|Z_j|$ and the conditional independence of the $Z_j$ to obtain
$$E(\Lambda_n^2\mid\mathcal F_t) \le \sum_{j=1}^nE(Z_j^2\mid\mathcal F_t) +2\sum_{1\le i<j\le n}E(|Z_i||Z_j|\mid\mathcal F_t) =n\beta_{2,t}+n(n-1)c_t^2.$$
By \eqref{eq:uniform-learned-interpolation},
$\beta_{2,t}\le B^{1/(\gamma-1)}c_t^{\alpha_\gamma}
\le B^{1/(\gamma-1)}n^{-\alpha_\gamma}$, while
$(n-1)c_t^2\le n^{-1}$. Hence
$$\frac1nE(\Lambda_n^2\mid\mathcal F_t) \le B^{1/(\gamma-1)}n^{-\alpha_\gamma}+n^{-1}.$$
Since $\delta_\gamma=\alpha_\gamma^2<\alpha_\gamma<1$, both
$n^{-\alpha_\gamma}$ and $n^{-1}$ are at most $n^{-\delta_\gamma}$.
The two cases therefore give
\begin{equation}
\label{eq:uniform-learned-second-local} \frac1nE(\Lambda_n^2\mid\mathcal F_t) \le K_Bn^{-\delta_\gamma},\qquad K_B:=\max\{2L_B,B^{1/(\gamma-1)}+1\},
\end{equation}
on $\{\beta_{\gamma,t}\le B\}$, with the same deterministic $K_B$ for every $t$ and $n\ge1$.

Finally, for $\varepsilon,\eta>0$, conditional Markov's inequality and
\eqref{eq:uniform-learned-second-local} imply, on $\{\beta_{\gamma,t}\le B\}$,
$$P\left\{\left|\frac{\Delta_{t,n}(\phi_t)}{\sqrt n}\right|>\varepsilon \middle|\mathcal F_t\right\} \le\frac{E(\Lambda_n^2\mid\mathcal F_t)}{n\varepsilon^2} \le\frac{K_Bn^{-\delta_\gamma}}{\varepsilon^2}.$$
Thus, if the conditional probability exceeds $\eta$, either
$\beta_{\gamma,t}>B$ or $K_Bn^{-\delta_\gamma}>\eta\varepsilon^2$. Taking the outer probability gives
\begin{equation}
\label{eq:uniform-learned-joint-tail} P\left[P\left\{\left|\frac{\Delta_{t,n}(\phi_t)}{\sqrt n}\right|>\varepsilon \middle|\mathcal F_t\right\}>\eta\right] \le P(\beta_{\gamma,t}>B) +\mathbf1\{K_Bn^{-\delta_\gamma}>\eta\varepsilon^2\}.
\end{equation}
Let $(t_k,n_k)$ be any deterministic sequence with $t_k\to\infty$ and $n_k\to\infty$.
For fixed $B$, the indicator in \eqref{eq:uniform-learned-joint-tail} is eventually zero because $K_B$ is independent of $t_k$ and $\delta_\gamma>0$.
The upper limit of the left side is therefore bounded by
$\limsup_kP(\beta_{\gamma,t_k}>B)\le\limsup_{t\to\infty}P(\beta_{\gamma,t}>B)$.
Since $\beta_{\gamma,t}=O_p(1)$,
$$\lim_{B\to\infty}\limsup_{t\to\infty}P(\beta_{\gamma,t}>B)=0.$$
Letting $B\to\infty$ proves the conditional convergence along the chosen joint sequence, without a relative growth restriction.
\end{proof}

Lemma~\ref{lem:uniform-target-imbalance} then controls the imbalance of other functions of the covariates by separating a martingale term from the imbalance of the learned score.
\begin{lemma}
\label{lem:uniform-target-imbalance}
For some $\gamma>2$, suppose
$$\beta_{\gamma,t}=E\{|\phi_t(X)|^\gamma\mid\mathcal F_t\}<\infty\quad\text{almost surely for each }t,\qquad \beta_{\gamma,t}=O_p(1).$$
Also assume $c_t^{-1}=O_p(1)$, where $c_t=E\{|\phi_t(X)|\mid\mathcal F_t\}$ and $1/0=+\infty$.
The following statements hold along every deterministic sequence with $t\to\infty$ and $n\to\infty$, without a restriction on their relative rates.
\begin{enumerate}
\item[(i)]
Let $h_t(x)$ be jointly measurable in the training history and $x$. Suppose $\eta_{2,t}=E\{h_t^2(X)\mid\mathcal F_t\}$ is finite almost surely for each $t$ and $\eta_{2,t}=O_p(1)$. Then
$$\frac{\Delta_{t,n}(h_t)}{\sqrt n}=O_{p(\cdot\mid\mathcal F_t)}(1).$$
\item[(ii)]
For every fixed measurable $h$ with $E|h(X)|<\infty$,
$$\frac{\Delta_{t,n}(h)}n=o_{p(\cdot\mid\mathcal F_t)}(1).$$
\end{enumerate}
\end{lemma}

\begin{proof}
Fix $t$ and work under the conditional law given $\mathcal F_t$. For a target $h_t$ with $\eta_{2,t}<\infty$, define the coefficient and adjusted target by
\begin{equation}\label{eq:scalar-target-projection}
d_t=\begin{cases}
E[h_t(X)\operatorname{sgn}\{\phi_t(X)\}\mid\mathcal F_t]/c_t,&c_t>0,\\
0,&c_t=0,
\end{cases}
\qquad w_t=h_t-d_t\phi_t.
\end{equation}
Conditional Cauchy--Schwarz gives $|d_t|\le\eta_{2,t}^{1/2}/c_t<\infty$ when $c_t>0$, while $d_t=0$ when $c_t=0$. Also, $\beta_{2,t}\le\beta_{\gamma,t}^{2/\gamma}<\infty$ almost surely. Thus the adjusted target is square-integrable under the conditional law:
$$\sigma_{\phi_t}^2(h_t):=E\{w_t^2(X)\mid\mathcal F_t\}
\le2\eta_{2,t}+2d_t^2\beta_{2,t}<\infty.$$
Introduce the history
$$\mathcal G_{t,j}:=\sigma\{\mathcal F_t,X_{t,\ell},T_{t,\ell}:1\le\ell\le j\},\qquad \mathcal G_{t,0}=\mathcal F_t.$$
Write $\Delta_{t,0}(\phi_t)=0$ and retain $\kappa_\rho=2\rho-1$. The scalar assignment rule gives
\begin{equation}\label{eq:scalar-assignment-drift}
E(2T_{t,j}-1\mid\mathcal G_{t,j-1},X_{t,j})
=-\kappa_\rho\operatorname{sgn}\{\Delta_{t,j-1}(\phi_t)\}
\operatorname{sgn}\{\phi_t(X_{t,j})\}.
\end{equation}
When $c_t>0$, the choice of $d_t$ ensures
$$E[w_t(X)\operatorname{sgn}\{\phi_t(X)\}\mid\mathcal F_t]
=E[h_t(X)\operatorname{sgn}\{\phi_t(X)\}\mid\mathcal F_t]-d_tc_t=0.$$
When $c_t=0$, $\phi_t(X)=0$ conditionally almost surely, so this expectation is also zero and the assignments are fair coins. Set
$$D_{t,j}:=(2T_{t,j}-1)w_t(X_{t,j}),\qquad M_{t,n}:=\sum_{j=1}^nD_{t,j},\qquad M_{t,0}=0.$$
Since $X_{t,j}$ is conditionally independent of $\mathcal G_{t,j-1}$ given $\mathcal F_t$, conditioning first on $X_{t,j}$ in \eqref{eq:scalar-assignment-drift} yields
\begin{equation}\label{eq:scalar-target-martingale-mean}
E(D_{t,j}\mid\mathcal G_{t,j-1})
=-\kappa_\rho\operatorname{sgn}\{\Delta_{t,j-1}(\phi_t)\}
E[w_t(X)\operatorname{sgn}\{\phi_t(X)\}\mid\mathcal F_t]=0.
\end{equation}
Also, since $(2T_{t,j}-1)^2=1$, 
$$E(D_{t,j}^2\mid\mathcal G_{t,j-1})=E\{w_t^2(X)\mid\mathcal F_t\}=\sigma_{\phi_t}^2(h_t).$$
Thus, conditional on the training history, $D_{t,j}$ are square-integrable martingale differences and $M_{t,n}$ is their partial sum. For $\ell<j$, the tower property gives
$E(D_{t,\ell}D_{t,j}\mid\mathcal F_t)
=E\{D_{t,\ell}E(D_{t,j}\mid\mathcal G_{t,j-1})\mid\mathcal F_t\}=0$.
Expanding the square and using $h_t=w_t+d_t\phi_t$ therefore gives
\begin{equation}
\label{eq:scalar-exact-decomposition}
\Delta_{t,n}(h_t)=M_{t,n}+d_t\Delta_{t,n}(\phi_t),\qquad
E(M_{t,n}^2\mid\mathcal F_t)=n\sigma_{\phi_t}^2(h_t).
\end{equation}

\textit{Part (i).}
Fix $B,H>0$ and $\kappa>0$, and set
$\mathcal E_t:=\{\beta_{\gamma,t}\le B,\eta_{2,t}\le H,c_t\ge\kappa\}$.
Conditional Cauchy--Schwarz gives
$|E[h_t(X)\operatorname{sgn}\{\phi_t(X)\}\mid\mathcal F_t]|^2\le\eta_{2,t}$.
Also $\beta_{2,t}\le\beta_{\gamma,t}^{2/\gamma}$ by conditional Jensen's inequality. Hence, on $\mathcal E_t$,
\begin{equation}\label{eq:scalar-target-coefficient-bounds}
d_t^2\le H/\kappa^2,\qquad
\sigma_{\phi_t}^2(h_t)\le2\eta_{2,t}+2d_t^2\beta_{2,t}
\le2H(1+B^{2/\gamma}/\kappa^2).
\end{equation}
Apply $(x+y)^2\le2x^2+2y^2$ to \eqref{eq:scalar-exact-decomposition}. The score bound \eqref{eq:uniform-learned-second-local} and \eqref{eq:scalar-target-coefficient-bounds} imply
$$\begin{aligned}
E\{\Delta_{t,n}^2(h_t)/n\mid\mathcal F_t\}
&\le2\sigma_{\phi_t}^2(h_t)+2d_t^2K_Bn^{-\delta_\gamma}\\
&\le4H(1+B^{2/\gamma}/\kappa^2)+2HK_B\kappa^{-2}n^{-\delta_\gamma}.
\end{aligned}$$
Here $K_B$ and $\delta_\gamma$ are the same constant and exponent as in Lemma~\ref{lem:uniform-learned-feature-imbalance}. Since $n^{-\delta_\gamma}\le1$, we obtain, on $\mathcal E_t$ and for every $n\ge1$,
\begin{equation}\label{eq:uniform-target-second-moment}
E\{\Delta_{t,n}^2(h_t)/n\mid\mathcal F_t\}\le C_{B,H,\kappa}
\end{equation}
where
$C_{B,H,\kappa}:=4H(1+B^{2/\gamma}/\kappa^2)+2HK_B/\kappa^2.$
This finite deterministic constant depends only on $B,H,\kappa,\gamma,\rho$, independently of $t,n$ and the target subject to these bounds.
To obtain conditional tightness, fix $C,\eta>0$. Conditional Markov's inequality and \eqref{eq:uniform-target-second-moment} give, on $\mathcal E_t$,
$$P\{|\Delta_{t,n}(h_t)|>C\sqrt n\mid\mathcal F_t\}
\le\frac{E\{\Delta_{t,n}^2(h_t)/n\mid\mathcal F_t\}}{C^2}
\le\frac{C_{B,H,\kappa}}{C^2}.$$
Thus, if this conditional probability exceeds $\eta$, either $\mathcal E_t$ fails or $C_{B,H,\kappa}>\eta C^2$. Taking probability over the training data yields
\begin{equation}\label{eq:scalar-target-tightness-tail}
P\left[P\{|\Delta_{t,n}(h_t)|>C\sqrt n\mid\mathcal F_t\}>\eta\right]
\le P(\mathcal E_t^c)+\mathbf1\{C_{B,H,\kappa}>\eta C^2\}.
\end{equation}
Moreover, by the definition of $\mathcal E_t$,
$$P(\mathcal E_t^c)\le P(\beta_{\gamma,t}>B)+P(\eta_{2,t}>H)+P(c_t^{-1}>\kappa^{-1}).$$
Let $(t_k,n_k)$ be any deterministic sequence with $t_k\to\infty$ and $n_k\to\infty$. For fixed $B,H,\kappa$, the indicator in \eqref{eq:scalar-target-tightness-tail} is zero once $C^2\ge C_{B,H,\kappa}/\eta$, independently of $t_k,n_k$. Taking the upper limit along the joint sequence and then letting $C\to\infty$ therefore bounds the resulting outer probability by $\limsup_{t\to\infty}P(\mathcal E_t^c)$. Finally let $B,H\to\infty$ and $\kappa\downarrow0$. The assumptions $\beta_{\gamma,t}=O_p(1)$, $\eta_{2,t}=O_p(1)$ and $c_t^{-1}=O_p(1)$ make this upper bound arbitrarily small. Hence, for every $\eta>0$,
$$\lim_{C\to\infty}\limsup_{k\to\infty}
P\left[P\{|\Delta_{t_k,n_k}(h_{t_k})|>C\sqrt{n_k}\mid\mathcal F_{t_k}\}>\eta\right]=0.$$
proving part (i).

\textit{Part (ii).}
For a fixed $L>0$, define $h_L(x):=h(x)\mathbf1(|h(x)|\le L)$. Its second moment is at most $L^2$, so \eqref{eq:uniform-target-second-moment} applies with $H=L^2$.
Split $\Delta_{t,n}(h)=\Delta_{t,n}(h_L)+\Delta_{t,n}(h-h_L)$. For $\epsilon>0$, the triangle inequality and conditional Markov inequalities of orders two and one give
$$\begin{aligned}
P\{|\Delta_{t,n}(h)|>n\epsilon\mid\mathcal F_t\}
&\le P\{|\Delta_{t,n}(h_L)|>n\epsilon/2\mid\mathcal F_t\}\\
&\quad+P\{|\Delta_{t,n}(h-h_L)|>n\epsilon/2\mid\mathcal F_t\}\\
&\le\frac{4C_{B,L^2,\kappa}}{n\epsilon^2}
+\frac{2E\{|h(X)|\mathbf1(|h(X)|>L)\}}{\epsilon}
\end{aligned}$$
on $\{\beta_{\gamma,t}\le B,c_t\ge\kappa\}$. For the last term, we used
$$E\{|\Delta_{t,n}(h-h_L)|\mid\mathcal F_t\}
\le\sum_{j=1}^nE\{|h-h_L|(X_{t,j})\mid\mathcal F_t\}
=nE\{|h(X)|\mathbf1(|h(X)|>L)\}.$$
Since $E|h(X)|<\infty$, dominated convergence gives $E\{|h(X)|\mathbf1(|h(X)|>L)\}\to0$ as $L\to\infty$.
Fix $B,\kappa,\epsilon,\eta>0$.
First choose $L$ and then $n$ sufficiently large that
$$\frac{2E\{|h(X)|\mathbf1(|h(X)|>L)\}}{\epsilon}<\frac{\eta}{2},
\qquad \frac{4C_{B,L^2,\kappa}}{n\epsilon^2}<\frac{\eta}{2}.$$
Once $L$ is fixed, $C_{B,L^2,\kappa}$ is independent of $t,n$, so the second choice is uniform in $t$.
The conditional probability $P\{|\Delta_{t,n}(h)|/n>\epsilon\mid\mathcal F_t\}$ is therefore below $\eta$ on $\{\beta_{\gamma,t}\le B,c_t\ge\kappa\}$.
Consequently, for all sufficiently large $n$, 
$$P\!\left[P\{|\Delta_{t,n}(h)|/n>\epsilon\mid\mathcal F_t\}>\eta\right]
\le P(\beta_{\gamma,t}>B)+P(c_t<\kappa).$$
Take the upper limit along any deterministic joint sequence $t,n\to\infty$, and then let $B\to\infty$ and $\kappa\downarrow0$, 
the assumptions $\beta_{\gamma,t}=O_p(1)$ and $c_t^{-1}=O_p(1)$ make the right side arbitrarily small.
Thus, for every $\epsilon,\eta>0$, the outer probability tends to zero, proving part (ii).
\end{proof}

\subsection{Lemma~\ref{lem:general-arm-lln}}
The next lemma supplies an empirical law of large numbers for quantities that depend on the first stage data and the potential outcomes.
\begin{lemma}
\label{lem:general-arm-lln}
Consider the two-stage design in Section~\ref{sec:baselineframework}, with fixed $1/2<\rho<1$. Suppose Assumptions~\ref{assum:Q-assumption}, \ref{assum:mu-moment}, and~\ref{assum:score-noncollapse} hold. Consider any deterministic joint sequence $n_0,n\to\infty$.
For $a\in\{0,1\}$, let $H_{n_0,a}$ be a measurable function of $\mathcal D_{n_0}$ and a fresh pair $(X,Y(a))$, where the fresh pair is independent of the first stage data. Write $H_{n_0,a,i}$ for the same quantity evaluated at $(X_i,Y_i(a))$. Suppose
$$E\{|H_{n_0,a}|\mid\mathcal D_{n_0}\}<\infty\quad\text{almost surely for each }n_0,$$

and, for every $\epsilon>0$,
\begin{equation}
\label{eq:general-arm-tail-condition}
\lim_{K\to\infty}\limsup_{n_0\to\infty}P\!\left(E\{|H_{n_0,a}|\mathbf1(|H_{n_0,a}|>K)\mid\mathcal D_{n_0}\}>\epsilon\right)=0.
\end{equation}
Then 
\begin{equation}
\label{eq:general-arm-lln}
\frac1{N_a}\sum_{i:T_i=a}H_{n_0,a,i}-E(H_{n_0,a}\mid\mathcal D_{n_0})
\xrightarrow{p}0.
\end{equation}
\end{lemma}

\begin{proof}
Define$$\widetilde{\mathcal G}_{n_0,j}
:=\sigma\{\mathcal D_{n_0},(X_i,T_i,Y_i):1\le i\le j\},\qquad\widetilde{\mathcal G}_{n_0,0}=\sigma(\mathcal D_{n_0}).$$

\textit{Step 1: Empirical means of bounded quantities.}
Fix $a\in\{0,1\}$ and $K>0$. Set $H_{n_0,a}^{(K)}=H_{n_0,a}\mathbf1(|H_{n_0,a}|\le K)$, and let $H_{n_0,a,i}^{(K)}$ denote its value for the corresponding unit. Define $$h_{n_0,a}^{(K)}(x):=E\{H_{n_0,a}^{(K)}\mid\mathcal D_{n_0},X=x\},
\qquad |h_{n_0,a}^{(K)}(x)|\le K.$$
The tower property gives $E\{h_{n_0,a}^{(K)}(X)\mid\mathcal D_{n_0}\}=E\{H_{n_0,a}^{(K)}\mid\mathcal D_{n_0}\}$.
Write the outcome residual summand as $$D_i:=\mathbf1(T_i=a)\{H_{n_0,a,i}^{(K)}-h_{n_0,a}^{(K)}(X_i)\}.$$
Using $\mathbf{1}(T_i=a)=\{1+(2a-1)(2T_i-1)\}/2$ and the definition of $D_i$, we first expand each summand as
\begin{align*}
\mathbf{1}(T_i=a)H_{n_0,a,i}^{(K)}
&=\mathbf{1}(T_i=a)\{H_{n_0,a,i}^{(K)}-h_{n_0,a}^{(K)}(X_i)\}+\mathbf{1}(T_i=a)h_{n_0,a}^{(K)}(X_i)\\
&=D_i+\frac{1+(2a-1)(2T_i-1)}{2}h_{n_0,a}^{(K)}(X_i)\\
&=D_i+\frac12 h_{n_0,a}^{(K)}(X_i)+\frac{2a-1}{2}(2T_i-1)h_{n_0,a}^{(K)}(X_i).
\end{align*}
Summing over $i$, dividing by $n$, and using
$E\{H_{n_0,a}^{(K)}\mid\mathcal D_{n_0}\}=E\{h_{n_0,a}^{(K)}(X)\mid\mathcal D_{n_0}\}$, we obtain
\begin{equation}
\begin{aligned}
\label{eq:truncated_expansion}
&\frac1n\sum_{i=1}^n\mathbf{1}(T_i=a)H_{n_0,a,i}^{(K)}-\frac12 E\{H_{n_0,a}^{(K)}\mid\mathcal D_{n_0}\}\\
=&\frac1n\sum_{i=1}^n D_i+\frac{1}{2n}\sum_{i=1}^n h_{n_0,a}^{(K)}(X_i)-\frac12 E\{h_{n_0,a}^{(K)}(X)\mid\mathcal D_{n_0}\}+\frac{2a-1}{2n}\sum_{i=1}^n(2T_i-1)h_{n_0,a}^{(K)}(X_i)\\
=&\frac1n\sum_{i=1}^n D_i+\frac12\left[\frac1n\sum_{i=1}^n h_{n_0,a}^{(K)}(X_i)-E\{h_{n_0,a}^{(K)}(X)\mid\mathcal D_{n_0}\}\right]+\frac{2a-1}{2n}\Delta_n(h_{n_0,a}^{(K)}),
\end{aligned}
\end{equation}

We control the three terms in \eqref{eq:truncated_expansion} in order. Conditional independence of the coin and the current potential outcomes gives $$E\{H_{n_0,a,i}^{(K)}\mid\widetilde{\mathcal G}_{n_0,i-1},X_i,T_i\}=h_{n_0,a}^{(K)}(X_i).$$
Hence $E(D_i\mid\widetilde{\mathcal G}_{n_0,i-1},X_i,T_i)=0$ and $E(D_i\mid\widetilde{\mathcal G}_{n_0,i-1})=0$.
When $T_i=a$, $H_{n_0,a,i}^{(K)}$ is a function of the observed outcome; when $T_i\ne a$, $D_i=0$. Thus $D_i$ is adapted to the observed history. Moreover,
\begin{align*}
E(D_i^2\mid\widetilde{\mathcal G}_{n_0,i-1},X_i,T_i)
&=\mathbf1(T_i=a)\operatorname{Var}(H_{n_0,a,i}^{(K)}\mid\mathcal D_{n_0},X_i)\le E\{(H_{n_0,a,i}^{(K)})^2\mid\mathcal D_{n_0},X_i\}\le K^2.
\end{align*}
The $D_i$ are therefore square-integrable martingale differences. For $j<i$,
$$E(D_jD_i\mid\mathcal D_{n_0})=E\{D_jE(D_i\mid\widetilde{\mathcal G}_{n_0,i-1})\mid\mathcal D_{n_0}\}=0,$$
so $$E\!\left[\left(\frac1n\sum_{i=1}^nD_i\right)^2\middle|\mathcal D_{n_0}\right]=\frac1{n^2}\sum_{i=1}^nE(D_i^2\mid\mathcal D_{n_0})\le\frac{K^2}{n}.$$
For the second term, conditional independence of the fresh covariates gives
\begin{align*}
E\!\left[\left\{\frac12\left(\frac1n\sum_{i=1}^nh_{n_0,a}^{(K)}(X_i)-E[h_{n_0,a}^{(K)}(X)\mid\mathcal D_{n_0}]\right)\right\}^2\middle|\mathcal D_{n_0}\right]=\frac1{4n}\operatorname{Var}\{h_{n_0,a}^{(K)}(X)\mid\mathcal D_{n_0}\}\le\frac{K^2}{4n}.
\end{align*}
For fixed $K$, both deterministic bounds tend to zero. Conditional Chebyshev's inequality and Lemma~\ref{lem:conditional-small-to-unconditional} therefore make the first two terms $o_p(1)$.

For the imbalance term, this target satisfies $E\{[h_{n_0,a}^{(K)}(X)]^2\mid\mathcal D_{n_0}\}\le K^2.$ Proposition~\ref{prop:hatm-moment} and Assumption~\ref{assum:score-noncollapse} verify the score conditions of Lemma~\ref{lem:uniform-target-imbalance}. Then Lemma~\ref{lem:uniform-target-imbalance} and Lemma~\ref{lem:conditional-big-to-unconditional} give
$$\frac{\Delta_n(h_{n_0,a}^{(K)})}{\sqrt n}=O_p(1),\qquad\frac{\Delta_n(h_{n_0,a}^{(K)})}{n}=\frac1{\sqrt n}O_p(1)=o_p(1).$$
Combining the three terms proves, for every fixed $K$,
\begin{equation}
\label{eq:general-arm-bounded-lln}
\frac1n\sum_{i=1}^n\mathbf1(T_i=a)H_{n_0,a,i}^{(K)}-\frac12E\{H_{n_0,a}^{(K)}\mid\mathcal D_{n_0}\}\xrightarrow{p}0.
\end{equation}

\textit{Step 2: Remove the truncation.}
Recall that $$H_{n_0,a}-H_{n_0,a}^{(K)}=H_{n_0,a}\mathbf{1}(|H_{n_0,a}|>K).$$ Thus, the difference between the untruncated and truncated centered expressions is
\begin{align*}
&\left[\frac1n\sum_{i=1}^n\mathbf{1}(T_i=a)H_{n_0,a,i}-\frac12 E(H_{n_0,a}\mid\mathcal D_{n_0})\right]-\left[\frac1n\sum_{i=1}^n\mathbf{1}(T_i=a)H_{n_0,a,i}^{(K)}-\frac12 E(H_{n_0,a}^{(K)}\mid\mathcal D_{n_0})\right]\\
=&\frac1n\sum_{i=1}^n\mathbf{1}(T_i=a)H_{n_0,a,i}\mathbf{1}(|H_{n_0,a,i}|>K)-\frac12 E\{H_{n_0,a}\mathbf{1}(|H_{n_0,a}|>K)\mid\mathcal D_{n_0}\}.
\end{align*}
By the triangle inequality and
$|E(U\mid\mathcal D_{n_0})|
\le E(|U|\mid\mathcal D_{n_0})$,
its absolute value is bounded as follows:
\begin{align*}
&\left|\frac1n\sum_{i=1}^n\mathbf{1}(T_i=a)H_{n_0,a,i}\mathbf{1}(|H_{n_0,a,i}|>K)-\frac12 E\{H_{n_0,a}\mathbf{1}(|H_{n_0,a}|>K)\mid\mathcal D_{n_0}\}\right|\\
\le&\frac1n\sum_{i=1}^n\mathbf{1}(T_i=a)|H_{n_0,a,i}|\mathbf{1}(|H_{n_0,a,i}|>K)+\frac12 E\{|H_{n_0,a}|\mathbf{1}(|H_{n_0,a}|>K)\mid\mathcal D_{n_0}\}.
\end{align*}
For $\delta>0$, define the event measurable with respect to the first stage data
$$\mathcal E_{n_0}(K,\delta)
:=\left\{E[|H_{n_0,a}|\mathbf1(|H_{n_0,a}|>K)\mid\mathcal D_{n_0}]\le\delta\right\}.$$
Consequently, conditional Markov's inequality gives,
on $\mathcal E_{n_0}(K,\delta)$,
\begin{align*}
&P\left(\frac1n\sum_{i=1}^n\mathbf{1}(T_i=a)|H_{n_0,a,i}|\mathbf{1}(|H_{n_0,a,i}|>K)+\frac12 E\left[|H_{n_0,a}|\mathbf{1}(|H_{n_0,a}|>K)\,\middle|\,\mathcal D_{n_0}\right]>\epsilon\,\middle|\,\mathcal D_{n_0}\right)\\
\le&\frac{\delta+\delta/2}{\epsilon}=\frac{3\delta}{2\epsilon}.
\end{align*}

On $\mathcal E_{n_0}(K,\delta)^c$, the same conditional probability is bounded by $1$.
Since $\mathcal E_{n_0}(K,\delta)$ is $\mathcal D_{n_0}$-measurable, taking expectations over the first stage data therefore yields
\begin{equation}
\begin{aligned}
&P\left(\frac1n\sum_{i=1}^n\mathbf{1}(T_i=a)|H_{n_0,a,i}|\mathbf{1}(|H_{n_0,a,i}|>K)+\frac12 E\left[|H_{n_0,a}|\mathbf{1}(|H_{n_0,a}|>K)\,\middle|\,\mathcal D_{n_0}\right]>\epsilon\right)\\
=&E\left[P\left(\frac1n\sum_{i=1}^n\mathbf{1}(T_i=a)|H_{n_0,a,i}|\mathbf{1}(|H_{n_0,a,i}|>K)+\frac12 E\left[|H_{n_0,a}|\mathbf{1}(|H_{n_0,a}|>K)\,\middle|\,\mathcal D_{n_0}\right]>\epsilon\,\middle|\,\mathcal D_{n_0}\right)\right]\\
\le& E\left[\frac{3\delta}{2\epsilon}\mathbf{1}\{\mathcal E_{n_0}(K,\delta)\}+\mathbf{1}\{\mathcal E_{n_0}(K,\delta)^c\}\right]=\frac{3\delta}{2\epsilon}P\{\mathcal E_{n_0}(K,\delta)\}+P\{\mathcal E_{n_0}(K,\delta)^c\}\\
\le& P\{\mathcal E_{n_0}(K,\delta)^c\}+\frac{3\delta}{2\epsilon}.
\end{aligned}
\label{eq:general-arm-truncation-bound}
\end{equation}
We now combine the bounded result with the truncation bound. Fix $K,\delta,\epsilon>0$.
By the triangle inequality, the union bound gives
\begin{align*}
&P\left(\left|\frac1n\sum_{i=1}^n\mathbf{1}(T_i=a)H_{n_0,a,i}-\frac12 E(H_{n_0,a}\mid\mathcal D_{n_0})\right|>\epsilon\right)\\
\le& P\left(\left|\frac1n\sum_{i=1}^n\mathbf{1}(T_i=a)H_{n_0,a,i}^{(K)}-\frac12 E(H_{n_0,a}^{(K)}\mid\mathcal D_{n_0})\right|>\frac{\epsilon}{2}\right)\\
&+P\left(\frac1n\sum_{i=1}^n\mathbf{1}(T_i=a)|H_{n_0,a,i}|\mathbf{1}(|H_{n_0,a,i}|>K)+\frac12 E\left[|H_{n_0,a}|\mathbf{1}(|H_{n_0,a}|>K)\,\middle|\,\mathcal D_{n_0}\right]>\frac{\epsilon}{2}\right).
\end{align*}
Applying \eqref{eq:general-arm-truncation-bound} to the second probability, with $\epsilon/2$ in place of $\epsilon$, we have
\begin{align*}
&P\left(\left|\frac1n\sum_{i=1}^n\mathbf{1}(T_i=a)H_{n_0,a,i}-\frac12 E(H_{n_0,a}\mid\mathcal D_{n_0})\right|>\epsilon\right)\\
\le& P\left(\left|\frac1n\sum_{i=1}^n\mathbf{1}(T_i=a)H_{n_0,a,i}^{(K)}-\frac12 E(H_{n_0,a}^{(K)}\mid\mathcal D_{n_0})\right|>\frac{\epsilon}{2}\right)+P\{\mathcal E_{n_0}(K,\delta)^c\}+\frac{3\delta}{\epsilon}.
\end{align*}
For every fixed $K$, \eqref{eq:general-arm-bounded-lln} implies that the first probability on the right side tends to zero as $n_0,n\to\infty$. 
For every fixed $\delta>0$, the second term tends to zero as $K\to\infty$ by \eqref{eq:general-arm-tail-condition}. Letting $\delta\downarrow0$ then proves
\begin{equation}
\label{eq:general-arm-weighted-lln}
\frac1n\sum_{i=1}^n\mathbf1(T_i=a)H_{n_0,a,i}-\frac12E(H_{n_0,a}\mid\mathcal D_{n_0})=o_p(1).
\end{equation}
Since $N_a/n\xrightarrow{p}1/2$ and $E\{|H_{n_0,a}|\mid\mathcal D_{n_0}\}\le K+E\{|H_{n_0,a}|\mathbf1(|H_{n_0,a}|>K)\mid\mathcal D_{n_0}\},$ use \eqref{eq:general-arm-tail-condition}, we have
$$|E(H_{n_0,a}\mid\mathcal D_{n_0})|\le E(|H_{n_0,a}|\mid\mathcal D_{n_0})=O_p(1).$$
Thus
\begin{align*}
&\frac1{N_a}\sum_{i:T_i=a}H_{n_0,a,i}-E(H_{n_0,a}\mid\mathcal D_{n_0})\\
=&\frac n{N_a}\left\{\frac1n\sum_{i=1}^n\mathbf1(T_i=a)H_{n_0,a,i}-\frac12E(H_{n_0,a}\mid\mathcal D_{n_0})\right\}\\
&+\left(\frac n{2N_a}-1\right)E(H_{n_0,a}\mid\mathcal D_{n_0})=o_p(1).
\end{align*}
Which proofs \eqref{eq:general-arm-lln}.
\end{proof}

\section{Proofs for Section~\ref{sec:MainTheoreticalProperties}}

\subsection{Proposition~\ref{prop:hatm-moment} with its Proof}
\begin{proposition}
    \label{prop:hatm-moment}
    Suppose Assumptions~\ref{assum:Q-assumption} and~\ref{assum:mu-moment} hold. Under the first stage simple randomization, we have$$E\!\left[|\widehat{m}_{n_0}(X)|^\gamma\mid\mathcal D_{n_0}\right]=O_p(1).$$
\end{proposition}
\begin{proof}
Lemma~\ref{lem:pilot_arm_means} gives $\bar{Y}_a^{(1)}=O_p(1)$ for $a=0,1$. 
Since these means are $\mathcal D_{n_0}$-measurable,
for $\gamma>1$, the definition of $\widehat{m}_{n_0}$ and the inequality $|z_1+\cdots+z_4|^\gamma\le4^{\gamma-1}\sum_{k=1}^4|z_k|^\gamma$ gives 
\begin{equation*}
E\!\left[|\widehat{m}_{n_0}(X)|^\gamma\mid\mathcal D_{n_0}\right]\le C_\gamma\Big[E\!\left\{|\widehat{\mu}_{n_0,1}(X)|^\gamma\mid\mathcal D_{n_0}\right\}+E\!\left\{|\widehat{\mu}_{n_0,0}(X)|^\gamma\mid\mathcal D_{n_0}\right\}+|\bar{Y}_1^{(1)}|^\gamma+|\bar{Y}_0^{(1)}|^\gamma\Big],
\end{equation*}
where $C_\gamma<\infty$ is a constant depending only on $\gamma$. Hence $E\!\left[|\widehat{m}_{n_0}(X)|^\gamma\mid\mathcal D_{n_0}\right]=O_p(1),$ follows from Assumption~\ref{assum:mu-moment} and the preceding bound on the first stage arm means.
\end{proof}

\subsection{Proposition~\ref{prop:hatm-l2-consistency} with its Proof}
\begin{proposition}
    \label{prop:hatm-l2-consistency}
    Suppose Assumptions~\ref{assum:Q-assumption} and~\ref{assum:mu-l2-consistency} hold. Under the first stage simple randomization, we have
    $$\left[E\!\left\{\bigl(\widehat{m}_{n_0}(X)-m(X)\bigr)^2\mid\mathcal D_{n_0}\right\}\right]^{1/2}=o_p(1).$$
\end{proposition}
\begin{proof}
For $a=0,1$, let
$e_{n_0,a}(x):=\widehat{\mu}_{n_0,a}(x)-\mu_a(x)$ and
$\zeta_{n_0,a}:=\bar{Y}_a^{(1)}-E\{Y(a)\}$.
By Assumption~\ref{assum:mu-l2-consistency}, we may choose a positive deterministic sequence $r_t\to0$ as $t\to \infty$, such that,
for $a=0,1$,
$$\left[E\!\left\{\bigl(\widehat{\mu}_{n_0,a}(X)-\mu_a(X)\bigr)^2\mid\mathcal D_{n_0}\right\}\right]^{1/2}
=O_p(r_{n_0}).$$ 
Set $a_{n_0}=r_{n_0}+{n_0}^{-1/2}$, then $a_{n_0}\to0$.
By Lemma~\ref{lem:pilot_arm_means} and Assumption~\ref{assum:Q-assumption}, $\zeta_{n_0,a}=O_p(n_0^{-1/2})$. Since
$$\widehat{m}_{n_0}(x)-m(x)=\frac{1}{2}\left[\{e_{n_0,1}(x)-\zeta_{n_0,1}\}+\{e_{n_0,0}(x)-\zeta_{n_0,0}\}\right],$$
the conditional $L_2(P_X)$ triangle inequality gives
\begin{align*}
\left[E\!\left\{\bigl(\widehat{m}_{n_0}(X)-m(X)\bigr)^2\mid\mathcal D_{n_0}\right\}\right]^{1/2}
&\le\frac12\sum_{a=0}^1\left(\left[E\!\left\{e_{n_0,a}^2(X)\mid\mathcal D_{n_0}\right\}\right]^{1/2}+|\zeta_{n_0,a}|\right)\\
&=O_p(r_{n_0}+n_0^{-1/2})=O_p(a_{n_0})=o_p(1).
\end{align*}
\end{proof}

\subsection{Proof of Theorem~\ref{theorem:balancing property}}
\begin{proof}
Fix an arbitrary deterministic joint sequence $n_0,n\to\infty$.
Apply the Lemma~\ref{lem:uniform-learned-feature-imbalance} and Lemma~\ref{lem:uniform-target-imbalance} with
$t=n_0$, $\mathcal F_t=\sigma(\mathcal D_{n_0})$,
and $\phi_t=\widehat{m}_{n_0}$,
using $\Delta_n$ for the second stage imbalance.
Proposition~\ref{prop:hatm-moment} verifies the moment hypothesis of Lemma~\ref{lem:uniform-learned-feature-imbalance}. Hence
\begin{equation}
\label{eq:two-stage-learned-balance}
n^{-1/2}\Delta_n(\widehat{m}_{n_0})=o_{p(\cdot\mid\mathcal D_{n_0})}(1).
\end{equation}
Lemma~\ref{lem:conditional-small-to-unconditional} then yields (i). 

For (ii), let $\delta_{n_0}:=\widehat{m}_{n_0}-m$ and $\eta_{n_0}:=\delta_{n_0}/a_{n_0}$,
where $a_{n_0}=r_{n_0}+n_0^{-1/2}\to0$ is the positive deterministic sequence introduced in the proof of Proposition~\ref{prop:hatm-l2-consistency}.
The normalized target is jointly measurable and has an almost surely finite conditional second moment. 
By Proposition~\ref{prop:hatm-l2-consistency},
$$E\{\eta_{n_0}^2(X)\mid\mathcal D_{n_0}\} =a_{n_0}^{-2}
E\{\delta_{n_0}^2(X)\mid\mathcal D_{n_0}\}=O_p(1).$$
Assumption~\ref{assum:score-noncollapse} gives
$c_{n_0}^{-1}=O_p(1)$, where
$c_{n_0}:=E\{|\widehat{m}_{n_0}(X)|\mid\mathcal D_{n_0}\}$
and $1/0=+\infty$.
Together with Proposition~\ref{prop:hatm-moment},
these bounds verify the conditions of
Lemma~\ref{lem:uniform-target-imbalance}(i)
with $h_t=\eta_{n_0}$.
Hence $n^{-1/2}\Delta_n(\eta_{n_0})
=O_{p(\cdot\mid\mathcal D_{n_0})}(1),$ and linearity gives 
\begin{equation}
\label{eq:two-stage-learning-error-balance}
n^{-1/2}\Delta_n(\delta_{n_0})=a_{n_0}\,n^{-1/2}\Delta_n(\eta_{n_0})=O_{p(\cdot\mid\mathcal D_{n_0})}(a_{n_0}).
\end{equation}
Since $m=\widehat{m}_{n_0}-\delta_{n_0}$ and $a_{n_0}\to0$,
combining \eqref{eq:two-stage-learned-balance}
and \eqref{eq:two-stage-learning-error-balance} yields
\begin{equation}
\label{eq:two-stage-oracle-balance}
n^{-1/2}\Delta_n(m)=n^{-1/2}\Delta_n(\widehat{m}_{n_0})-n^{-1/2}\Delta_n(\delta_{n_0})=o_{p(\cdot\mid\mathcal D_{n_0})}(1).
\end{equation}
Applying Lemma~\ref{lem:conditional-small-to-unconditional} to \eqref{eq:two-stage-oracle-balance} gives
$n^{-1/2}\Delta_n(m)=o_p(1),$ which proves (ii) along every joint sequence $n_0,n\to\infty$.
\end{proof}

\subsection{Proof of Theorem~\ref{theo:clt for tau_m}}
\begin{proof}
We apply Lemma~\ref{lem:uniform-target-imbalance} conditionally on $\mathcal D_{n_0}$, taking $t=n_0$ and $\phi_t=\widehat{m}_{n_0}$. Proposition~\ref{prop:hatm-moment} and Assumption~\ref{assum:score-noncollapse} verify its moment and noncollapse conditions on the balancing score. For $a=0,1$, write $v_a(x):=\operatorname{var}\{Y(a)\mid X=x\}$. Assumption~\ref{assum:Q-assumption} gives $E\{m_a^2(X)\}<\infty$ and $E\{v_a(X)\}<\infty$.
All limits below are taken along an arbitrary deterministic joint sequence of sample sizes.

\medskip\noindent\textit{Step 1: Decomposition and imbalance bounds.}
Let
$$A_i:=Y_i(1)-E\{Y_i(1)\},\qquad B_i:=Y_i(0)-E\{Y_i(0)\}.$$
For the constant target $h\equiv1$, write $\Delta_n(1):=\sum_{i=1}^n(2T_i-1)=N_1-N_0$. Then$$N_{1}=\frac n2\left(1+\frac{\Delta_n(1)}{n}\right),\qquad N_{0}=\frac n2\left(1-\frac{\Delta_n(1)}{n}\right).$$
On the event $N_1N_0>0$, whose conditional probability tends to one, we have
\begin{align}
\sqrt n\{\widehat{\tau}(\widehat{m}_{n_0})-\tau\}&=\frac{\sqrt n}{N_{1}}\sum_{i=1}^nT_iA_i-\frac{\sqrt n}{N_{0}}\sum_{i=1}^n(1-T_i)B_i\notag\\
&=\left(1+\frac{\Delta_n(1)}{n}\right)^{-1}\frac2{\sqrt n}\sum_{i=1}^nT_iA_i-\left(1-\frac{\Delta_n(1)}{n}\right)^{-1}\frac2{\sqrt n}\sum_{i=1}^n(1-T_i)B_i\notag\\
&=\left(1+\frac{\Delta_n(1)}{n}\right)^{-1}\left(1-\frac{\Delta_n(1)}{n}\right)^{-1}\{\xi_n+J_n\},\label{eq:two-stage-ate-exact}
\end{align}
where$$\xi_n:=\frac2{\sqrt n}\sum_{i=1}^n\{T_iA_i-(1-T_i)B_i\}, \quad J_n:=-\frac{\Delta_n(1)}{\sqrt n}\frac2n\sum_{i=1}^n\{T_iA_i+(1-T_i)B_i\}.$$
We first show that $\Delta_n(1)/n$ and $J_n$ are negligible. Taking $h\equiv1$ in parts (i)--(ii) of Lemma~\ref{lem:uniform-target-imbalance} gives$$\frac{\Delta_n(1)}{\sqrt n}=O_{p(\cdot\mid\mathcal D_{n_0})}(1),\qquad \frac{\Delta_n(1)}{n}=o_{p(\cdot\mid\mathcal D_{n_0})}(1).$$
Next we will show $J_n=o_{p(\cdot\mid\mathcal D_{n_0})}(1)$, we only prove the treatment part, the control part is similar. Since
$$\frac1n\sum_{i=1}^nT_iA_i=\frac1n\sum_{i=1}^nT_i\varepsilon_{i1}+\frac1n\sum_{i=1}^nT_im_1(X_i).$$
For the residual part, let $$\mathcal G_{n_0,n}=\sigma(\mathcal D_{n_0},X_1,\ldots,X_n,T_1,\ldots,T_n).$$
The second stage assignments use only the first stage data, the covariates and independent randomization coins. 
Thus, given $\mathcal G_{n_0,n}$, the residual pairs are independent across units, and 
$$E\left\{\frac1n\sum_{i=1}^nT_i\varepsilon_{i1}\mid\mathcal G_{n_0,n}\right\}=0,
\quad 
\Var\left\{\frac1n\sum_{i=1}^nT_i\varepsilon_{i1}\mid\mathcal G_{n_0,n}\right\}\le\frac1{n^2}\sum_{i=1}^nv_1(X_i).$$
For every $\epsilon>0$, conditional Chebyshev's inequality and the zero conditional means give
$$\begin{aligned}
P\!\left(\left|\frac1n\sum_{i=1}^nT_i\varepsilon_{i1}\right|>\epsilon\middle|\mathcal G_{n_0,n}\right)
&\le\frac1{\epsilon^2}\Var\!\left(\frac1n\sum_{i=1}^nT_i\varepsilon_{i1}\middle|\mathcal G_{n_0,n}\right)
\le\frac{\sum_{i=1}^nv_1(X_i)}{n^2\epsilon^2}.
\end{aligned}$$
Since $\sigma(\mathcal D_{n_0})\subseteq\mathcal G_{n_0,n}$, the tower property then yields
$$\begin{aligned}
P\!\left(\left|\frac1n\sum_{i=1}^nT_i\varepsilon_{i1}\right|>\epsilon\middle|\mathcal D_{n_0}\right)
&=E\!\left[P\!\left(\left|\frac1n\sum_{i=1}^nT_i\varepsilon_{i1}\right|>\epsilon\middle|\mathcal G_{n_0,n}\right)\middle|\mathcal D_{n_0}\right]\\
&\le\frac1{n^2\epsilon^2}E\!\left[\sum_{i=1}^nv_1(X_i)\middle|\mathcal D_{n_0}\right]
=\frac{E\{v_1(X)\}}{n\epsilon^2}\longrightarrow0.
\end{aligned}$$
The last equality uses the independence of the second stage covariates from the first stage data, and the limit follows from $E\{v_1(X)\}<\infty$. Thus
$$\frac1n\sum_{i=1}^nT_i\varepsilon_{i1}=o_{p(\cdot\mid\mathcal D_{n_0})}(1).$$
For the prognostic component,$$\frac1n\sum_{i=1}^nT_im_1(X_i)=\frac1{2n}\sum_{i=1}^nm_1(X_i)+\frac1{2n}\sum_{i=1}^n(2T_i-1)m_1(X_i).$$
The first term is $o_{p(\cdot\mid\mathcal D_{n_0})}(1)$ by the ordinary law of large numbers, because the second stage covariates are independent of $\mathcal D_{n_0}$ and $E\{m_1(X)\}=0$. The second term is $o_{p(\cdot\mid\mathcal D_{n_0})}(1)$ by Lemma~\ref{lem:uniform-target-imbalance}(ii) with $h=m_1$. Therefore
$$\frac1n\sum_{i=1}^nT_iA_i=o_{p(\cdot\mid\mathcal D_{n_0})}(1).$$
The same argument with $B_i=\varepsilon_{i0}+m_0(X_i)$ gives$$\frac1n\sum_{i=1}^n(1-T_i)B_i=o_{p(\cdot\mid\mathcal D_{n_0})}(1).$$
Combining these displays with the bound for $\Delta_n(1)/\sqrt n$ gives
$$J_n=o_{p(\cdot\mid\mathcal D_{n_0})}(1).$$

Having controlled $J_n$, we next decompose $\xi_n$ to identify the contribution of the imbalance
of the ACCEPO. Write $g(x):=m_1(x)-m_0(x)$ for the centered conditional treatment effect, recall the definition of $A_i$ and $B_i$, we have
\begin{align}
\xi_n&=\frac2{\sqrt n}\sum_{i=1}^n\{T_i\varepsilon_{i1}-(1-T_i)\varepsilon_{i0}\}\notag\\
&\quad+\frac1{\sqrt n}\sum_{i=1}^n\{m_1(X_i)+m_0(X_i)\}(2T_i-1)+\frac1{\sqrt n}\sum_{i=1}^n\{m_1(X_i)-m_0(X_i)\}\notag\\
&=:S_{n,\varepsilon}+S_{n,m}+S_{n,g}.\label{eq:two-stage-leading-decomposition}
\end{align}
The middle term in \eqref{eq:two-stage-leading-decomposition} is exactly $2\Delta_n(m)/\sqrt n$, by \eqref{eq:two-stage-oracle-balance}, $S_{n,m}=o_{p(\cdot\mid\mathcal D_{n_0})}(1).$ 
It remains to study the sum of the residual term $S_{n,\varepsilon}$ and the covariate term $S_{n,g}$.

\medskip\noindent\textit{Step 2: Conditional normal approximation for the residual term.}
Write
$$Z_{n,i}:=\frac2{\sqrt n}\{T_i\varepsilon_{i1}-(1-T_i)\varepsilon_{i0}\},\qquad S_{n,\varepsilon}=\sum_{i=1}^nZ_{n,i}.$$
Conditional on $\mathcal G_{n_0,n}$, the $Z_{n,i}$ are independent and centered, with finite second moments by Assumption~\ref{assum:Q-assumption}. We verify the conditional variance and Lindeberg conditions of Lemma~\ref{lem:conditional-lindeberg-feller} to obtain the approximation in \eqref{eq:two-stage-residual-cf-error}. Their total conditional variance is
$$V_{n,\varepsilon}:=\sum_{i=1}^nE(Z_{n,i}^2\mid\mathcal G_{n_0,n})=\frac4n\sum_{i=1}^n\{T_iv_1(X_i)+(1-T_i)v_0(X_i)\}.$$
For the treatment part,$$\frac1n\sum_{i=1}^nT_iv_1(X_i)=\frac1{2n}\sum_{i=1}^nv_1(X_i)+\frac1{2n}\sum_{i=1}^n(2T_i-1)v_1(X_i).$$
The first term converges to $E\{v_1(X)\}/2$ by the ordinary law of large numbers, and the second term is $o_{p(\cdot\mid\mathcal D_{n_0})}(1)$ by Lemma~\ref{lem:uniform-target-imbalance}(ii) with $h=v_1$. Therefore
$$\frac1n\sum_{i=1}^nT_iv_1(X_i)\xrightarrow{p(\cdot\mid\mathcal D_{n_0})}\frac12E\{v_1(X)\}.$$
Similarly,$$\frac1n\sum_{i=1}^n(1-T_i)v_0(X_i)\xrightarrow{p(\cdot\mid\mathcal D_{n_0})}\frac12E\{v_0(X)\}.$$
Thus$$V_{n,\varepsilon}\xrightarrow{p(\cdot\mid\mathcal D_{n_0})}2E\{v_1(X)\}+2E\{v_0(X)\}=2\Var(\varepsilon_{i1})+2\Var(\varepsilon_{i0}):=V_\varepsilon.$$

Assumption~\ref{assum:Q-assumption} ensures that $0<V_\varepsilon<\infty$. With the variance limit established, we next verify the Lindeberg condition. For each $\eta>0$, the Lindeberg quantity for the conditional array is
$$L_{n,\eta}:=\sum_{i=1}^nE\!\left[Z_{n,i}^2\mathbf1\{|Z_{n,i}|>\eta\}\mid\mathcal G_{n_0,n}\right].$$
We show that $L_{n,\eta}=o_{p(\cdot\mid\mathcal D_{n_0})}(1)$. Since $T_i\in\{0,1\}$, the normalized summand $Z_{n,i}$ equals $2\varepsilon_{i1}/\sqrt n$ when $T_i=1$ and $-2\varepsilon_{i0}/\sqrt n$ when $T_i=0$. Thus only one residual is selected, and $|Z_{n,i}|>\eta$ when the selected residual has absolute value exceeding $\eta\sqrt n/2$. Moreover, by the assignment mechanism, the conditional law of $\varepsilon_{ia}$ given $\mathcal G_{n_0,n}$ is its conditional law given $X_i$, therefore
$$\begin{aligned}
L_{n,\eta}&=\frac4n\sum_{i=1}^n\Big[
T_iE\{\varepsilon_{i1}^2\mathbf1(|\varepsilon_{i1}|>\eta\sqrt n/2)\mid X_i\}
+(1-T_i)E\{\varepsilon_{i0}^2\mathbf1(|\varepsilon_{i0}|>\eta\sqrt n/2)\mid X_i\}\Big]\\
&\le\frac4n\sum_{i=1}^n\sum_{a=0}^1E\{\varepsilon_{ia}^2\mathbf1(|\varepsilon_{ia}|>\eta\sqrt n/2)\mid X_i\}.
\end{aligned}$$
Taking expectation conditional on $\mathcal D_{n_0}$ and using the independence of the second stage covariates and potential outcomes from $\mathcal D_{n_0}$ gives$$E(L_{n,\eta}\mid\mathcal D_{n_0})\le 4E\{\varepsilon_{i1}^2I(|\varepsilon_{i1}|>\eta\sqrt n/2)\}+4E\{\varepsilon_{i0}^2I(|\varepsilon_{i0}|>\eta\sqrt n/2)\}.$$
The right side tends to zero by Assumption~\ref{assum:Q-assumption} and dominated convergence. For every $\delta>0$, conditional Markov's inequality gives
$$P(L_{n,\eta}>\delta\mid\mathcal D_{n_0})\le\delta^{-1}E(L_{n,\eta}\mid\mathcal D_{n_0})\longrightarrow0.$$
Consequently, $L_{n,\eta}=o_{p(\cdot\mid\mathcal D_{n_0})}(1)$ for every $\eta>0$. Lemma~\ref{lem:conditional-small-to-unconditional} gives $V_{n,\varepsilon}\xrightarrow{p}V_\varepsilon$ and $L_{n,\eta}\xrightarrow{p}0$. Along the fixed joint sequence of sample sizes, apply Lemma~\ref{lem:conditional-lindeberg-feller} with $\mathcal F_r=\sigma(\mathcal D_{n_0})$, $\mathcal H_r=\mathcal G_{n_0,n}$, $k_r=n$, and $X_{r,i}=Z_{n,i}$. Define
$$\psi_n(u):=E\{\exp(iuS_{n,\varepsilon})\mid\mathcal G_{n_0,n}\},\qquad
\psi_\varepsilon(u):=\exp\left(-\frac12u^2V_\varepsilon\right).$$
The lemma gives, for every fixed $u\in\mathbb R$,
\begin{equation}
\label{eq:two-stage-residual-cf-error}
E\{|\psi_n(u)-\psi_\varepsilon(u)|\mid\mathcal D_{n_0}\}=o_p(1).
\end{equation}
And \eqref{eq:two-stage-residual-cf-error} will allow us to combine the residual and covariate terms in \eqref{eq:two-stage-leading-conditional-cf}, even though they need not be independent.

\medskip\noindent\textit{Step 3: Combine the residual and covariate terms.}
We now establish the conditional characteristic function limit in \eqref{eq:two-stage-leading-conditional-cf}. The covariate term satisfies an ordinary i.i.d. central limit theorem, but combining it with the residual term requires accounting for their dependence.
For the covariate term, $E\{g(X)\}=0$ and $E\{g^2(X)\}<\infty$. Since $X_1,\ldots,X_n$ are i.i.d. and independent of $\mathcal D_{n_0}$, the ordinary central limit theorem gives$$S_{n,g}=\frac1{\sqrt n}\sum_{i=1}^ng(X_i)\xrightarrow{d}N\left(0,\Var\{g(X)\}\right).$$

Using $|\exp(iuS_{n,g})|=1$, the tower property and \eqref{eq:two-stage-residual-cf-error} give
\begin{align*}
&E\{\exp(iu(S_{n,\varepsilon}+S_{n,g}))\mid\mathcal D_{n_0}\}\\
=&E\left[\exp(iuS_{n,g})E\{\exp(iuS_{n,\varepsilon})\mid\mathcal G_{n_0,n}\}\mid \mathcal D_{n_0}\right]\\
=&\psi_\varepsilon(u)E\{\exp(iuS_{n,g})\mid \mathcal D_{n_0}\}+o_p(1).
\end{align*}

Since $S_{n,g}$ is independent of $\mathcal D_{n_0}$, the preceding ordinary central limit theorem gives$$E\{\exp(iuS_{n,g})\mid\mathcal D_{n_0}\}=E\{\exp(iuS_{n,g})\}\longrightarrow\exp\left[-\frac12u^2\Var\{g(X)\}\right].$$
Therefore, for every fixed $u\in\mathbb R$,
\begin{equation}\label{eq:two-stage-leading-conditional-cf}
E\{\exp(iu(S_{n,\varepsilon}+S_{n,g}))\mid\mathcal D_{n_0}\}\xrightarrow{p}\exp\left[-\frac12u^2\{V_\varepsilon+\Var(g(X))\}\right].
\end{equation}
Since the conditional characteristic functions have modulus at most one, taking expectations and using the tower property gives
$$E\{\exp(iu(S_{n,\varepsilon}+S_{n,g}))\}\longrightarrow\exp\left[-\frac12u^2\{V_\varepsilon+\Var(g(X))\}\right].$$
The characteristic function continuity theorem \citep[Theorem~3.3.17(ii)]{Durrett2019PTE} therefore yields
$$S_{n,\varepsilon}+S_{n,g}\xrightarrow{d}N\left(0,V_\varepsilon+\Var\{g(X)\}\right).$$
The limiting variance satisfies$$V_\varepsilon+\Var\{g(X)\}=2\Var(\varepsilon_{i1})+2\Var(\varepsilon_{i0})+\Var\{m_1(X_i)-m_0(X_i)\}=V^*.$$
\medskip\noindent\textit{Step 4: Derive the ATE limit.}
We return to the exact representation in \eqref{eq:two-stage-ate-exact}. Applying Lemma~\ref{lem:conditional-small-to-unconditional} to the conditional bounds from Step 1, we obtain $\Delta_n(1)/n=o_p(1)$, hence $\left(1+\Delta_n(1)/n\right)^{-1}\left(1-\Delta_n(1)/n\right)^{-1}=1+o_p(1)$. For $\xi_n$, $S_{n,m}=o_p(1)$ and the Step 3 bound $S_{n,\varepsilon}+S_{n,g}=O_p(1)$ give, by \eqref{eq:two-stage-leading-decomposition}, $\xi_n=S_{n,\varepsilon}+S_{n,g}+o_p(1)=O_p(1)$. Finally, $J_n=o_p(1)$. Substituting these bounds into \eqref{eq:two-stage-ate-exact} yields
\begin{equation}\label{eq:two-stage-ate-expansion}
\sqrt n\{\widehat{\tau}(\widehat{m}_{n_0})-\tau\}
=S_{n,\varepsilon}+S_{n,g}+o_p(1).
\end{equation}
Slutsky's theorem now gives
$$\sqrt n\{\widehat{\tau}(\widehat{m}_{n_0})-\tau\}\xrightarrow{d}N(0,V^*).$$

The preceding limit proves the theorem. For the pooling argument in Theorem~\ref{theo:pool}, we also need the conditional characteristic function limit in \eqref{eq:two-stage-ate-conditional-cf}. We obtain it by combining the expansion \eqref{eq:two-stage-ate-expansion} with the conditional limit \eqref{eq:two-stage-leading-conditional-cf}. For every fixed $u\in\mathbb R$, the inequality $|e^{iux}-e^{iuy}|\le\min\{2,|u||x-y|\}$ gives
$$\begin{aligned}
&\left|
E\!\left[e^{iu\sqrt n\{\widehat{\tau}(\widehat{m}_{n_0})-\tau\}}\middle|\mathcal D_{n_0}\right]-E\{\exp(iu(S_{n,\varepsilon}+S_{n,g}))\mid\mathcal D_{n_0}\}\right|\\
\le& E\!\left[\min\left\{2,\,|u|\left|\sqrt n\{\widehat{\tau}(\widehat{m}_{n_0})-\tau\}-S_{n,\varepsilon}-S_{n,g}\right|\right\}\middle|\mathcal D_{n_0}\right]=o_p(1).
\end{aligned}$$
Combining this bound with \eqref{eq:two-stage-leading-conditional-cf} gives
\begin{equation}
\label{eq:two-stage-ate-conditional-cf}
E\!\left[e^{iu\sqrt n\{\widehat{\tau}(\widehat{m}_{n_0})-\tau\}}\middle|\mathcal D_{n_0}\right]
\xrightarrow{p}e^{-u^2V^*/2}.
\end{equation}
\end{proof}

\subsection{Proof of Theorem~\ref{theo:plugin_consistency}}
\begin{proof}
Fix an arbitrary deterministic sequence $n_0,n\to\infty$. Recall $N_a=\sum_{i=1}^n\mathbf1\{T_i=a\}$, for $a=0,1$ with $N_a>0$, the plug-in variance estimator is
\begin{equation}
\label{eq:restate-plugin-form}
\widehat{V}_n=2\widehat{\sigma}_{\varepsilon,1}^2+2\widehat{\sigma}_{\varepsilon,0}^2
+s_n^2(\widehat{\mu}_{n_0,1}-\widehat{\mu}_{n_0,0}),
\end{equation}
where 
$$\begin{aligned}
\widehat{\sigma}_{\varepsilon,a}^2=\frac1{N_a}\sum_{i:T_i=a}
\{Y_i-\widehat{\mu}_{n_0,a}(X_i)\}^2,
\quad 
s_n(h):=\left[\frac1n\sum_{i=1}^n
\left\{h(X_i)-\frac1n\sum_{j=1}^nh(X_j)\right\}^2\right]^{1/2},
\end{aligned}$$
and $s_n(h)$ is the sample standard deviation with divisor $n$. 

We first prove consistency of the residual variance estimates.
Set $e_{n_0,a}(x):=\widehat{\mu}_{n_0,a}(x)-\mu_a(x)$, since $\varepsilon_{ia}=Y_i(a)-\mu_a(X_i)$, we have $Y_i-\widehat{\mu}_{n_0,a}(X_i)=\varepsilon_{ia}-e_{n_0,a}(X_i)$. 
Expanding its square gives
\begin{equation}
\label{eq:plugin-consistency-residual-term}
\widehat{\sigma}_{\varepsilon,a}^2
=\frac1{N_a}\sum_{i:T_i=a}\varepsilon_{ia}^2
-\frac2{N_a}\sum_{i:T_i=a}\varepsilon_{ia}e_{n_0,a}(X_i)
+\frac1{N_a}\sum_{i:T_i=a}e_{n_0,a}^2(X_i).
\end{equation}

We first show that the first term converges to $\Var(\varepsilon_{ia})$ and the third term vanishes, then we use the Cauchy--Schwarz inequality to control the second term.

First, Proposition~\ref{prop:hatm-moment} and Assumption~\ref{assum:score-noncollapse} verify the score moment and nonvanishing conditions of Lemma~\ref{lem:uniform-target-imbalance}, with $t=n_0$ and $\phi_{n_0}=\widehat{m}_{n_0}$. Its part (ii), applied to $h\equiv1$, and Lemma~\ref{lem:conditional-small-to-unconditional} give $N_a/n=\{1+(2a-1)\Delta_n(1)/n\}/2\xrightarrow{p}1/2$, thus $n/N_a\xrightarrow{p}2$.

For the first term in \eqref{eq:plugin-consistency-residual-term}, recall $v_a(x)=E(\varepsilon_{ia}^2\mid X_i=x)$. Assumption~\ref{assum:Q-assumption} gives $Ev_a(X)=E(\varepsilon_{ia}^2)\le E\{Y_i^2(a)\}<\infty$. Then we can have
$$\begin{aligned}
\frac1n\sum_{i:T_i=a}\varepsilon_{ia}^2
=\frac1n\sum_{i:T_i=a}v_a(X_i)
+\frac1n\sum_{i:T_i=a}\{\varepsilon_{ia}^2-v_a(X_i)\}.
\end{aligned}$$
Since the treatment indicator $\mathbf1\{T_i=a\}=\{1+(2a-1)(2T_i-1)\}/2$, we have
$$\begin{aligned}
\frac1n\sum_{i:T_i=a}v_a(X_i)
&=\frac1n\sum_{i=1}^n\mathbf1\{T_i=a\}v_a(X_i)\\
&=\frac1{2n}\sum_{i=1}^n\{1+(2a-1)(2T_i-1)\}v_a(X_i)\\
&=\frac1{2n}\sum_{i=1}^nv_a(X_i)
+\frac{2a-1}{2n}\sum_{i=1}^n(2T_i-1)v_a(X_i).
\end{aligned}$$
Denote $\Delta_n(v_a)=\sum_{i=1}^n(2T_i-1)v_a(X_i)$, thus we have
\begin{equation}
\label{eq:plugin-residual-term}
\begin{aligned}
\frac1n\sum_{i:T_i=a}\varepsilon_{ia}^2
=\frac1{2n}\sum_{i=1}^nv_a(X_i)+\frac{2a-1}{2n}\Delta_n(v_a)+\frac1n\sum_{i:T_i=a}\{\varepsilon_{ia}^2-v_a(X_i)\}.
\end{aligned}
\end{equation}
The first term converges to $E(\varepsilon_{ia}^2)/2$ by the ordinary law of large numbers for the i.i.d. covariates. Since $v_a\ge0$ is integrable, Lemma~\ref{lem:uniform-target-imbalance}(ii) and Lemma~\ref{lem:conditional-small-to-unconditional} make the second term $o_p(1)$.
It remains to control the last term.

Let $\mathcal G_{n_0,n}=\sigma(\mathcal D_{n_0},X_{1:n},T_{1:n})$. As in the proof of Theorem~\ref{theo:clt for tau_m}, the residuals are independent across units conditional on $\mathcal G_{n_0,n}$, and $E(\varepsilon_{ia}^2\mid\mathcal G_{n_0,n})=v_a(X_i)$. For a fixed $L>0$, we define
$$f_L(u):=u^2\mathbf1\{|u|\le L\},\qquad
r_L(u):=u^2\mathbf1\{|u|>L\}.$$
Using the identity $\varepsilon_{ia}^2=f_L(\varepsilon_{ia})+r_L(\varepsilon_{ia})$, we have 
$$\begin{aligned}
\varepsilon_{ia}^2-v_a(X_i)
=\left[f_L(\varepsilon_{ia})-E\{f_L(\varepsilon_{ia})\mid\mathcal G_{n_0,n}\}\right]+\left[r_L(\varepsilon_{ia})-E\{r_L(\varepsilon_{ia})\mid\mathcal G_{n_0,n}\}\right].
\end{aligned}$$
For the part inside $[-L,L]$, $0\le f_L(u)\le L^2$. 
Conditional on $\mathcal G_{n_0,n}$, the treatment indices are fixed and the centered summands are independent with mean zero, thus
$$\begin{aligned}
&E\!\left[\left\{\frac1n\sum_{i:T_i=a}
\left[f_L(\varepsilon_{ia})-E\{f_L(\varepsilon_{ia})\mid\mathcal G_{n_0,n}\}\right]\right\}^{\!2}
\middle|\mathcal G_{n_0,n}\right]\\
=&\frac1{n^2}\sum_{i:T_i=a}\Var\{f_L(\varepsilon_{ia})\mid\mathcal G_{n_0,n}\}
\le\frac{N_aL^4}{n^2}\le\frac{L^4}{n}.
\end{aligned}$$
For the part outside $[-L,L]$, use $r_L\ge0$ and the triangle inequality:
$$\begin{aligned}
&E\left|\frac1n\sum_{i:T_i=a}
\left[r_L(\varepsilon_{ia})-E\{r_L(\varepsilon_{ia})\mid\mathcal G_{n_0,n}\}\right]\right|\\
\le&\frac1n\sum_{i=1}^nE\!\left[\mathbf1\{T_i=a\}r_L(\varepsilon_{ia})\right]+\frac1n\sum_{i=1}^nE\!\left[\mathbf1\{T_i=a\}
E\{r_L(\varepsilon_{ia})\mid\mathcal G_{n_0,n}\}\right]\\
=&\frac2n\sum_{i=1}^nE\!\left[\mathbf1\{T_i=a\}r_L(\varepsilon_{ia})\right]
\le2E\{r_L(\varepsilon_{ia})\}.
\end{aligned}$$
Combining the bounds for the inside and outside parts yields
$$E\left|\frac1n\sum_{i:T_i=a}\{\varepsilon_{ia}^2-v_a(X_i)\}\right|
\le\frac{L^2}{\sqrt n}+2E\!\left[\varepsilon_{ia}^2\mathbf1\{|\varepsilon_{ia}|>L\}\right].$$
This bound is independent of $n_0$. For each fixed $L$, taking the limit superior along the joint sequence $n_0,n\to\infty$ removes the first term. 
Letting $L\to\infty$ then makes the tail expectation vanish by dominated convergence and $E(\varepsilon_{ia}^2)<\infty$. 
Since the left side does not depend on $L$, it tends to zero. 
Markov's inequality therefore yields $\frac1n\sum_{i:T_i=a}\{\varepsilon_{ia}^2-v_a(X_i)\}=o_p(1)$. Combining the three terms in~\eqref{eq:plugin-residual-term}, use $N_a/n\xrightarrow{p}1/2$ and $E(\varepsilon_{ia}\mid X_i)=0$, we have
$$\frac1{N_a}\sum_{i:T_i=a}\varepsilon_{ia}^2
\xrightarrow{p}E(\varepsilon_{ia}^2)=\Var(\varepsilon_{ia}).$$

For the third term in \eqref{eq:plugin-consistency-residual-term}, Assumption~\ref{assum:mu-l2-consistency} gives $E\{e_{n_0,a}^2(X)\mid\mathcal D_{n_0}\}=o_p(1)$. 
Independence of the second stage covariates from the first stage data gives
$E\{n^{-1}\sum_{i=1}^ne_{n_0,a}^2(X_i)\mid\mathcal D_{n_0}\}=E\{e_{n_0,a}^2(X)\mid\mathcal D_{n_0}\}$.
For every $\eta>0$, conditional Markov's inequality gives
$$P\!\left(\frac1n\sum_{i=1}^ne_{n_0,a}^2(X_i)>\eta\,\middle|\,\mathcal D_{n_0}\right)
\le\frac{E\{e_{n_0,a}^2(X)\mid\mathcal D_{n_0}\}}{\eta}\xrightarrow{p}0.$$
Lemma~\ref{lem:conditional-small-to-unconditional} then yields $n^{-1}\sum_{i=1}^ne_{n_0,a}^2(X_i)=o_p(1)$. Since $n/N_a=O_p(1)$,
$$\frac1{N_a}\sum_{i:T_i=a}e_{n_0,a}^2(X_i)\le\frac n{N_a}\frac1n\sum_{i=1}^ne_{n_0,a}^2(X_i)=o_p(1).$$
With the first and third terms in \eqref{eq:plugin-consistency-residual-term} controlled, then the second term is bounded by
\begin{equation*}
\left|\frac2{N_a}\sum_{i:T_i=a}\varepsilon_{ia}e_{n_0,a}(X_i)\right|
\le2\left\{\frac1{N_a}\sum_{i:T_i=a}\varepsilon_{ia}^2\right\}^{1/2}
\left\{\frac1{N_a}\sum_{i:T_i=a}e_{n_0,a}^2(X_i)\right\}^{1/2}=o_p(1),
\end{equation*}
Substituting these limits into \eqref{eq:plugin-consistency-residual-term} proves $\widehat{\sigma}_{\varepsilon,a}^2\xrightarrow{p}\Var(\varepsilon_{ia})$ for $a=0,1$.

We now turn to the sample variance term. Since $s_n^2(h)$ is the sample variance of $h(X_1),\ldots,h(X_n)$ with divisor $n$, write $\bar{h}_n:=n^{-1}\sum_{i=1}^nh(X_i)$ for the sample mean, we have
\begin{equation}
\label{eq:sample-variance-identity}
\begin{aligned}
s_n^2(h)&=\frac1n\sum_{i=1}^n\{h(X_i)-\bar{h}_n\}^2=\frac1n\sum_{i=1}^nh^2(X_i)-2\bar{h}_n\frac1n\sum_{i=1}^nh(X_i)+\bar{h}_n^2\\
&=\frac1n\sum_{i=1}^nh^2(X_i)-\bar{h}_n^2.
\end{aligned}
\end{equation}

Conditional Jensen's inequality and Assumption~\ref{assum:Q-assumption} give $E\{\mu_a^2(X)\}\le E\{Y^2(a)\}<\infty$, so the second moment of $\mu_1-\mu_0$ is finite. 
Applying the ordinary law of large numbers to the first and second empirical moments of $\mu_1-\mu_0$ yields 
$s_n^2(\mu_1-\mu_0)\xrightarrow{p}\Var\{\mu_1(X)-\mu_0(X)\}$, in particular, $s_n(\mu_1-\mu_0)=O_p(1)$.

For the regression error, reusing \eqref{eq:sample-variance-identity}, we have
$$\begin{aligned}
s_n^2(e_{n_0,1}-e_{n_0,0})
&=\frac1n\sum_{i=1}^n[e_{n_0,1}(X_i)-e_{n_0,0}(X_i)]^2-\left\{\frac1n\sum_{i=1}^n[e_{n_0,1}(X_i)-e_{n_0,0}(X_i)]\right\}^2\\
&\le\frac1n\sum_{i=1}^n[e_{n_0,1}(X_i)-e_{n_0,0}(X_i)]^2\\
&\le\frac1n\sum_{i=1}^n\left[e_{n_0,1}^2(X_i)
+2|e_{n_0,1}(X_i)e_{n_0,0}(X_i)|+e_{n_0,0}^2(X_i)\right]\\
&\le\frac2n\sum_{i=1}^n[e_{n_0,1}^2(X_i)+e_{n_0,0}^2(X_i)]=o_p(1).
\end{aligned}$$
Thus $s_n(e_{n_0,1}-e_{n_0,0})=o_p(1)$. 

Since 
$\widehat{\mu}_{n_0,1}-\widehat{\mu}_{n_0,0}=(\mu_1-\mu_0)+(e_{n_0,1}-e_{n_0,0})$, 
and for simplicity, we take $h=\mu_1-\mu_0$ and $k=e_{n_0,1}-e_{n_0,0}$. The sample mean satisfies $\overline{h+k}_n=\bar{h}_n+\bar{k}_n$, then $(h+k)(X_i)-\overline{h+k}_n=\{h(X_i)-\bar{h}_n\}+\{k(X_i)-\bar{k}_n\}$. Squaring and averaging yields
$$\begin{aligned}
s_n^2(h+k)
&=\frac1n\sum_{i=1}^n\left[\{h(X_i)-\bar{h}_n\}+\{k(X_i)-\bar{k}_n\}\right]^2\\
&=s_n^2(h)+\frac2n\sum_{i=1}^n\{h(X_i)-\bar{h}_n\}\{k(X_i)-\bar{k}_n\}+s_n^2(k).
\end{aligned}$$
Applying the Cauchy--Schwarz inequality, we obtain
$$\begin{aligned}
|s_n^2(h+k)-s_n^2(h)|
&\le2\left|\frac1n\sum_{i=1}^n\{h(X_i)-\bar{h}_n\}\{k(X_i)-\bar{k}_n\}\right|+s_n^2(k)\\
&\le2\left\{\frac1n\sum_{i=1}^n[h(X_i)-\bar{h}_n]^2\right\}^{1/2}
\left\{\frac1n\sum_{i=1}^n[k(X_i)-\bar{k}_n]^2\right\}^{1/2}
+s_n^2(k)\\
&=2s_n(h)s_n(k)+s_n^2(k).
\end{aligned}$$
Since $\widehat{\mu}_{n_0,1}-\widehat{\mu}_{n_0,0}=(\mu_1-\mu_0)+(e_{n_0,1}-e_{n_0,0})$, $h=\mu_1-\mu_0$ and $k=e_{n_0,1}-e_{n_0,0}$, we obtain
\begin{equation*}
\label{eq:plug-in-sample-variance-term}
\begin{aligned}
&\left|s_n^2(\widehat{\mu}_{n_0,1}-\widehat{\mu}_{n_0,0})-s_n^2(\mu_1-\mu_0)\right|\\
\le&2s_n(\mu_1-\mu_0)s_n(e_{n_0,1}-e_{n_0,0})
+s_n^2(e_{n_0,1}-e_{n_0,0})=o_p(1),
\end{aligned}
\end{equation*}
Since $m_1-m_0=\mu_1-\mu_0-\tau$, Combining the three variance component limits in \eqref{eq:restate-plugin-form} concludes the proof:
$$\widehat{V}_n\xrightarrow{p}2\Var(\varepsilon_{i1})+2\Var(\varepsilon_{i0})
+\Var\{m_1(X)-m_0(X)\}=V^*.$$
Therefore Theorem~\ref{theo:clt for tau_m} and Slutsky's theorem give 
$$\frac{\sqrt n\{\widehat{\tau}(\widehat{m}_{n_0})-\tau\}}{\sqrt{\widehat{V}_n}}\xrightarrow{d}N(0,1).$$
The stated confidence interval coverage follows by continuity of the standard normal distribution. 
\end{proof}

\subsection{Proof of Theorem~\ref{theo:general-score-ate}}
\begin{proof}
The proof first accounts for the random arm sizes and isolates the imbalance of the true ACCEPO target in the ATE expansion. We then adjust this target directly to remove the allocation drift and express the leading ATE error as a martingale sum. Its conditional variance and tail bounds yield the standardized normal limit.

Fix an arbitrary deterministic joint sequence $n_0,n\to\infty$. Proposition~\ref{prop:hatm-moment} and Assumption~\ref{assum:score-noncollapse} give
\[
E\{|\widehat{m}_{n_0}(X)|^\gamma\mid\mathcal D_{n_0}\}=O_p(1),
\qquad
\{E[|\widehat{m}_{n_0}(X)|\mid\mathcal D_{n_0}]\}^{-1}=O_p(1),
\]
where $\gamma>2$, the first conditional moment is finite almost surely for each $n_0$, and $1/0=+\infty$. Thus the score conditions in Lemmas~\ref{lem:uniform-learned-feature-imbalance} and~\ref{lem:uniform-target-imbalance} hold with $t=n_0$, $\mathcal F_{n_0}=\sigma(\mathcal D_{n_0})$, and balancing score $\widehat{m}_{n_0}$. Conditional Jensen's inequality also gives
\[
E\{\widehat{m}_{n_0}^2(X)\mid\mathcal D_{n_0}\}
\le\{E[|\widehat{m}_{n_0}(X)|^\gamma\mid\mathcal D_{n_0}]\}^{2/\gamma}
=O_p(1).
\]

\textit{Step 1: ATE expansion and consistency.}
Recall $\varepsilon_{ia}=Y_i(a)-\mu_a(X_i)$,
$m_a(x)=\mu_a(x)-E\{Y(a)\}$, and $g(x)=m_1(x)-m_0(x)$.
Thus $Y_i(a)-E\{Y(a)\}=\varepsilon_{ia}+m_a(X_i)$,
$E(\varepsilon_{ia}\mid X_i)=0$, and $E\{g(X)\}=0$.
Their second moments are finite by Assumption~\ref{assum:Q-assumption}.
Retain the notation from the proof of Theorem~\ref{theo:clt for tau_m}:
$$A_i:=Y_i(1)-E\{Y(1)\},\qquad
B_i:=Y_i(0)-E\{Y(0)\}.$$
The count identities are
$$N_1=\frac{n+\Delta_n(1)}2,\qquad
N_0=\frac{n-\Delta_n(1)}2.$$
On the event $N_1N_0>0$, the same algebra gives
\begin{equation}\label{eq:general-exact-ate}
\begin{aligned}
\sqrt n\{\widehat{\tau}(\widehat{m}_{n_0})-\tau\}
&=\frac{\sqrt n}{N_1}\sum_{i=1}^nT_iA_i
  -\frac{\sqrt n}{N_0}\sum_{i=1}^n(1-T_i)B_i\\
&=\left(1+\frac{\Delta_n(1)}n\right)^{-1}
  \frac2{\sqrt n}\sum_{i=1}^nT_iA_i
  -\left(1-\frac{\Delta_n(1)}n\right)^{-1}
  \frac2{\sqrt n}\sum_{i=1}^n(1-T_i)B_i\\
&=\left(1+\frac{\Delta_n(1)}n\right)^{-1}
  \left(1-\frac{\Delta_n(1)}n\right)^{-1}
  \{\xi_n+J_n\},
\end{aligned}
\end{equation}
where
$$\xi_n:=\frac2{\sqrt n}\sum_{i=1}^n
\{T_iA_i-(1-T_i)B_i\},\qquad
J_n:=-\frac{\Delta_n(1)}{\sqrt n}\frac2n\sum_{i=1}^n
\{T_iA_i+(1-T_i)B_i\}.$$

The bounds on the count imbalance and arm averages in the first part of that proof use only
Assumptions~\ref{assum:Q-assumption}, \ref{assum:mu-moment}, and~\ref{assum:score-noncollapse}.
Applying Lemmas~\ref{lem:conditional-small-to-unconditional}
and~\ref{lem:conditional-big-to-unconditional} to those conditional bounds gives
$$\frac{\Delta_n(1)}{\sqrt n}=O_p(1),\qquad
\frac{\Delta_n(1)}n=o_p(1), \qquad 
\frac1n\sum_{i=1}^nT_iA_i=o_p(1),\qquad
\frac1n\sum_{i=1}^n(1-T_i)B_i=o_p(1).$$
Consequently, $J_n=o_p(1)$, $\{1\pm\Delta_n(1)/n\}^{-1}=1+o_p(1)$ and $N_a/n \xrightarrow{p} 1/2$. To analyze $\xi_n$, retain
$$S_{n,\varepsilon}:=\frac2{\sqrt n}\sum_{i=1}^n
\{T_i\varepsilon_{i1}-(1-T_i)\varepsilon_{i0}\},\qquad
S_{n,g}:=\frac1{\sqrt n}\sum_{i=1}^ng(X_i).$$
Substituting $A_i=\varepsilon_{i1}+m_1(X_i)$ and
$B_i=\varepsilon_{i0}+m_0(X_i)$ yields
\begin{align*}
\xi_n
&=\frac2{\sqrt n}\sum_{i=1}^n
 \left[T_i\{\varepsilon_{i1}+m_1(X_i)\}
 -(1-T_i)\{\varepsilon_{i0}+m_0(X_i)\}\right]\\
&=S_{n,\varepsilon}
 +\frac2{\sqrt n}\sum_{i=1}^n
 \{T_im_1(X_i)-(1-T_i)m_0(X_i)\}\\
&=S_{n,\varepsilon}
 +\frac1{\sqrt n}\sum_{i=1}^n
 \{g(X_i)+2(2T_i-1)m(X_i)\}\\
&=S_{n,\varepsilon}+S_{n,g}
 +\frac{2\Delta_n(m)}{\sqrt n}.
\end{align*}
The calculation establishing zero conditional means and conditional independence for the outcome residuals in the proof of Theorem~\ref{theo:clt for tau_m}, followed by the tower property, gives
$$E(S_{n,\varepsilon}^2)\le4E\{v_1(X)+v_0(X)\}<\infty,$$ where $v_a(x):=\operatorname{Var}\{Y(a)\mid X=x\}$.
Since $g(X_i)$ are i.i.d. and centered,
$E(S_{n,g}^2)=E\{g^2(X)\}<\infty$.
Thus $S_{n,\varepsilon}=O_p(1)$ and $S_{n,g}=O_p(1)$.
Moreover, $E\{m^2(X)\}<\infty$, so
Lemma~\ref{lem:uniform-target-imbalance}(i) with $h=m$ and
Lemma~\ref{lem:conditional-big-to-unconditional} give
$\Delta_n(m)/\sqrt n=O_p(1)$.
Hence $\xi_n=O_p(1)$.
Combining this bound with $J_n=o_p(1)$ and
$\{1\pm\Delta_n(1)/n\}^{-1}=1+o_p(1)$ yields
\begin{equation}\label{eq:general-leading-decomposition}
\sqrt n\{\widehat{\tau}(\widehat{m}_{n_0})-\tau\}=\xi_n+o_p(1)=S_{n,\varepsilon}+S_{n,g}+\frac{2\Delta_n(m)}{\sqrt n}+o_p(1).
\end{equation}
The right side is $O_p(1)$. Hence $\widehat{\tau}(\widehat{m}_{n_0})-\tau=O_p(n^{-1/2})$, proving consistency. We now study the normal limit of the leading sum.

\textit{Step 2: Drift adjustment and martingale representation.}
The remaining term that depends on the design in \eqref{eq:general-leading-decomposition} is the imbalance of $m$. We adjust $m$ directly using the coefficient in \eqref{eq:general-score-coefficient}. We have 
\[
|d_{\widehat{m}_{n_0}}(m)|
\le\frac{E|m(X)|}{E[|\widehat{m}_{n_0}(X)|\mid\mathcal D_{n_0}]}
=O_p(1),
\]
because $E\{m^2(X)\}<\infty$ by Assumption~\ref{assum:Q-assumption}. Theorem~\ref{theorem:balancing property}(i) gives $\Delta_n(\widehat{m}_{n_0})/\sqrt n=o_p(1)$, thus we have
\begin{align*}
\frac{2\Delta_n(m)}{\sqrt n}
&=\frac2{\sqrt n}\Delta_n\{m-d_{\widehat{m}_{n_0}}(m)\widehat{m}_{n_0}\}
 +\frac{2d_{\widehat{m}_{n_0}}(m)\Delta_n(\widehat{m}_{n_0})}{\sqrt n}\\
&=\frac2{\sqrt n}\Delta_n\{m-d_{\widehat{m}_{n_0}}(m)\widehat{m}_{n_0}\}+o_p(1).
\end{align*}
Substituting this decomposition into \eqref{eq:general-leading-decomposition} gives
$$\sqrt n\{\widehat{\tau}(\widehat{m}_{n_0})-\tau\}=S_{n,\varepsilon}+S_{n,g}
+\frac2{\sqrt n}\Delta_n\{m-d_{\widehat{m}_{n_0}}(m)\widehat{m}_{n_0}\}+o_p(1).$$
We combine these three leading terms into a single martingale sum so that their joint fluctuation, including the cross terms in the variance, can be analyzed directly. Define
\begin{equation}\label{eq:general-martingale-increment}
M_i:={}2\{T_i\varepsilon_{i1}-(1-T_i)\varepsilon_{i0}\}+g(X_i)+2(2T_i-1)\{m(X_i)-d_{\widehat{m}_{n_0}}(m)\widehat{m}_{n_0}(X_i)\}.
\end{equation}
Consequently,
\begin{equation}\label{eq:general-ate-martingale}
\sqrt n\{\widehat{\tau}(\widehat{m}_{n_0})-\tau\}
=\frac1{\sqrt n}\sum_{i=1}^nM_i+o_p(1).
\end{equation}

We now verify $M_i$ is a martingale difference sequence. Let
$$\widetilde{\mathcal G}_{n_0,j}:=\sigma\{\mathcal D_{n_0},(X_i,T_i,Y_i):1\le i\le j\},\qquad j=0,\ldots,n,$$
where $Y_i=T_iY_i(1)+(1-T_i)Y_i(0)$ and
$\widetilde{\mathcal G}_{n_0,0}=\sigma(\mathcal D_{n_0})$.
Only the selected outcome residual appears in $M_i$, so $M_i$ is measurable with respect to $\widetilde{\mathcal G}_{n_0,i}$.
With $\operatorname{sgn}(0)=0$ and $\Delta_0(\widehat{m}_{n_0})=0$, the biased-coin rule gives
\begin{equation}\label{eq:general-allocation-mean}
E(2T_i-1\mid\widetilde{\mathcal G}_{n_0,i-1},X_i)
=-(2\rho-1)\operatorname{sgn}\{\Delta_{i-1}(\widehat{m}_{n_0})\}
\operatorname{sgn}\{\widehat{m}_{n_0}(X_i)\}.
\end{equation}
The randomization coin is independent of the current potential outcomes given the past and $X_i$. Hence
$E(\varepsilon_{ia}\mid\widetilde{\mathcal G}_{n_0,i-1},X_i,T_i)=0$ for $a=0,1$. Then the tower property gives, 
\begin{align*}
E(T_i\varepsilon_{i1}\mid\widetilde{\mathcal G}_{n_0,i-1})
&=E\!\left[T_i E(\varepsilon_{i1}\mid\widetilde{\mathcal G}_{n_0,i-1},X_i,T_i)
\mid\widetilde{\mathcal G}_{n_0,i-1}\right]=0,\\
E\{(1-T_i)\varepsilon_{i0}\mid\widetilde{\mathcal G}_{n_0,i-1}\}
&=E\!\left[(1-T_i)E(\varepsilon_{i0}\mid\widetilde{\mathcal G}_{n_0,i-1},X_i,T_i)
\mid\widetilde{\mathcal G}_{n_0,i-1}\right]=0.
\end{align*}
The centered covariate term also satisfies $E\{g(X_i)\mid\widetilde{\mathcal G}_{n_0,i-1}\}=E\{g(X)\}=0.$ For the adjusted target term, by the definition of the coefficient  $d_{\widehat{m}_{n_0}}(m)$,
\begin{equation}\label{eq:general-drift-cancellation}
\begin{aligned}
&E\!\left[\{m(X)-d_{\widehat{m}_{n_0}}(m)\widehat{m}_{n_0}(X)\}
\operatorname{sgn}\{\widehat{m}_{n_0}(X)\}\mid\mathcal D_{n_0}\right]\\
&=E[m(X)\operatorname{sgn}\{\widehat{m}_{n_0}(X)\}\mid\mathcal D_{n_0}]-d_{\widehat{m}_{n_0}}(m)E[|\widehat{m}_{n_0}(X)|\mid\mathcal D_{n_0}]
=0.
\end{aligned}
\end{equation}
Applying the tower property gives
\begin{align*}
&E\!\left[(2T_i-1)\{m(X_i)-d_{\widehat{m}_{n_0}}(m)\widehat{m}_{n_0}(X_i)\}
 \mid\widetilde{\mathcal G}_{n_0,i-1}\right]\\
=&E\!\left[
 E\!\left[(2T_i-1)\{m(X_i)-d_{\widehat{m}_{n_0}}(m)\widehat{m}_{n_0}(X_i)\}
 \mid\widetilde{\mathcal G}_{n_0,i-1},X_i\right]
 \middle|\widetilde{\mathcal G}_{n_0,i-1}\right]\\
=&E\!\left[
 \{m(X_i)-d_{\widehat{m}_{n_0}}(m)\widehat{m}_{n_0}(X_i)\}
 E(2T_i-1\mid\widetilde{\mathcal G}_{n_0,i-1},X_i)
 \middle|\widetilde{\mathcal G}_{n_0,i-1}\right]\\
=&-(2\rho-1)\operatorname{sgn}\{\Delta_{i-1}(\widehat{m}_{n_0})\}\times E\!\left[
 \{m(X_i)-d_{\widehat{m}_{n_0}}(m)\widehat{m}_{n_0}(X_i)\}
 \operatorname{sgn}\{\widehat{m}_{n_0}(X_i)\}
 \mid\widetilde{\mathcal G}_{n_0,i-1}\right]\\
=&-(2\rho-1)\operatorname{sgn}\{\Delta_{i-1}(\widehat{m}_{n_0})\}\times E\!\left[
 \{m(X)-d_{\widehat{m}_{n_0}}(m)\widehat{m}_{n_0}(X)\}
 \operatorname{sgn}\{\widehat{m}_{n_0}(X)\}
 \mid\mathcal D_{n_0}\right]=0.
\end{align*}

Combining the three conditional mean calculations gives $ E(M_i\mid\widetilde{\mathcal G}_{n_0,i-1})=0$
under the law conditional on the first stage data. Integrability under that law follows from
\begin{equation*}
E(|M_i|\mid\mathcal D_{n_0})\le2E|\varepsilon_1|+2E|\varepsilon_0|+E|g(X)|+2E|m(X)|+2|d_{\widehat{m}_{n_0}}(m)|E[|\widehat{m}_{n_0}(X)|\mid\mathcal D_{n_0}]
<\infty.
\end{equation*}
Thus $M_i$ is a martingale difference sequence, and its partial sums form a martingale, under the law conditional on the first stage data. We calculate its conditional second moments next.

\textit{Step 3: Predictable variance.}
We compute the conditional second moment directly from the three terms in \eqref{eq:general-martingale-increment}. The variance in the theorem is
\begin{align*}
V_{n_0}
&=V^*+4E\!\left[\{m(X)-d_{\widehat{m}_{n_0}}(m)\widehat{m}_{n_0}(X)\}^2
\mid\mathcal D_{n_0}\right]\\
&=2E\{v_1(X)+v_0(X)\}+E\{g^2(X)\}+4E\!\left[\{m(X)-d_{\widehat{m}_{n_0}}(m)\widehat{m}_{n_0}(X)\}^2
\mid\mathcal D_{n_0}\right].
\end{align*}
Here $Eg(X)=0$ gives $E\{g^2(X)\}=\operatorname{Var}\{m_1(X)-m_0(X)\}$, and $E\{v_a(X)\}=E(\varepsilon_a^2)$. The score moment bound and the coefficient bound in Step 2 imply
\begin{equation}\label{eq:general-variance-bounds}
\begin{aligned}
0<V^*\le V_{n_0}\le V^*+8E\{m^2(X)\}
+8d_{\widehat{m}_{n_0}}^2(m)E\{\widehat{m}_{n_0}^2(X)\mid\mathcal D_{n_0}\}=O_p(1).
\end{aligned}
\end{equation}

In the square of $M_i$, the cross products of the outcome residual term with the other terms have conditional expectation zero. Indeed, those other terms are fixed given the past, $X_i$, and $T_i$, whereas the selected outcome residual has conditional mean zero given that information. Since $T_i(1-T_i)=0$ and apply \eqref{eq:general-allocation-mean}, the squared outcome residual term gives
\begin{align*}
&E\!\left[4\{T_i\varepsilon_{i1}-(1-T_i)\varepsilon_{i0}\}^2
\mid\widetilde{\mathcal G}_{n_0,i-1}\right]\\
=&4E\{T_iv_1(X_i)+(1-T_i)v_0(X_i)\mid\widetilde{\mathcal G}_{n_0,i-1}\}\\
=&2E\{v_1(X)+v_0(X)\}
+2E\!\left[(2T_i-1)\{v_1(X_i)-v_0(X_i)\}
\mid\widetilde{\mathcal G}_{n_0,i-1}\right]\\
=&2E\{v_1(X)+v_0(X)\}
-2(2\rho-1)\operatorname{sgn}\{\Delta_{i-1}(\widehat{m}_{n_0})\}\times E\!\left[\{v_1(X)-v_0(X)\}
\operatorname{sgn}\{\widehat{m}_{n_0}(X)\}\mid\mathcal D_{n_0}\right].
\end{align*}

For the remaining two terms, expand the square and use $(2T_i-1)^2=1$:
\begin{align*}
&E\!\left[
\bigl[g(X_i)+2(2T_i-1)\{m(X_i)-d_{\widehat{m}_{n_0}}(m)\widehat{m}_{n_0}(X_i)\}\bigr]^2
\mid\widetilde{\mathcal G}_{n_0,i-1}\right]\\
=&E\{g^2(X)\}
+4E\!\left[\{m(X)-d_{\widehat{m}_{n_0}}(m)\widehat{m}_{n_0}(X)\}^2
\mid\mathcal D_{n_0}\right]\\
&+4E\!\left[(2T_i-1)g(X_i)
\{m(X_i)-d_{\widehat{m}_{n_0}}(m)\widehat{m}_{n_0}(X_i)\}
\mid\widetilde{\mathcal G}_{n_0,i-1}\right]\\
=&E\{g^2(X)\}
+4E\!\left[\{m(X)-d_{\widehat{m}_{n_0}}(m)\widehat{m}_{n_0}(X)\}^2
\mid\mathcal D_{n_0}\right]\\
&-4(2\rho-1)\operatorname{sgn}\{\Delta_{i-1}(\widehat{m}_{n_0})\} E\!\left[g(X)\{m(X)-d_{\widehat{m}_{n_0}}(m)\widehat{m}_{n_0}(X)\}
\operatorname{sgn}\{\widehat{m}_{n_0}(X)\}\mid\mathcal D_{n_0}\right].
\end{align*}
The last equality again uses \eqref{eq:general-allocation-mean}. Combining these calculations yields
\begin{equation}\label{eq:general-predictable-variance}
E(M_i^2\mid\widetilde{\mathcal G}_{n_0,i-1})
=V_{n_0}-(2\rho-1)C_{n_0}\operatorname{sgn}\{\Delta_{i-1}(\widehat{m}_{n_0})\},
\end{equation}
where the $\mathcal D_{n_0}$-measurable coefficient
\begin{align*}
C_{n_0}
&:=2E\!\left[\{v_1(X)-v_0(X)\}
\operatorname{sgn}\{\widehat{m}_{n_0}(X)\}\mid\mathcal D_{n_0}\right]\\
&\quad+4E\!\left[g(X)\{m(X)-d_{\widehat{m}_{n_0}}(m)\widehat{m}_{n_0}(X)\}
\operatorname{sgn}\{\widehat{m}_{n_0}(X)\}\mid\mathcal D_{n_0}\right].
\end{align*} 
To bound it, use $v_a\ge0$ and $|\operatorname{sgn}(z)|\le1$ to obtain
\begin{align*}
|C_{n_0}|
&\le2E|v_1(X)-v_0(X)|+4E\!\left[|g(X)|\,|m(X)-d_{\widehat{m}_{n_0}}(m)\widehat{m}_{n_0}(X)|
\mid\mathcal D_{n_0}\right],\\
&\le 2E\{v_1(X)+v_0(X)\}+4E\!\left[|g(X)|\,|m(X)-d_{\widehat{m}_{n_0}}(m)\widehat{m}_{n_0}(X)|
\mid\mathcal D_{n_0}\right].
\end{align*}
Conditional Cauchy--Schwarz gives
\begin{align*}
&E\!\left[|g(X)|\,|m(X)-d_{\widehat{m}_{n_0}}(m)\widehat{m}_{n_0}(X)|
\mid\mathcal D_{n_0}\right]\\
\le&\{E[g^2(X)\mid\mathcal D_{n_0}]\}^{1/2}
\left[E\!\left\{[m(X)-d_{\widehat{m}_{n_0}}(m)\widehat{m}_{n_0}(X)]^2
\mid\mathcal D_{n_0}\right\}\right]^{1/2}\\
=&\{E[g^2(X)]\}^{1/2}
\left[E\!\left\{[m(X)-d_{\widehat{m}_{n_0}}(m)\widehat{m}_{n_0}(X)]^2
\mid\mathcal D_{n_0}\right\}\right]^{1/2}.
\end{align*}
Applying $4ab\le a^2+4b^2$, thus we have
\begin{equation}\label{eq:general-variance-drift-bound}
\begin{aligned}
|C_{n_0}|
&\le2E\{v_1(X)+v_0(X)\}+4\{E[g^2(X)]\}^{1/2}
\left[E\!\left\{[m(X)-d_{\widehat{m}_{n_0}}(m)\widehat{m}_{n_0}(X)]^2
\mid\mathcal D_{n_0}\right\}\right]^{1/2}\\
&\le2E\{v_1(X)+v_0(X)\}+E\{g^2(X)\}+4E\!\left[\{m(X)-d_{\widehat{m}_{n_0}}(m)\widehat{m}_{n_0}(X)\}^2
\mid\mathcal D_{n_0}\right]\\
&=V_{n_0}.
\end{aligned}
\end{equation}
Combining \eqref{eq:general-predictable-variance} and \eqref{eq:general-variance-drift-bound}, we have
$$E(M_i^2\mid\widetilde{\mathcal G}_{n_0,i-1})\le V_{n_0}+(2\rho-1)|C_{n_0}|\le2\rho V_{n_0}.$$
For each row $(n_0,n)$, define the standardized increments
$$\widetilde{M}_i:=\frac{M_i}{\sqrt{nV_{n_0}}},\qquad i=1,\ldots,n.$$
Since $V_{n_0}$ is $\mathcal D_{n_0}$-measurable and positive,
$$E(\widetilde{M}_i^{\,2}\mid\widetilde{\mathcal G}_{n_0,i-1})\le\frac{V_{n_0}+(2\rho-1)|C_{n_0}|}{nV_{n_0}}\le\frac{2\rho}{n}.$$
Taking expectations gives $E(\widetilde{M}_i^{\,2})\le2\rho/n$, so the standardized increments are square-integrable under the joint law. Their conditional mean is zero by Step 2, and hence they form a martingale difference array under that law. Dividing \eqref{eq:general-ate-martingale} by $\sqrt{V_{n_0}}$, with $V_{n_0}\ge V^*>0$, gives
\begin{equation}\label{eq:general-studentized-expansion}
\frac{\sqrt n\{\widehat{\tau}(\widehat{m}_{n_0})-\tau\}}{\sqrt{V_{n_0}}}=\sum_{i=1}^n\widetilde{M}_i+o_p(1).
\end{equation}

\textit{Step 4: Vanishing of the variance correction.}
In \eqref{eq:general-predictable-variance}, the variance correction is proportional to $\operatorname{sgn}\{\Delta_{i-1}(\widehat{m}_{n_0})\}$. We show that the average of these signs tends to zero. The same signs appear in the conditional mean of the score increments. Similarly with \eqref{eq:general-allocation-mean},
$$E\{(2T_i-1)\widehat{m}_{n_0}(X_i)\mid\widetilde{\mathcal G}_{n_0,i-1}\}
=-(2\rho-1)E\{|\widehat{m}_{n_0}(X)|\mid\mathcal D_{n_0}\}
\operatorname{sgn}\{\Delta_{i-1}(\widehat{m}_{n_0})\}.$$
Define:
\begin{align*}
L_j
&:=\sum_{i=1}^j\left[(2T_i-1)\widehat{m}_{n_0}(X_i)
-E\{(2T_i-1)\widehat{m}_{n_0}(X_i)\mid\widetilde{\mathcal G}_{n_0,i-1}\}\right]\\
&=\Delta_j(\widehat{m}_{n_0})
+(2\rho-1)E\{|\widehat{m}_{n_0}(X)|\mid\mathcal D_{n_0}\}
\sum_{i=1}^j\operatorname{sgn}\{\Delta_{i-1}(\widehat{m}_{n_0})\}.
\end{align*}
Conditional on the first stage data, $L_j$ is a square-integrable martingale, with $L_0=0$. We first show that both $L_j/j$ and $\Delta_j(\widehat{m}_{n_0})/j$ are $o_p(1)$. 
For $1\le\ell<i$, orthogonality of the martingale increments follows explicitly from
\begin{equation*}
E\{(L_\ell-L_{\ell-1})(L_i-L_{i-1})\mid\mathcal D_{n_0}\}=E\!\left[(L_\ell-L_{\ell-1})
E\{L_i-L_{i-1}\mid\widetilde{\mathcal G}_{n_0,i-1}\}
\mid\mathcal D_{n_0}\right]=0.
\end{equation*}
Thus the cross terms in $E(L_j^2\mid\mathcal D_{n_0})$ vanish. By definition,
$$L_i-L_{i-1}=(2T_i-1)\widehat{m}_{n_0}(X_i)-E\{(2T_i-1)\widehat{m}_{n_0}(X_i)\mid\widetilde{\mathcal G}_{n_0,i-1}\},$$
so
$$E\{(L_i-L_{i-1})^2\mid\widetilde{\mathcal G}_{n_0,i-1}\}
=\operatorname{Var}\{(2T_i-1)\widehat{m}_{n_0}(X_i)
\mid\widetilde{\mathcal G}_{n_0,i-1}\}.$$
Using the tower property and the conditional mean displayed above, we obtain
\begin{align*}
E(L_j^2\mid\mathcal D_{n_0})
&=\sum_{i=1}^jE\{(L_i-L_{i-1})^2\mid\mathcal D_{n_0}\}\\
&=\sum_{i=1}^jE\!\left[
E\{(L_i-L_{i-1})^2\mid\widetilde{\mathcal G}_{n_0,i-1}\}
\mid\mathcal D_{n_0}\right]\\
&=\sum_{i=1}^jE\!\left[
\operatorname{Var}\{(2T_i-1)\widehat{m}_{n_0}(X_i)
\mid\widetilde{\mathcal G}_{n_0,i-1}\}
\mid\mathcal D_{n_0}\right]\\
&=jE\{\widehat{m}_{n_0}^2(X)\mid\mathcal D_{n_0}\}\\
&\quad-(2\rho-1)^2\{E[|\widehat{m}_{n_0}(X)|\mid\mathcal D_{n_0}]\}^2
\sum_{i=1}^jE[\operatorname{sgn}^2\{\Delta_{i-1}(\widehat{m}_{n_0})\}
\mid\mathcal D_{n_0}]\\
&\le jE\{\widehat{m}_{n_0}^2(X)\mid\mathcal D_{n_0}\}\\
&\le j\{E[|\widehat{m}_{n_0}(X)|^\gamma\mid\mathcal D_{n_0}]\}^{2/\gamma}.
\end{align*}
Conditional Chebyshev's inequality therefore gives, for every $\epsilon>0$,
$$P\{|L_j|/j>\epsilon\mid\mathcal D_{n_0}\}
\le\min\left\{1,
\frac{\{E[|\widehat{m}_{n_0}(X)|^\gamma\mid\mathcal D_{n_0}]\}^{2/\gamma}}
{j\epsilon^2}\right\}=o_p(1),$$
because the conditional score moment is $O_p(1)$ and $j\to\infty$. Lemma~\ref{lem:conditional-small-to-unconditional} gives $L_j/j=o_p(1)$.
Lemma~\ref{lem:uniform-learned-feature-imbalance} and Lemma~\ref{lem:conditional-small-to-unconditional} give
$\Delta_j(\widehat{m}_{n_0})/\sqrt j=o_p(1)$, hence
$\Delta_j(\widehat{m}_{n_0})/j=o_p(1)$.
Solving the definition of $L_j$ for the sign average yields
\begin{equation}\label{eq:general-sign-average}
\frac1j\sum_{i=1}^j\operatorname{sgn}\{\Delta_{i-1}(\widehat{m}_{n_0})\}
=\frac{L_j/j-\Delta_j(\widehat{m}_{n_0})/j}
{(2\rho-1)E[|\widehat{m}_{n_0}(X)|\mid\mathcal D_{n_0}]}
=o_p(1).
\end{equation}
Assumption~\ref{assum:score-noncollapse} makes the inverse denominator $O_p(1)$ and the probability of a zero denominator tend to zero. This argument holds along every deterministic joint sequence $n_0,j\to\infty$.

The martingale central limit theorem used below requires the cumulative conditional variance to converge at each fixed fraction of the row. For fixed $u\in(0,1]$, put $j=\lfloor nu\rfloor$. Since $V_{n_0}$ is $\mathcal D_{n_0}$-measurable, substituting \eqref{eq:general-predictable-variance}, we obtain
\begin{align*}
\sum_{i=1}^jE(\widetilde{M}_i^{\,2}\mid\widetilde{\mathcal G}_{n_0,i-1})
&=\frac1{nV_{n_0}}\sum_{i=1}^jE(M_i^2\mid\widetilde{\mathcal G}_{n_0,i-1})\\
&=\frac1{nV_{n_0}}\sum_{i=1}^j
\left[V_{n_0}-(2\rho-1)C_{n_0}\operatorname{sgn}\{\Delta_{i-1}(\widehat{m}_{n_0})\}\right]\\
&=\frac{jV_{n_0}}{nV_{n_0}}
-\frac{(2\rho-1)C_{n_0}}{nV_{n_0}}
\sum_{i=1}^j\operatorname{sgn}\{\Delta_{i-1}(\widehat{m}_{n_0})\}\\
&=\frac jn-(2\rho-1)\frac{C_{n_0}}{V_{n_0}}\frac jn
\left[\frac1j\sum_{i=1}^j\operatorname{sgn}\{\Delta_{i-1}(\widehat{m}_{n_0})\}\right]
\xrightarrow{p}u.
\end{align*}
Indeed, $j/n\to u$, $|C_{n_0}|/V_{n_0}\le1$ by \eqref{eq:general-variance-drift-bound}, and the sign average is $o_p(1)$ by \eqref{eq:general-sign-average}. At $u=0$ the sum is zero. No limit for $V_{n_0}$ is required.

\textit{Step 5: Lindeberg condition.}
Recall $A_i:=Y_i(1)-E\{Y(1)\}, B_i:=Y_i(0)-E\{Y(0)\},g+2m=2m_1,$ and $g-2m=-2m_0$, \eqref{eq:general-martingale-increment} gives
\begin{align*}
M_i&=2\{T_i\varepsilon_{i1}-(1-T_i)\varepsilon_{i0}\}+2T_im_1(X_i)-2(1-T_i)m_0(X_i)-2(2T_i-1)d_{\widehat{m}_{n_0}}(m)\widehat{m}_{n_0}(X_i)\\
&=2T_iA_i-2(1-T_i)B_i-2(2T_i-1)d_{\widehat{m}_{n_0}}(m)\widehat{m}_{n_0}(X_i)..
\end{align*}
For real $s,t,u$ and $L>0$,
$$(s+t+u)^2\mathbf1(|s+t+u|>L)\le 9s^2\mathbf1(|s|>L/3)+9t^2\mathbf1(|t|>L/3)+9u^2\mathbf1(|u|>L/3).$$
Applying this inequality to $M_i$ gives
$$
\begin{aligned}
M_i^2\mathbf1(|M_i|>L)
&\le 36T_iA_i^2\mathbf1(|A_i|>L/6)+36(1-T_i)B_i^2\mathbf1(|B_i|>L/6)\\
&+36d_{\widehat{m}_{n_0}}^2(m)\widehat{m}_{n_0}^2(X_i)\mathbf1\!\left\{|d_{\widehat{m}_{n_0}}(m)\widehat{m}_{n_0}(X_i)|>L/6\right\}.
\end{aligned}
$$
For the last term, use
$t^2\mathbf1(|t|>L/6)\le(L/6)^{2-\gamma}|t|^\gamma$.
Consequently, for every $\epsilon>0$, the lower bound
$V_{n_0}\ge V^*>0$ gives
\begin{align*}
&\sum_{i=1}^n E\!\left[\widetilde{M}_i^{\,2}\mathbf1(|\widetilde{M}_i|>\epsilon)\mid\widetilde{\mathcal G}_{n_0,i-1}\right]\\
\le&\frac1{nV^*}\sum_{i=1}^n E\!\left[M_i^2\mathbf1(|M_i|>\epsilon\sqrt{nV^*})\mid\widetilde{\mathcal G}_{n_0,i-1}\right]\\
\le&\frac{36}{V^*}\sum_{a=0}^1 E\!\left[\{Y(a)-E Y(a)\}^2\mathbf1\!\left\{|Y(a)-E Y(a)|>\frac{\epsilon\sqrt{nV^*}}6\right\}\right]\\
&+\frac{36}{V^*}\left(\frac{\epsilon\sqrt{nV^*}}6\right)^{2-\gamma}|d_{\widehat{m}_{n_0}}(m)|^\gamma E\!\left[|\widehat{m}_{n_0}(X)|^\gamma\mid\mathcal D_{n_0}\right]
=o_p(1).
\end{align*}
This proves the Lindeberg condition along every deterministic joint sequence $n_0,n\to\infty$, without a restriction on their relative rates.

\textit{Step 6: Normal limit.}
Along the arbitrary deterministic joint sequence fixed above, each pair $(n_0,n)$ specifies a row of the martingale array, with the first stage data included in its initial information. Under the joint law of the first stage and second stage data, apply the Lindeberg–Feller theorem for martingale arrays \citep[Theorem~8.2.4]{Durrett2019PTE} to the standardized increments $\widetilde{M}_i$. Their square integrability was verified in Step 3, and Steps 4--5 verify the required prefix variance and Lindeberg conditions. 
The theorem gives
$$\sum_{i=1}^n\widetilde{M}_i=\frac1{\sqrt{nV_{n_0}}}\sum_{i=1}^nM_i
\xrightarrow{d}N(0,1).$$
Finally, \eqref{eq:general-studentized-expansion} yields
$$\frac{\sqrt n\{\widehat{\tau}(\widehat{m}_{n_0})-\tau\}}{\sqrt{V_{n_0}}}=\sum_{i=1}^n\widetilde{M}_i+o_p(1)\xrightarrow{d}N(0,1).$$
Together with the consistency established in Step 1, this proves the theorem. No limit of $V_{n_0}$ or restriction on the relative rates of $n_0$ and $n$ is needed.
\end{proof}

\subsection{Proof of Theorem~\ref{theo:general-score-inference}}
\begin{proof}
Fix an arbitrary deterministic joint sequence $n_0,n\to\infty$. We first rewrite the total variance using outcome moments in each arm, then prove consistency of the coefficient and variance estimators. The final step combines these results with Theorem~\ref{theo:general-score-ate} to obtain feasible inference.

\textit{Step 1: An estimable representation of the total variance.}
Define
\begin{equation}\label{eq:general-potential-residuals}
R_a:=Y(a)-E\{Y(a)\}-d_{\widehat{m}_{n_0}}(m)\widehat{m}_{n_0}(X),
\qquad a=0,1,
\end{equation}
and write $R_{ia}$ for the same expression for unit $i$; the index $a$ denotes the treatment arm. Use $\varepsilon_a=Y(a)-\mu_a(X)$ and $m_a(X)=\mu_a(X)-E\{Y(a)\},$ we have  $$R_a=\varepsilon_a+m_a(X)-d_{\widehat{m}_{n_0}}(m)\widehat{m}_{n_0}(X),$$ Conditional on $X$ and the first stage data, the adjusted $m_a$ term is fixed,
$E(\varepsilon_a\mid X,\mathcal D_{n_0})=0$, and
$E(\varepsilon_a^2\mid X,\mathcal D_{n_0})=v_a(X)$.
Expanding the square and conditioning first on $X$ therefore gives
\begin{equation}\label{eq:general-residual-second-moment}
\begin{aligned}
E\{R_a^2\mid\mathcal D_{n_0}\}
&=E(\varepsilon_a^2)
+2E\!\left[\{m_a(X)-d_{\widehat{m}_{n_0}}(m)\widehat{m}_{n_0}(X)\}
 E(\varepsilon_a\mid X,\mathcal D_{n_0})\mid\mathcal D_{n_0}\right]\\
&\quad+E\!\left[\{m_a(X)-d_{\widehat{m}_{n_0}}(m)\widehat{m}_{n_0}(X)\}^2
\mid\mathcal D_{n_0}\right]\\
&=E\{v_a(X)\}
+E\!\left[\{m_a(X)-d_{\widehat{m}_{n_0}}(m)\widehat{m}_{n_0}(X)\}^2
\mid\mathcal D_{n_0}\right].
\end{aligned}
\end{equation}
Here $E\{v_a(X)\}=E(\varepsilon_a^2)$ is the average conditional outcome variance, finite by Assumption~\ref{assum:Q-assumption}.
Using $m=(m_1+m_0)/2$ and $g=m_1-m_0$, the pointwise square identity is
\begin{align*}
&2\sum_{a=0}^1\{m_a(X)-d_{\widehat{m}_{n_0}}(m)\widehat{m}_{n_0}(X)\}^2\\
&=2\{g(X)/2+m(X)-d_{\widehat{m}_{n_0}}(m)\widehat{m}_{n_0}(X)\}^2+2\{-g(X)/2+m(X)-d_{\widehat{m}_{n_0}}(m)\widehat{m}_{n_0}(X)\}^2\\
&=g^2(X)+4\{m(X)-d_{\widehat{m}_{n_0}}(m)\widehat{m}_{n_0}(X)\}^2.
\end{align*}
Since $Eg(X)=0$, we have $E\{g^2(X)\}=\operatorname{Var}\{m_1(X)-m_0(X)\}$. Multiplying \eqref{eq:general-residual-second-moment} by $2$ and summing over the arms yields
\begin{equation}
\label{eq:general-variance-residual-form}
\begin{aligned}
2\sum_{a=0}^1E\{R_a^2\mid\mathcal D_{n_0}\}
&=2E\{v_1(X)+v_0(X)\}+E\{g^2(X)\}\\
&\quad+4E\!\left[\{m(X)-d_{\widehat{m}_{n_0}}(m)\widehat{m}_{n_0}(X)\}^2
\mid\mathcal D_{n_0}\right]\\
&=V^*+4E\!\left[\{m(X)-d_{\widehat{m}_{n_0}}(m)\widehat{m}_{n_0}(X)\}^2
\mid\mathcal D_{n_0}\right]
=V_{n_0}.
\end{aligned}
\end{equation}
Thus
$$V_{n_0}
=2\sum_{a=0}^1E\!\left[
\{Y(a)-E\{Y(a)\}-d_{\widehat{m}_{n_0}}(m)\widehat{m}_{n_0}(X)\}^2
\mid\mathcal D_{n_0}\right].$$

\textit{Step 2: Consistency of the drift coefficient estimator.}
Recall :
$$\widehat{d}_{\widehat{m}_{n_0},n}(m)
=\frac{\displaystyle\frac12\sum_{a=0}^1\frac1{N_a}
\sum_{i:T_i=a}(Y_i-\bar{Y}_a)\operatorname{sgn}\{\widehat{m}_{n_0}(X_i)\}}
{\displaystyle\frac1n\sum_{i=1}^n|\widehat{m}_{n_0}(X_i)|}.$$
We first show that its numerator estimates the corresponding conditional moment, apply Lemma~\ref{lem:general-arm-lln}, separately for each arm, to the following random variables:
$$H_{n_0,a}=Y(a),\qquad
H_{n_0,a}=\operatorname{sgn}\{\widehat{m}_{n_0}(X)\},\qquad
H_{n_0,a}=Y(a)\operatorname{sgn}\{\widehat{m}_{n_0}(X)\}.$$
The first and third have absolute value at most $|Y(a)|$; their conditional tail expectations are bounded by the fixed tail expectation of $|Y(a)|$, which tends to zero. The second is bounded by one. Thus all three satisfy the lemma's tail condition, giving
\begin{align*}
\bar{Y}_a&=E\{Y(a)\}+o_p(1),\\
\frac1{N_a}\sum_{i:T_i=a}\operatorname{sgn}\{\widehat{m}_{n_0}(X_i)\}
&=E[\operatorname{sgn}\{\widehat{m}_{n_0}(X)\}\mid\mathcal D_{n_0}]+o_p(1),\\
\frac1{N_a}\sum_{i:T_i=a}Y_i\operatorname{sgn}\{\widehat{m}_{n_0}(X_i)\}
&=E[Y(a)\operatorname{sgn}\{\widehat{m}_{n_0}(X)\}\mid\mathcal D_{n_0}]+o_p(1).
\end{align*}
Thus we have
\begin{align*}
&\frac12\sum_{a=0}^1\frac1{N_a}\sum_{i:T_i=a}
(Y_i-\bar{Y}_a)\operatorname{sgn}\{\widehat{m}_{n_0}(X_i)\}\\
=&\frac12\sum_{a=0}^1E[\{Y(a)-E Y(a)\}
\operatorname{sgn}\{\widehat{m}_{n_0}(X)\}\mid\mathcal D_{n_0}]+o_p(1)\\
=&E[m(X)\operatorname{sgn}\{\widehat{m}_{n_0}(X)\}\mid\mathcal D_{n_0}]+o_p(1).
\end{align*}

Since the second stage covariates are i.i.d. conditionally on the first stage data. Thus, for every $\epsilon>0$, conditional Chebyshev's inequality gives
$$
\begin{aligned}
P\!\left(\left|\frac1n\sum_{i=1}^n|\widehat{m}_{n_0}(X_i)|
-E[|\widehat{m}_{n_0}(X)|\mid\mathcal D_{n_0}]\right|>\epsilon
\middle|\mathcal D_{n_0}\right)
&\le\frac{\operatorname{Var}(|\widehat{m}_{n_0}(X)|\mid\mathcal D_{n_0})}{n\epsilon^2}\\
&\le\frac{E\{\widehat{m}_{n_0}^2(X)\mid\mathcal D_{n_0}\}}{n\epsilon^2}=o_p(1)
\end{aligned}$$
Lemma~\ref{lem:conditional-small-to-unconditional} gives $\frac1n\sum_{i=1}^n|\widehat{m}_{n_0}(X_i)| -E[|\widehat{m}_{n_0}(X)|\mid\mathcal D_{n_0}]=o_p(1).$ Assumption~\ref{assum:score-noncollapse} gives
$$\{E[|\widehat{m}_{n_0}(X)|\mid\mathcal D_{n_0}]\}^{-1}=O_p(1),
\qquad
\left\{\frac1n\sum_{i=1}^n|\widehat{m}_{n_0}(X_i)|\right\}^{-1}=O_p(1),$$
with $1/0=+\infty$. Both denominators are thus positive with probability tending to one. 

Since $E|m(X)|\le\{E[m^2(X)]\}^{1/2}<\infty$,
$$|d_{\widehat{m}_{n_0}}(m)|
\le E|m(X)|\{E[|\widehat{m}_{n_0}(X)|\mid\mathcal D_{n_0}]\}^{-1}=O_p(1).$$
The numerator calculation above equals
$d_{\widehat{m}_{n_0}}(m)E[|\widehat{m}_{n_0}(X)|\mid\mathcal D_{n_0}]+o_p(1)$.
Substituting this into the estimator yields
\begin{align*}
\widehat{d}_{\widehat{m}_{n_0},n}(m)-d_{\widehat{m}_{n_0}}(m)
&=\frac{d_{\widehat{m}_{n_0}}(m)E[|\widehat{m}_{n_0}(X)|\mid\mathcal D_{n_0}]+o_p(1)}
{\displaystyle\frac1n\sum_{i=1}^n|\widehat{m}_{n_0}(X_i)|}
-d_{\widehat{m}_{n_0}}(m)\\
&=\frac{d_{\widehat{m}_{n_0}}(m)\left\{E[|\widehat{m}_{n_0}(X)|\mid\mathcal D_{n_0}]
-\displaystyle\frac1n\sum_{i=1}^n|\widehat{m}_{n_0}(X_i)|\right\}+o_p(1)}
{\displaystyle\frac1n\sum_{i=1}^n|\widehat{m}_{n_0}(X_i)|}\\
&=\{O_p(1)o_p(1)+o_p(1)\}O_p(1)=o_p(1).
\end{align*}
Since $\widehat{d}_{\widehat{m}_{n_0},n}(m)=d_{\widehat{m}_{n_0}}(m)+o_p(1)=O_p(1)$, we obtain
\begin{equation}\label{eq:general-coefficient-consistency}
\widehat{d}_{\widehat{m}_{n_0},n}(m)-d_{\widehat{m}_{n_0}}(m)=o_p(1),
\qquad\widehat{d}_{\widehat{m}_{n_0},n}(m)=O_p(1).
\end{equation}

\textit{Step 3: Tail control and arm averages of the squared residuals.}
We verify the tail condition in Lemma~\ref{lem:general-arm-lln} for $H_{n_0,a}=R_a^2$.
For real $s,t$, $L>0$, and $\gamma>2$,
$$\begin{aligned}
(s+t)^2\mathbf1(|s+t|>L)
&\le4s^2\mathbf1(|s|>L/2)+4t^2\mathbf1(|t|>L/2),\\
t^2\mathbf1(|t|>L/2)&\le(L/2)^{2-\gamma}|t|^\gamma.
\end{aligned}$$
Applying these inequalities to \eqref{eq:general-potential-residuals} gives
\begin{equation}\label{eq:general-residual-tail}
\begin{aligned}
&E\{R_a^2\mathbf1(|R_a|>L)\mid\mathcal D_{n_0}\}\\
\le&4E\!\left[\{Y(a)-E Y(a)\}^2
\mathbf1\{|Y(a)-E Y(a)|>L/2\}\right]\\
&+4E\!\left[
 d_{\widehat{m}_{n_0}}^2(m)\widehat{m}_{n_0}^2(X)
 \mathbf1\{|d_{\widehat{m}_{n_0}}(m)\widehat{m}_{n_0}(X)|>L/2\}
 \mid\mathcal D_{n_0}\right]\\
\le&4E\!\left[\{Y(a)-E Y(a)\}^2
\mathbf1\{|Y(a)-E Y(a)|>L/2\}\right]\\
&+4(L/2)^{2-\gamma}|d_{\widehat{m}_{n_0}}(m)|^\gamma
E\{|\widehat{m}_{n_0}(X)|^\gamma\mid\mathcal D_{n_0}\}.
\end{aligned}
\end{equation}
The first term tends to zero as $L\to\infty$ by Assumption~\ref{assum:Q-assumption}.
For the second term, \begin{align*}
&P\!\left(L^{2-\gamma}|d_{\widehat{m}_{n_0}}(m)|^\gamma
E\{|\widehat{m}_{n_0}(X)|^\gamma\mid\mathcal D_{n_0}\}>\eta\right)\\
&=P\!\left(|d_{\widehat{m}_{n_0}}(m)|^\gamma
E\{|\widehat{m}_{n_0}(X)|^\gamma\mid\mathcal D_{n_0}\}>\eta L^{\gamma-2}\right)
\end{align*} for $\eta>0$.
Tightness of $|d_{\widehat{m}_{n_0}}(m)|^\gamma E\{|\widehat{m}_{n_0}(X)|^\gamma\mid\mathcal D_{n_0}\}$ and $\gamma>2$ therefore give
\begin{equation}\label{eq:general-residual-ui}
\lim_{L\to\infty}\limsup_{n_0\to\infty}
P\!\left[E\{R_a^2\mathbf1(|R_a|>L)\mid\mathcal D_{n_0}\}>\eta\right]=0
\qquad(\eta>0).
\end{equation}
Also, $E(R_a^2\mid\mathcal D_{n_0})
\le2E[\{Y(a)-E Y(a)\}^2]
+2d_{\widehat{m}_{n_0}}^2(m)E\{\widehat{m}_{n_0}^2(X)\mid\mathcal D_{n_0}\}
=O_p(1),$
and this conditional moment is finite almost surely. Applying Lemma~\ref{lem:general-arm-lln} with $H_{n_0,a}=R_a^2$, yields
\begin{equation}\label{eq:general-residual-arm-lln}
\frac1{N_a}\sum_{i:T_i=a}R_{ia}^2
=E(R_a^2\mid\mathcal D_{n_0})+o_p(1)=O_p(1).
\end{equation}

We also need the mean within each arm of $\widehat{m}_{n_0}^2(X_i)$ to control coefficient replacement in Step 4.
$$E\!\left[\widehat{m}_{n_0}^2(X)\mathbf1\{\widehat{m}_{n_0}^2(X)>K\}
\mid\mathcal D_{n_0}\right]
\le K^{1-\gamma/2}E\{|\widehat{m}_{n_0}(X)|^\gamma\mid\mathcal D_{n_0}\}.$$
Taking the upper limit as $n_0\to\infty$, then $K\to\infty$, verifies the tail condition for $H_{n_0,a}=\widehat{m}_{n_0}^2(X)$. Thus Lemma~\ref{lem:general-arm-lln} gives
\begin{equation}\label{eq:general-score-arm-second}
\frac1{N_a}\sum_{i:T_i=a}\widehat{m}_{n_0}^2(X_i)
=E\{\widehat{m}_{n_0}^2(X)\mid\mathcal D_{n_0}\}+o_p(1)=O_p(1).
\end{equation}

\textit{Step 4: Estimated residuals and feasible inference.}
For an observed unit in arm $a$, define
$$\widehat{R}_{ia}:=Y_i-\bar{Y}_a
-\widehat{d}_{\widehat{m}_{n_0},n}(m)\widehat{m}_{n_0}(X_i),
\qquad T_i=a.$$
Then $\widehat{V}_{n_0,n}=2\sum_{a=0}^1N_a^{-1}\sum_{i:T_i=a}\widehat{R}_{ia}^{\,2}$.
We first control the error from replacing $R_{ia}$ by $\widehat{R}_{ia}$. For $T_i=a$,
$$\widehat{R}_{ia}-R_{ia}
=-\{\bar{Y}_a-E Y(a)\}
-\{\widehat{d}_{\widehat{m}_{n_0},n}(m)-d_{\widehat{m}_{n_0}}(m)\}\widehat{m}_{n_0}(X_i).$$
Using $(s+t)^2\le2s^2+2t^2$, Step 2, and \eqref{eq:general-score-arm-second}, we obtain
\begin{align*}
\frac1{N_a}\sum_{i:T_i=a}(\widehat{R}_{ia}-R_{ia})^2
&\le2\{\bar{Y}_a-E Y(a)\}^2\\
&\quad+2\{\widehat{d}_{\widehat{m}_{n_0},n}(m)-d_{\widehat{m}_{n_0}}(m)\}^2
\frac1{N_a}\sum_{i:T_i=a}\widehat{m}_{n_0}^2(X_i)
=o_p(1).
\end{align*}
Next, expand
$\widehat{R}_{ia}^{\,2}-R_{ia}^2=2R_{ia}(\widehat{R}_{ia}-R_{ia})+(\widehat{R}_{ia}-R_{ia})^2$.
The empirical Cauchy--Schwarz inequality gives
\begin{align*}
\left|\frac1{N_a}\sum_{i:T_i=a}
(\widehat{R}_{ia}^{\,2}-R_{ia}^2)\right|
&\le2\left\{\frac1{N_a}\sum_{i:T_i=a}R_{ia}^2\right\}^{1/2}
\left\{\frac1{N_a}\sum_{i:T_i=a}(\widehat{R}_{ia}-R_{ia})^2\right\}^{1/2}\\
&\quad+\frac1{N_a}\sum_{i:T_i=a}(\widehat{R}_{ia}-R_{ia})^2
=o_p(1),
\end{align*}
where \eqref{eq:general-residual-arm-lln} supplies the boundedness in probability of the first factor.
Combining this replacement error with \eqref{eq:general-residual-arm-lln} and
\eqref{eq:general-variance-residual-form} yields
\begin{align*}
\widehat{V}_{n_0,n}-V_{n_0}
&=2\sum_{a=0}^1\frac1{N_a}\sum_{i:T_i=a}
(\widehat{R}_{ia}^{\,2}-R_{ia}^2)+2\sum_{a=0}^1\left\{\frac1{N_a}\sum_{i:T_i=a}R_{ia}^2
-E(R_a^2\mid\mathcal D_{n_0})\right\}
=o_p(1).
\end{align*}
Since $V_{n_0}\ge V^*>0$,
$$\left|\frac{\widehat{V}_{n_0,n}}{V_{n_0}}-1\right|
=\frac{|\widehat{V}_{n_0,n}-V_{n_0}|}{V_{n_0}}
\le\frac{|\widehat{V}_{n_0,n}-V_{n_0}|}{V^*}=o_p(1).$$
Thus $\widehat{V}_{n_0,n}/V_{n_0}\xrightarrow{p}1$ and
$P(\widehat{V}_{n_0,n}>0)\to1$. Theorem~\ref{theo:general-score-ate} and Slutsky's theorem give
$$\frac{\sqrt n\{\widehat{\tau}(\widehat{m}_{n_0})-\tau\}}
{\sqrt{\widehat{V}_{n_0,n}}}
=\frac{\sqrt n\{\widehat{\tau}(\widehat{m}_{n_0})-\tau\}}
{\sqrt{V_{n_0}}}
\left(\frac{\widehat{V}_{n_0,n}}{V_{n_0}}\right)^{-1/2}
\xrightarrow{d}N(0,1).$$
The stated confidence interval coverage follows by continuity of the standard normal distribution. 
\end{proof}

\section{Proofs for Section~\ref{sec:extensions}}
\subsection{Proof of Theorem~\ref{theo:pool}}
\begin{proof}
Fix a deterministic sequence $n_0,n\to\infty$ satisfying
Assumption~\ref{assum:lambda}, and write $$A_{n_0}:=\sqrt{n_0}(\widehat{\tau}_1-\tau),\qquad B_{n_0,n}:=\sqrt n(\widehat{\tau}(\widehat{m}_{n_0})-\tau).$$
By Assumption~\ref{assum:pilotate},$$A_{n_0}\xrightarrow{d}Z_1,\qquad Z_1\sim N(0,V_1).$$
We first show the joint convergence of $A_{n_0}$ and $B_{n_0,n}$.
Since $A_{n_0}$ is $\mathcal D_{n_0}$-measurable, the tower
property gives, for every $s,t\in\mathbb R$,
\begin{align*}
E\{\exp(isA_{n_0}+itB_{n_0,n})\}
&=E\left[E\{\exp(isA_{n_0})\exp(itB_{n_0,n})\mid\mathcal D_{n_0}\}\right]\\
&=E\left[\exp(isA_{n_0})E\{\exp(itB_{n_0,n})\mid\mathcal D_{n_0}\}\right].
\end{align*}
By \eqref{eq:two-stage-ate-conditional-cf}, for every fixed $t\in\mathbb R$,$$E\{\exp(itB_{n_0,n})\mid\mathcal D_{n_0}\}\xrightarrow{p}E\{\exp(itZ_2)\}=\exp\left(-\frac12t^2V^*\right).$$
Let $$C_{n_0,n}(t):=E\{\exp(itB_{n_0,n})\mid\mathcal D_{n_0}\}, \quad
\varphi_2(t):=\exp\left(-\frac12t^2V^*\right).$$
Because $|C_{n_0,n}(t)|\le1$ and $|\varphi_2(t)|\le1$, the convergence $C_{n_0,n}(t)\xrightarrow{p}\varphi_2(t)$ also gives $E|C_{n_0,n}(t)-\varphi_2(t)|\to0.$
Therefore,
\begin{align*}
&\left|E\{\exp(isA_{n_0}+itB_{n_0,n})\}-\varphi_2(t)E\{\exp(isA_{n_0})\}\right|\\
=&\left|E\left[\exp(isA_{n_0})\{C_{n_0,n}(t)-\varphi_2(t)\}\right]\right|\\
\le& E|C_{n_0,n}(t)-\varphi_2(t)|\to0.
\end{align*}
Using the convergence of $A_{n_0}$, we also have$$E\{\exp(isA_{n_0})\}\to E\{\exp(isZ_1)\}=\exp\left(-\frac12s^2V_1\right).$$
Combining the previous displays yields$$E\{\exp(isA_{n_0}+itB_{n_0,n})\}\to\exp\left(-\frac12s^2V_1\right)\exp\left(-\frac12t^2V^*\right).$$
The characteristic function convergence theorem \citep[Theorem~3.10.5]{Durrett2019PTE} therefore yields
\begin{equation}
(A_{n_0},B_{n_0,n})\xrightarrow{d}(Z_1,Z_2),
\label{eq:pool-joint-limit}
\end{equation}
where $Z_1\sim N(0,V_1)$ and $Z_2\sim N(0,V^*)$ are independent.

Assumption~\ref{assum:pilotate} and Theorem~\ref{theo:plugin_consistency} give $\widehat{V}_1\xrightarrow{p}V_1$ and $\widehat{V}_n\xrightarrow{p}V^*$.
Since $V_1,V^*>0$ and $\widehat{\lambda}\to\lambda\in[0,1]$,
\begin{align*}
&\widehat{\lambda}\widehat{V}_n+(1-\widehat{\lambda})\widehat{V}_1
-\{\lambda V^*+(1-\lambda)V_1\}\\
=&\widehat{\lambda}(\widehat{V}_n-V^*)
+(1-\widehat{\lambda})(\widehat{V}_1-V_1)
+(\widehat{\lambda}-\lambda)(V^*-V_1)
\xrightarrow{p}0.
\end{align*}
The continuous mapping theorem also gives
$\widehat{w}\xrightarrow{p}w_\lambda^*$.
Applying Slutsky's theorem to \eqref{eq:pool-joint-limit}, we obtain
\begin{equation}
\label{eq:pool-limit-expansion}
\begin{aligned}
\sqrt N(\widehat{\tau}_{\mathrm{pool}}-\tau)
&=\frac{\widehat{w}}{\sqrt{\widehat{\lambda}}}A_{n_0}+
\frac{1-\widehat{w}}{\sqrt{1-\widehat{\lambda}}}B_{n_0,n}\\
&\xrightarrow{d}
\frac{\sqrt\lambda V^*}
{\lambda V^*+(1-\lambda)V_1}Z_1
+
\frac{\sqrt{1-\lambda}V_1}
{\lambda V^*+(1-\lambda)V_1}Z_2.
\end{aligned}
\end{equation}
This argument applies also at $\lambda=0$ and $\lambda=1$.
By independence of $Z_1$ and $Z_2$, the limit is centered normal with variance
$$\frac{\lambda(V^*)^2V_1+(1-\lambda)V_1^2V^*}{\{\lambda V^*+(1-\lambda)V_1\}^2}
=\frac{V_1V^*}{\lambda V^*+(1-\lambda)V_1}
=V_\lambda^*.$$
The same variance consistency results imply
$\widehat{V}_{\mathrm{pool}}\xrightarrow{p}V_\lambda^*>0$.
Thus Slutsky's theorem and \eqref{eq:pool-limit-expansion} give $$\frac{\sqrt N(\widehat{\tau}_{\mathrm{pool}}-\tau)}
{\sqrt{\widehat{V}_{\mathrm{pool}}}}
\xrightarrow{d}N(0,1).$$
The stated confidence interval coverage follows by continuity of the standard normal distribution.

For optimality, first suppose that $\lambda\in(0,1)$ and fix $w\in[0,1]$. By \eqref{eq:pool-joint-limit},
\begin{align*}
\sqrt N\{\widehat{\tau}(w)-\tau\}
&=\frac{w}{\sqrt{\widehat{\lambda}}}A_{n_0}
+\frac{1-w}{\sqrt{1-\widehat{\lambda}}}B_{n_0,n}\\
&\xrightarrow{d}\frac{w}{\sqrt\lambda}Z_1
+\frac{1-w}{\sqrt{1-\lambda}}Z_2.
\end{align*}
Its limiting variance is $V_\lambda(w):=w^2V_1/\lambda+(1-w)^2V^*/(1-\lambda)$. Then we have
\begin{equation}
\label{eq:pool-optimality-identity}
V_\lambda(w)
=\left\{\frac{V_1}{\lambda}+\frac{V^*}{1-\lambda}\right\}(w-w_\lambda^*)^2+V_\lambda^*.
\end{equation}
Since the coefficient of the squared term is positive and $w_\lambda^*\in(0,1)$, \eqref{eq:pool-optimality-identity} is uniquely minimized at $w=w_\lambda^*$, with minimum $V_\lambda^*$. 
When $\lambda=0$, $w_\lambda^*=0$ and
$\widehat{\tau}(w_\lambda^*)=\widehat{\tau}(\widehat{m}_{n_0})$.
For any fixed $w>0$,
$\sqrt{n_0}\{\widehat{\tau}(w)-\tau\}\xrightarrow{d}wZ_1$;
since $N/n_0\to\infty$ and $V_1>0$, the $\sqrt N$-scaled error is not tight.
Similarly, when $\lambda=1$, $w_\lambda^*=1$ and
$\widehat{\tau}(w_\lambda^*)=\widehat{\tau}_1$,
whereas any fixed $w<1$ gives a non-tight $\sqrt N$-scaled error.
Thus only $w=w_\lambda^*$ yields a finite limiting variance at either endpoint.
\end{proof}

\subsection{Proposition~\ref{prop:dynamic-stagewise-l2-rate} with its Proof}
For the following proposition and the proof of Theorem~\ref{theo:multi-stage-dynamic}, 
we follow the terminology of Section~\ref{sec:dynamic-updating}.
We fix an arbitrary deterministic joint sequence of sample sizes satisfying $n_0\to\infty$ and $\min_{1\le k\le K}n_k\to\infty$.
By Assumption~\ref{assum:dynamic-learner-rate} and the finiteness of $K$, we may choose positive deterministic sequences $r_t,s_t\to0$ such that, for every $k=0,\ldots,K-1$ and $a=0,1$,
$$\left[E\!\left\{\bigl(\widehat{\mu}_a^{[k]}(X)-\mu_a(X)\bigr)^2\mid\mathcal H_k\right\}\right]^{1/2}=O_p(r_{M_k}),
\qquad
\widehat{\nu}_a^{[k]}-E\{Y(a)\}=O_p(s_{M_k}).$$

\begin{proposition}
\label{prop:dynamic-stagewise-l2-rate}
Suppose Assumptions~\ref{assum:Q-assumption} and \ref{assum:dynamic-learner-rate} hold, and set $a_t=r_t+s_t$. Then, for each fixed $k=0,\ldots,K-1$,
$$\begin{aligned}
E\!\left[|\widehat{m}^{[k]}(X)|^\gamma\mid\mathcal H_k\right]&=O_p(1),\\
\left[E\!\left\{\bigl(\widehat{m}^{[k]}(X)-m(X)\bigr)^2\mid\mathcal H_k\right\}\right]^{1/2}&=O_p(a_{M_k}),
\end{aligned}$$
\end{proposition}

\begin{proof}
Fix $k$. Assumption~\ref{assum:Q-assumption} gives $|E\{Y(a)\}|<\infty$, then Assumption~\ref{assum:dynamic-learner-rate} gives
$$\widehat{\nu}_a^{[k]}=E\{Y(a)\}+O_p(s_{M_k})=O_p(1),\qquad a\in\{0,1\}.$$
Using $|z_1+\cdots+z_4|^\gamma\le4^{\gamma-1}\sum_{j=1}^4|z_j|^\gamma$ and $\widehat{\nu}_a^{[k]}$ is $\mathcal H_k$-measurable, we obtain
$$E\!\left[|\widehat{m}^{[k]}(X)|^\gamma\mid\mathcal H_k\right]
\le 2^{\gamma-2}\sum_{a=0}^1
\left\{E\!\left[|\widehat{\mu}_a^{[k]}(X)|^\gamma\mid\mathcal H_k\right]+|\widehat{\nu}_a^{[k]}|^\gamma\right\}
=O_p(1).$$
For the learning error, since
$$\widehat{m}^{[k]}(x)-m(x)
=\frac12\sum_{a=0}^1\{\widehat{\mu}_a^{[k]}(x)-\mu_a(x)\}
-\frac12\sum_{a=0}^1[\widehat{\nu}_a^{[k]}-E\{Y(a)\}].$$
The conditional triangle inequality therefore yields
$$\begin{aligned}
&\left[E\!\left\{\bigl(\widehat{m}^{[k]}(X)-m(X)\bigr)^2\mid\mathcal H_k\right\}\right]^{1/2}\\
\le&\frac12\sum_{a=0}^1
\left[E\!\left\{\bigl(\widehat{\mu}_a^{[k]}(X)-\mu_a(X)\bigr)^2\mid\mathcal H_k\right\}\right]^{1/2}
+\frac12\sum_{a=0}^1|\widehat{\nu}_a^{[k]}-E\{Y(a)\}|\\
=&O_p(r_{M_k})+O_p(s_{M_k})=O_p(a_{M_k}).
\end{aligned}$$
\end{proof}

\subsection{Proof of Theorem~\ref{theo:multi-stage-dynamic}}
\begin{proof}
\medskip\noindent\textit{Step 1: Imbalance bounds within each stage.}
Fix an arbitrary deterministic sequence with $n_0\to\infty$ and
$\min_{1\le k\le K}n_k\to\infty$.
For a measurable target $h$, write
$$\Delta_k(h):=\sum_{i\in\mathcal I_k}(2T_i-1)h(X_i).$$
For each fixed $k$, conditional on $\mathcal H_{k-1}$, the score $\widehat{m}^{[k-1]}$ is fixed. 
Apply Lemma~\ref{lem:uniform-learned-feature-imbalance} and Lemma~\ref{lem:uniform-target-imbalance} with $t=M_{k-1}$, $\mathcal F_t=\mathcal H_{k-1}$ and $\phi=\widehat{m}^{[k-1]}$.
Proposition~\ref{prop:dynamic-stagewise-l2-rate} supplies the conditional $\gamma$-th moment bound and the convergence rate within each stage, Assumption~\ref{assum:dynamic-feature-regularity} supplies the nonvanishing conditions, thus we have
\begin{equation}
\label{eq:multistage-within-stage-learned-score-balance}
\frac{\Delta_k(\widehat{m}^{[k-1]})}{\sqrt{n_k}}
=o_{p(\cdot\mid\mathcal H_{k-1})}(1).
\end{equation}
Similar with the proof of Theorem~\ref{theorem:balancing property}, define
$$\delta_{k-1}(x):=\widehat{m}^{[k-1]}(x)-m(x),\quad
R_{k-1}:=\left[E\{\delta_{k-1}^2(X)\mid\mathcal H_{k-1}\}\right]^{1/2}.$$
By Proposition~\ref{prop:dynamic-stagewise-l2-rate}, $R_{k-1}=O_p(a_{M_{k-1}})$, set $\eta_{k-1}=\delta_{k-1}/a_{M_{k-1}}$. Thus we have
$$E\{\eta_{k-1}^2(X)\mid\mathcal H_{k-1}\}
=\frac{R_{k-1}^2}{a_{M_{k-1}}^2}=O_p(1).$$
Thus Lemma~\ref{lem:uniform-target-imbalance}(i) gives
$$\frac{\Delta_k(\eta_{k-1})}{\sqrt{n_k}}
=O_{p(\cdot\mid\mathcal H_{k-1})}(1).$$
By linearity of the imbalance,
\begin{equation}
\label{eq:multistage-within-stage-learned-error-balance}
\frac{\Delta_k(\delta_{k-1})}{\sqrt{n_k}}
=a_{M_{k-1}}\frac{\Delta_k(\eta_{k-1})}{\sqrt{n_k}}
=O_{p(\cdot\mid\mathcal H_{k-1})}(a_{M_{k-1}}).
\end{equation}

\medskip\noindent\textit{Step 2: Transfer the stagewise bounds to $\mathcal H_0$ then aggregate.}
We first transfer the stagewise bounds to $\mathcal H_0$. Since $\mathcal H_0\subseteq\mathcal H_{k-1}$, for every event $A$, the tower property gives
\begin{equation}
\label{eq:multi-stage-transfer-H0}
P(A\mid\mathcal H_0)=E\{P(A\mid\mathcal H_{k-1})\mid\mathcal H_0\}.
\end{equation}
Fix $\epsilon>0$ and let $B_{k,n}(\epsilon):=\{|\Delta_k(\widehat{m}^{[k-1]})|>\epsilon\sqrt{n_k}\}$. The first stagewise bound \eqref{eq:multistage-within-stage-learned-score-balance} gives $P\{B_{k,n}(\epsilon)\mid\mathcal H_{k-1}\}\xrightarrow{p}0$. Since conditional probability lies in $[0,1]$, so for every $\eta>0$, Markov's inequality and \eqref{eq:multi-stage-transfer-H0} yield
$$
\begin{aligned}
P\!\left[P\{B_{k,n}(\epsilon)\mid\mathcal H_0\}>\eta\right]
&\le\frac1\eta E\!\left[P\{B_{k,n}(\epsilon)\mid\mathcal H_0\}\right]\\
&=\frac1\eta E\!\left[P\{B_{k,n}(\epsilon)\mid\mathcal H_{k-1}\}\right]
\longrightarrow0.
\end{aligned}
$$
Since $\epsilon$ and $\eta$ are arbitrary, this proves
$\Delta_k(\widehat{m}^{[k-1]})/\sqrt{n_k}
=o_{p(\cdot\mid\mathcal H_0)}(1)$.

Let $A_{k,n}(C):=\{|\Delta_k(\delta_{k-1})|>C\sqrt{n_k}\,a_{M_{k-1}}\}$ and $Q_{C,n}:=P\{A_{k,n}(C)\mid\mathcal H_{k-1}\}$. The bound in \eqref{eq:multistage-within-stage-learned-error-balance} gives $\lim_{C\to\infty}\limsup P(Q_{C,n}>\eta)=0$ for every $\eta>0$. Since $0\le Q_{C,n}\le1$, and $Q_{C,n}\le\eta/2+\mathbf1\{Q_{C,n}>\eta/2\}$, thus
$E(Q_{C,n}\mid\mathcal H_0)\le\eta/2+P(Q_{C,n}>\eta/2\mid\mathcal H_0)$.
Use \eqref{eq:multi-stage-transfer-H0} again, we have
$$
\begin{aligned}
P\!\left[P\{A_{k,n}(C)\mid\mathcal H_0\}>\eta\right]
&=P\{E(Q_{C,n}\mid\mathcal H_0)>\eta\}\\
&\le P\!\left[P(Q_{C,n}>\eta/2\mid\mathcal H_0)>\eta/2\right]\\
&\le\frac2\eta P(Q_{C,n}>\eta/2).
\end{aligned}
$$
Taking the limit superior along the joint sequence and then letting $C\to\infty$ proves
$\Delta_k(\delta_{k-1})/\sqrt{n_k}
=O_{p(\cdot\mid\mathcal H_0)}(a_{M_{k-1}})$.
For the total contribution from learning error, the scales $a_{M_{k-1}}$ are positive and deterministic. Write $q_{k,n}=n_k/n$. The triangle inequality gives
$$\begin{aligned}
\left|\frac1{\sqrt n}\sum_{k=1}^K\Delta_k(\delta_{k-1})\right|
&\le\sum_{k=1}^K\sqrt{q_{k,n}}\,
\frac{|\Delta_k(\delta_{k-1})|}{\sqrt{n_k}}\\
&=\sum_{k=1}^K\sqrt{q_{k,n}}a_{M_{k-1}}\,
\frac{|\Delta_k(\delta_{k-1})|}{\sqrt{n_k}\,a_{M_{k-1}}}\\
&\le\left(\sum_{k=1}^K\sqrt{q_{k,n}}a_{M_{k-1}}\right)
\max_{1\le k\le K}
\frac{|\Delta_k(\delta_{k-1})|}{\sqrt{n_k}\,a_{M_{k-1}}}.
\end{aligned}$$
The maximum is conditionally tight given $\mathcal H_0$ because $K$ is fixed and a union bound applies to the transferred rate bounds. Hence the total imbalance of the learning error satisfies
$$\frac1{\sqrt n}\sum_{k=1}^K\Delta_k(\delta_{k-1})
=O_{p(\cdot\mid\mathcal H_0)}\!\left(
\sum_{k=1}^K\sqrt{q_{k,n}}a_{M_{k-1}}\right).$$
The transferred bounds on the imbalance of the learned score also give
$$\frac1{\sqrt n}\sum_{k=1}^K\Delta_k(\widehat{m}^{[k-1]})
=\sum_{k=1}^K\sqrt{q_{k,n}}\,
\frac{\Delta_k(\widehat{m}^{[k-1]})}{\sqrt{n_k}}
=o_{p(\cdot\mid\mathcal H_0)}(1),$$
since $0\le\sqrt{q_{k,n}}\le1$ and $K$ is fixed. 
Since $\Delta_k(m)=\Delta_k(\widehat{m}^{[k-1]})-\Delta_k(\delta_{k-1})$, we have
$$
\begin{aligned}
\frac1{\sqrt n}\sum_{k=1}^K\Delta_k(m)
&=\frac1{\sqrt n}\sum_{k=1}^K\Delta_k(\widehat{m}^{[k-1]})
-\frac1{\sqrt n}\sum_{k=1}^K\Delta_k(\delta_{k-1})\\
&=o_{p(\cdot\mid\mathcal H_0)}(1)
+O_{p(\cdot\mid\mathcal H_0)}\!\left(\sum_{k=1}^K\sqrt{q_{k,n}}a_{M_{k-1}}\right).
\end{aligned}
$$
Lemmas~\ref{lem:conditional-small-to-unconditional} and \ref{lem:conditional-big-to-unconditional} give the unconditional bound 
\begin{equation*}
\frac1{\sqrt n}\sum_{k=1}^K\sum_{i\in\mathcal I_k}(2T_i-1)m(X_i)=O_p\!\left(\sum_{k=1}^K \sqrt{q_{k,n}}a_{M_{k-1}}\right)+o_p(1).
\label{eq:dynamic-oracle-imbalance-bound}
\end{equation*}
To obtain the bound in~\eqref{eq:dynamic-oracle-imbalance-small}, Cauchy--Schwarz and $\sum_{k=1}^Kq_{k,n}=1$ give
$$
\begin{aligned}
\sum_{k=1}^K\sqrt{q_{k,n}}a_{M_{k-1}}
&\le\left(\sum_{k=1}^Kq_{k,n}\right)^{1/2}
\left(\sum_{k=1}^Ka_{M_{k-1}}^2\right)^{1/2}\\
&\le\sqrt K\max_{1\le k\le K}a_{M_{k-1}}
\le\sqrt K\sup_{t\ge n_0}a_t\longrightarrow0.
\end{aligned}
$$
The last inequality uses $M_{k-1}\ge n_0$. Since $K$ is fixed and $a_t\to0$, for every $\epsilon>0$ there is $t_\epsilon$ such that $a_t<\epsilon/\sqrt K$ for all $t\ge t_\epsilon$, once $n_0\ge t_\epsilon$, the final bound is at most $\epsilon$. 
This proves the imbalance bound in \eqref{eq:dynamic-oracle-imbalance-small}, as well as its version conditional on $\mathcal H_0$. 
No restriction on the relative growth of $n_0$ and the stage sizes is needed.

\medskip\noindent\textit{Step 3: ATE decomposition.}
With the $n$ main stage observations indexed as in Section~\ref{sec:dynamic-updating}, retain $A_i=Y_i(1)-E\{Y(1)\}$, $B_i=Y_i(0)-E\{Y(0)\}$, $\varepsilon_{ia}=Y_i(a)-\mu_a(X_i)$, $g=m_1-m_0$, $N_a=\sum_{i=1}^n\mathbf1\{T_i=a\}$ and $v_a(x)=E(\varepsilon_{ia}^2\mid X_i=x)$ from the proof of Theorem~\ref{theo:clt for tau_m}. Here, $D_n=\sum_{k=1}^K\Delta_k(1)=N_1-N_0$, define
$$
\xi_n:=\frac2{\sqrt n}\sum_{i=1}^n\{T_iA_i-(1-T_i)B_i\},
\qquad
J_n:=-\frac{D_n}{\sqrt n}\frac2n\sum_{i=1}^n\{T_iA_i+(1-T_i)B_i\}.
$$
With this indexing, $\widehat{\tau}_{\mathrm{dyn}}$ has the same difference-in-means form as \eqref{equ:tau_m}. The exact algebraic decomposition in \eqref{eq:two-stage-ate-exact} therefore applies with $D_n$ in place of $\Delta_n(1)$, on $\{N_0N_1>0\}$,
\begin{equation}\label{eq:dynamic-ate-exact}
\begin{aligned}
\sqrt n(\widehat{\tau}_{\mathrm{dyn}}-\tau)
&=\frac{\sqrt n}{N_1}\sum_{i=1}^nT_iA_i
-\frac{\sqrt n}{N_0}\sum_{i=1}^n(1-T_i)B_i\\
&=\frac{\xi_n+J_n}{1-(D_n/n)^2}.
\end{aligned}
\end{equation}
Over all $n$ main stage observations, reuse the notation
$$S_{n,\varepsilon}:=\frac2{\sqrt n}\sum_{i=1}^n
\{T_i\varepsilon_{i1}-(1-T_i)\varepsilon_{i0}\},
\quad
S_{n,g}:=\frac1{\sqrt n}\sum_{i=1}^ng(X_i).$$
The decomposition of the centered outcomes in \eqref{eq:two-stage-leading-decomposition} gives
$$
\xi_n=S_{n,\varepsilon}+S_{n,g}
+\frac2{\sqrt n}\sum_{k=1}^K\Delta_k(m)
=S_{n,\varepsilon}+S_{n,g}+o_p(1),
$$
where the last equality uses the bound in \eqref{eq:dynamic-oracle-imbalance-small} justified above.

Later stages use scores fitted from earlier outcomes, so we apply the conditioning argument for the residuals from Theorem~\ref{theo:clt for tau_m} separately within each stage. For stage $k$, set
$$
\mathcal G_{k,n}
:=\sigma\!\left(\mathcal H_{k-1},\{X_i,T_i:i\in\mathcal I_k\}\right).
$$
The sample of potential outcomes and covariates in stage $k$ is i.i.d. and independent of $\mathcal H_{k-1}$.
Similarly, conditional on $\mathcal G_{k,n}$, the residual pairs are independent across units, with means zero and variances $v_a(X_i)$.
Proposition~\ref{prop:dynamic-stagewise-l2-rate} and Assumption~\ref{assum:dynamic-feature-regularity} supply the moment and noncollapse conditions on the score.
Consequently, the arguments for count imbalance and centered arm averages used to control $J_n$ in Step~1 of the proof of Theorem~\ref{theo:clt for tau_m} apply within each stage, with $\sigma(\mathcal D_{n_0})$, $\mathcal G_{n_0,n}$, and $n$ replaced by $\mathcal H_{k-1}$, $\mathcal G_{k,n}$, and $n_k$, respectively. 
They first give conditional bounds given $\mathcal H_{k-1}$,
Lemmas~\ref{lem:conditional-small-to-unconditional} and~\ref{lem:conditional-big-to-unconditional} then yield
$$
\frac{\Delta_k(1)}{\sqrt{n_k}}=O_p(1),\qquad
\frac1{n_k}\sum_{i\in\mathcal I_k}T_iA_i=o_p(1),\qquad
\frac1{n_k}\sum_{i\in\mathcal I_k}(1-T_i)B_i=o_p(1).
$$
Since $K$ is fixed and $0\le q_{k,n}\le1$, these bounds imply
$$
\begin{aligned}
\frac{D_n}{\sqrt n}
&=\sum_{k=1}^K\sqrt{q_{k,n}}\frac{\Delta_k(1)}{\sqrt{n_k}}
=O_p(1),\\
\frac1n\sum_{i=1}^n\{T_iA_i+(1-T_i)B_i\}
&=\sum_{k=1}^Kq_{k,n}\frac1{n_k}
\sum_{i\in\mathcal I_k}\{T_iA_i+(1-T_i)B_i\}
=o_p(1).
\end{aligned}
$$
Hence $J_n=o_p(1)$ and $D_n/n=o_p(1)$. Also, $N_a/n=\{1+(2a-1)D_n/n\}/2\xrightarrow{p}1/2$, the denominator factor in \eqref{eq:dynamic-ate-exact} is now $1+o_p(1)$. 
Next, we will combine the $S_{n,\varepsilon}$ and $S_{n,g}$ to show $\xi_n=O_p(1)$ like in the proof of Theorem~\ref{theo:clt for tau_m}.

\medskip\noindent\textit{Step 4: Conditional characteristic functions within stages.}
Define$$
\begin{aligned}
Z_{k,n}^{\varepsilon}
&:=\frac2{\sqrt{n_k}}\sum_{i\in\mathcal I_k}
\{T_i\varepsilon_{i1}-(1-T_i)\varepsilon_{i0}\},\\
Z_{k,n}^{g}
&:=\frac1{\sqrt{n_k}}\sum_{i\in\mathcal I_k}g(X_i),
\quad Z_{k,n}:=Z_{k,n}^{\varepsilon}+Z_{k,n}^{g}.
\end{aligned}$$
These stage sums are $\mathcal H_k$-measurable, and
$S_{n,\varepsilon}+S_{n,g}=\sum_k\sqrt{q_{k,n}}Z_{k,n}$.
Similar with the Step~2 in the proof of Theorem~\ref{theo:clt for tau_m}, for every fixed $u\in\mathbb R$, we can have 
$$E\!\left[
\left|E(e^{iuZ_{k,n}^{\varepsilon}}\mid\mathcal G_{k,n})
-e^{-u^2V_\varepsilon/2}\right|
\middle|\mathcal H_{k-1}\right]=o_p(1),
\quad V_\varepsilon:=2E\{v_1(X)+v_0(X)\}.$$
The variables $g(X_i)$ in each stage are i.i.d., centered, independent of $\mathcal H_{k-1}$, and have finite variance. Since $Z_{k,n}^{g}$ is $\mathcal G_{k,n}$-measurable, similar with the Step~3 in the proof of Theorem~\ref{theo:clt for tau_m} leading to \eqref{eq:two-stage-leading-conditional-cf}, we can get
\begin{equation}\label{eq:dynamic-stage-conditional-cf}
\begin{aligned}
\psi_{k,n}(u)&:=E(e^{iuZ_{k,n}}\mid\mathcal H_{k-1})\\
&=e^{-u^2V_\varepsilon/2}E(e^{iuZ_{k,n}^{g}}\mid\mathcal H_{k-1})+o_p(1)\\
&\xrightarrow{p}e^{-u^2V^*/2},
\end{aligned}
\end{equation}
where the last step uses the ordinary CLT for $Z_{k,n}^{g}$ and $V^*=V_\varepsilon+E\{g^2(X)\}>0$.

\medskip\noindent\textit{Step 5: Allowing for varying stage weights.}
The contribution of stage $k$ to the full sum is $\sqrt{q_{k,n}}Z_{k,n}$, whose conditional characteristic function at $u$ is $\psi_{k,n}(u\sqrt{q_{k,n}})$. This argument can vary with the sample sizes even for fixed $u$, so we first extend the pointwise limit in \eqref{eq:dynamic-stage-conditional-cf} to uniform convergence in probability on each fixed compact interval.

We first bound the conditional second moment of $Z_{k,n}$ to control how much $\psi_{k,n}$ changes between nearby arguments. 
Since $E(Z_{k,n}^{\varepsilon}\mid\mathcal G_{k,n})=0$ and $Z_{k,n}^{g}$ is $\mathcal G_{k,n}$-measurable, the tower property gives
$$
E(Z_{k,n}^{\varepsilon}Z_{k,n}^{g}\mid\mathcal H_{k-1})
=E\!\left[Z_{k,n}^{g}E(Z_{k,n}^{\varepsilon}\mid\mathcal G_{k,n})
\mid\mathcal H_{k-1}\right]=0.
$$
Using conditional independence of the residuals and the i.i.d.\ covariate terms, we have
$$\begin{aligned}
E(Z_{k,n}^2\mid\mathcal H_{k-1})
&=\frac4{n_k}E\!\left[\sum_{i\in\mathcal I_k}
\{T_iv_1(X_i)+(1-T_i)v_0(X_i)\}\middle|\mathcal H_{k-1}\right]
+E\{g^2(X)\}\\
&\le4E\{v_1(X)+v_0(X)\}+E\{g^2(X)\}=:C_0<\infty.
\end{aligned}$$
For all real $s,t$, the inequality $|e^{ix}-e^{iy}|\le|x-y|$ and conditional Cauchy--Schwarz inequality give
$$|\psi_{k,n}(s)-\psi_{k,n}(t)|
\le |s-t|E(|Z_{k,n}|\mid\mathcal H_{k-1})
\le\sqrt{C_0}|s-t|.$$

Write $\varphi(t):=e^{-t^2V^*/2}$. Fix $U>0$ and choose a finite deterministic grid $\{t_j\}_{j=1}^J\subset[-U,U]$ such that every $t$ in the interval is within $d$ of a grid point $t_j$. Since $|\varphi'(t)|\le UV^*$ on this interval, then we have
$$
\begin{aligned}
|\psi_{k,n}(t)-\varphi(t)|
&\le |\psi_{k,n}(t)-\psi_{k,n}(t_j)|
+|\psi_{k,n}(t_j)-\varphi(t_j)|
+|\varphi(t_j)-\varphi(t)|\\
&\le\max_{1\le j\le J}|\psi_{k,n}(t_j)-\varphi(t_j)|
+(\sqrt{C_0}+UV^*)d.
\end{aligned}
$$
For any $\eta>0$, choose $d$ so that $(\sqrt{C_0}+UV^*)d<\eta/2$ and then hold the grid fixed. Taking the supremum over $t$, a finite union bound and the pointwise limits in \eqref{eq:dynamic-stage-conditional-cf} give
$$
P\!\left(\sup_{|t|\le U}|\psi_{k,n}(t)-\varphi(t)|>\eta\right)
\le\sum_{j=1}^J
P\!\left(|\psi_{k,n}(t_j)-\varphi(t_j)|>\eta/2\right)
\longrightarrow0.
$$
For fixed $u$, choose $U\ge\max(1,|u|)$. Since $0\le q_{k,n}\le1$, the argument $u\sqrt{q_{k,n}}$ remains in this interval, and hence
$R_{k,n}(u):=\left|
\psi_{k,n}(u\sqrt{q_{k,n}})-e^{-u^2q_{k,n}V^*/2}
\right|\xrightarrow{p}0.$
Since $0\le R_{k,n}(u)\le2$, for every $\eta>0$, let the sample sizes tend to infinity and then $\eta\downarrow0$, we have
\begin{equation}\label{eq:dynamic-stage-mean-error}
e_{k,n}(u):=E R_{k,n}(u)
\le\eta+2P\{R_{k,n}(u)>\eta\}\to0.
\end{equation}

\medskip\noindent\textit{Step 6: Aggregation and the ATE limit.}
The stage contributions need not be independent, because later scores are fitted using earlier outcomes. We combine them by conditioning on the preceding histories. For $0\le\ell\le K$, let
$$\Psi_{\ell,n}(u):=E\exp\left(iu\sum_{k=1}^{\ell}\sqrt{q_{k,n}}Z_{k,n}\right),
\qquad\Psi_{0,n}(u)=1.$$
Tower property gives
$$
\begin{aligned}
\Psi_{\ell,n}(u)
&=E\!\left[
\exp\left(iu\sum_{k<\ell}\sqrt{q_{k,n}}Z_{k,n}\right)
e^{iu\sqrt{q_{\ell,n}}Z_{\ell,n}}\right]\\
&=E\!\left[E\!\left\{
\exp\left(iu\sum_{k<\ell}\sqrt{q_{k,n}}Z_{k,n}\right)
e^{iu\sqrt{q_{\ell,n}}Z_{\ell,n}}
\middle|\mathcal H_{\ell-1}\right\}\right]\\
&=E\!\left[
\exp\left(iu\sum_{k<\ell}\sqrt{q_{k,n}}Z_{k,n}\right)
E\!\left(e^{iu\sqrt{q_{\ell,n}}Z_{\ell,n}}
\mid\mathcal H_{\ell-1}\right)\right]\\
&=E\!\left[
\exp\left(iu\sum_{k<\ell}\sqrt{q_{k,n}}Z_{k,n}\right)
\psi_{\ell,n}(u\sqrt{q_{\ell,n}})\right].
\end{aligned}
$$
The second equality is the tower property; the third takes the $\mathcal H_{\ell-1}$-measurable factor outside the inner conditional expectation, and the last uses the definition of $\psi_{\ell,n}$.
Since the Gaussian factor $e^{-u^2q_{\ell,n}V^*/2}$ is deterministic, the definition of $\Psi_{\ell-1,n}(u)$ gives
$$e^{-u^2q_{\ell,n}V^*/2}\Psi_{\ell-1,n}(u)
=E\!\left[
\exp\left(iu\sum_{k<\ell}\sqrt{q_{k,n}}Z_{k,n}\right)
e^{-u^2q_{\ell,n}V^*/2}\right].$$
Subtracting the preceding identities, we have
$$\begin{aligned}
&\left|\Psi_{\ell,n}(u)
-e^{-u^2q_{\ell,n}V^*/2}\Psi_{\ell-1,n}(u)\right|\\
=&\left|E\!\left[
\exp\left(iu\sum_{k<\ell}\sqrt{q_{k,n}}Z_{k,n}\right)
\left\{\psi_{\ell,n}(u\sqrt{q_{\ell,n}})
-e^{-u^2q_{\ell,n}V^*/2}\right\}\right]\right|\\
\le& E\!\left[
\left|\exp\left(iu\sum_{k<\ell}\sqrt{q_{k,n}}Z_{k,n}\right)\right|
\left|\psi_{\ell,n}(u\sqrt{q_{\ell,n}})
-e^{-u^2q_{\ell,n}V^*/2}\right|\right]\\
=&E\!\left|
\psi_{\ell,n}(u\sqrt{q_{\ell,n}})
-e^{-u^2q_{\ell,n}V^*/2}\right|
=e_{\ell,n}(u).
\end{aligned}$$
Subtract and add
$e^{-u^2q_{\ell,n}V^*/2}\Psi_{\ell-1,n}(u)$. The triangle inequality gives
\begin{equation}
\label{eq:dynamic-cf-recursion}
\begin{aligned}
&\left|\Psi_{\ell,n}(u)-\prod_{k=1}^{\ell}e^{-u^2q_{k,n}V^*/2}\right|\\
\le&
\left|\Psi_{\ell,n}(u)-e^{-u^2q_{\ell,n}V^*/2}\Psi_{\ell-1,n}(u)\right|+e^{-u^2q_{\ell,n}V^*/2}
\left|\Psi_{\ell-1,n}(u)-\prod_{k=1}^{\ell-1}e^{-u^2q_{k,n}V^*/2}\right|\\
\le& e_{\ell,n}(u)+e^{-u^2q_{\ell,n}V^*/2}
\left|\Psi_{\ell-1,n}(u)-\prod_{k=1}^{\ell-1}e^{-u^2q_{k,n}V^*/2}\right|.
\end{aligned}
\end{equation}
Since $\Psi_{0,n}(u)=1$, use \eqref{eq:dynamic-cf-recursion} to do iteration, the first two steps are
$$
\begin{aligned}
\left|\Psi_{1,n}(u)-e^{-u^2q_{1,n}V^*/2}\right|
&\le e_{1,n}(u),\\
\left|\Psi_{2,n}(u)-\prod_{k=1}^2e^{-u^2q_{k,n}V^*/2}\right|
&\le e_{2,n}(u)+e^{-u^2q_{2,n}V^*/2}e_{1,n}(u).
\end{aligned}
$$
At each subsequent step, the preceding errors are multiplied by the next Gaussian factor and the new error is added. Continuing the recursion in \eqref{eq:dynamic-cf-recursion} to $\ell=K$ therefore yields
$$
\begin{aligned}
\left|\Psi_{K,n}(u)-\prod_{k=1}^Ke^{-u^2q_{k,n}V^*/2}\right|
&\le\sum_{\ell=1}^K e_{\ell,n}(u)
\prod_{j=\ell+1}^Ke^{-u^2q_{j,n}V^*/2}\\
&\le\sum_{\ell=1}^K e_{\ell,n}(u)\longrightarrow0.
\end{aligned}
$$
The second inequality uses that every Gaussian factor lies in $(0,1]$, and the convergence uses fixed $K$. Since $\sum_kq_{k,n}=1$, the characteristic function continuity theorem gives
$$
S_{n,\varepsilon}+S_{n,g}
=\sum_{k=1}^K\sqrt{q_{k,n}}Z_{k,n}
\xrightarrow{d}N(0,V^*).
$$
This limit supplies the tightness needed for the denominator reduction. Together with $\xi_n=S_{n,\varepsilon}+S_{n,g}+o_p(1)$, $J_n=o_p(1)$, and $D_n/n=o_p(1)$, the exact ATE identity \eqref{eq:dynamic-ate-exact} now gives
$$
\sqrt n(\widehat{\tau}_{\mathrm{dyn}}-\tau)
=\frac{\xi_n+J_n}{1-(D_n/n)^2}
=S_{n,\varepsilon}+S_{n,g}+o_p(1)
\xrightarrow{d}N(0,V^*)
$$
by Slutsky's theorem, with the same oracle variance as in Theorem~\ref{theo:clt for tau_m}.
\end{proof}

\section{Proofs for Appendix~\ref{app:pfn-pseudo-theory}}
\label{app:pseudo-target-proofs}

\subsection{Proof of Theorem~\ref{theo:pseudo-target-ate}}
\begin{proof}
\setlength{\jot}{2pt}
Fix an arbitrary deterministic joint sequence $n_0,n\to\infty$.
All $L_2(P_X)$ norms below integrate over $X$ with the first stage data held fixed.
Assumption~\ref{assum:pseudo-target-learning} and
Lemma~\ref{lem:pilot_arm_means} give
\begin{equation}\label{eq:app-pseudo-score-error}
\|\widehat{m}_{n_0}-m_{n_0}^\dagger\|_{L_2(P_X)}
=o_p(1).
\end{equation}
The balance of the learned score follows from
Theorem~\ref{theorem:balancing property}(i).
Repeating the proof of part~(ii), with $m$ replaced by
$m_{n_0}^\dagger$, yields
\[
\frac{\Delta_n(m_{n_0}^\dagger)}{\sqrt n}
=\frac{\Delta_n(\widehat{m}_{n_0})}{\sqrt n}
+o_p(1)=o_p(1).
\]
This proves \eqref{eq:pseudo-score-balance}.

For the ATE limit, it suffices to compare the variance in
Theorem~\ref{theo:general-score-ate} with that in
\eqref{eq:pseudo-ate-clt}.
Proposition~\ref{prop:hatm-moment} and
Assumption~\ref{assum:score-noncollapse} give
$\|\widehat{m}_{n_0}\|_{L_2(P_X)}=O_p(1)$ and
$\{E[|\widehat{m}_{n_0}(X)|\mid\mathcal D_{n_0}]\}^{-1}=O_p(1)$.
Since $m-b_{n_0}^\dagger=m_{n_0}^\dagger$, the coefficient
definitions imply,
\[
\begin{aligned}
&d_{\widehat{m}_{n_0}}(m)
-d_{\widehat{m}_{n_0}}(b_{n_0}^\dagger)-1\\
&\quad=
\frac{E[\{m_{n_0}^\dagger(X)-\widehat{m}_{n_0}(X)\}
\operatorname{sgn}\{\widehat{m}_{n_0}(X)\}\mid\mathcal D_{n_0}]}
{E[|\widehat{m}_{n_0}(X)|\mid\mathcal D_{n_0}]}
=o_p(1).
\end{aligned}
\]
Together with \eqref{eq:app-pseudo-score-error}, this gives
\[
\left\|
m-d_{\widehat{m}_{n_0}}(m)\widehat{m}_{n_0}
-\{b_{n_0}^\dagger
-d_{\widehat{m}_{n_0}}(b_{n_0}^\dagger)\widehat{m}_{n_0}\}
\right\|_{L_2(P_X)}
=o_p(1).
\]
The two residual norms are $O_p(1)$ by
\eqref{eq:general-variance-bounds} and the triangle inequality,
so their squared norms differ by $o_p(1)$.
Hence the two variances differ by $o_p(1)$.
Both are bounded below by $V^*>0$; therefore
Theorem~\ref{theo:general-score-ate} and Slutsky's theorem
give \eqref{eq:pseudo-ate-clt}.
\end{proof}

\subsection{Proof of Corollary~\ref{cor:pseudo-target-variance}}
\begin{proof}
The first assertion follows from the variance comparison established
in the proof of Theorem~\ref{theo:pseudo-target-ate}.
By Theorem~\ref{theo:general-score-inference},
$\widehat{V}_{n_0,n}-V_{n_0}=o_p(1)$.
Combining these two conclusions proves the second assertion.
\end{proof}

\subsection{Proof of Proposition~\ref{prop:pfn-pseudo-target}}
\begin{proof}
Fix $a\in\{0,1\}$ and $n_0$, recall that
$$\widehat{\mu}_{n_0,a}(x)=f_{\widehat{\theta},a}(x;\mathcal D_{n_0}),
\qquad
\mu_{n_0,a}^\dagger(x)=E_{\mathcal D_{n_0}}\{f_{\widehat{\theta},a}(x;\mathcal D_{n_0})\}.$$
Assumption~\ref{assum:pfn-replacement-stability} and Tonelli's theorem make the latter a deterministic measurable function that is well defined for $P_X$-almost every $x$; set it to zero on the remaining null set.
Jensen's inequality gives
$$E\{(\mu_{n_0,a}^\dagger(X))^2\}
\le E\{f_{\widehat{\theta},a}^2(X;\mathcal D_{n_0})\}<\infty.$$

Let $\mathcal D_{n_0}^{\prime(j)}$ denote the first stage data obtained by independently replacing its $j$th observation, as in \eqref{eq:pfn-replacement-stability}.
For $P_X$-almost every fixed $x$, apply the Efron--Stein inequality for independent replacement \citep[Proposition~1]{BoucheronEtAl2005Moment} to $f_{\widehat{\theta},a}(x;\mathcal D_{n_0})$ as a function of the $n_0$ independent first stage observations.
Integrating over the independent query $X$ and using Tonelli's theorem and \eqref{eq:pfn-replacement-stability}, we obtain
\begin{align*}
&E\!\left[\{\widehat{\mu}_{n_0,a}(X)-\mu_{n_0,a}^\dagger(X)\}^2\right]\\
\le&\frac12\sum_{j=1}^{n_0}
 E\!\left[\{f_{\widehat{\theta},a}(X;\mathcal D_{n_0})
 -f_{\widehat{\theta},a}(X;\mathcal D_{n_0}^{\prime(j)})\}^2\right]
\le\frac{n_0}{2}C^2n_0^{-2\alpha}
=\frac{C^2}{2}n_0^{1-2\alpha}
\end{align*}
for all sufficiently large $n_0$. This proves \eqref{eq:pfn-efron-stein-bound}. By the tower property and Markov's inequality, for every $M>0$ and all sufficiently large $n_0$,
$$P\!\left(
 E\!\left[\{\widehat{\mu}_{n_0,a}(X)-\mu_{n_0,a}^\dagger(X)\}^2
 \mid\mathcal D_{n_0}\right]>M^2n_0^{1-2\alpha}
\right)\le\frac{C^2}{2M^2}.$$
Thus the conditional $L_2$ error is $O_p(n_0^{1/2-\alpha})$ for both arms.
Since $\alpha>1/2$, this verifies Assumption~\ref{assum:pseudo-target-learning} with $n_0^{1/2-\alpha}\to0$.
\end{proof}

\renewcommand{\refname}{References for the Supplementary Material}
\putbib
\end{bibunit}

\end{document}